\documentclass[longbibliography,showpacs,floatfix,superscriptaddress,twocolumn,aps,prl]{revtex4-2}

\usepackage{floatrow}
\usepackage[caption=false]{subfig}
\usepackage{graphicx,color}
\usepackage{amsfonts}
\usepackage[figuresright]{rotating}  
\usepackage{amssymb}
\usepackage{bm}
\usepackage{amsmath}
\usepackage{mathtools}
\usepackage{psfrag}
\usepackage{multirow}
\usepackage{units}
\usepackage{lipsum}
\usepackage{physics}
\usepackage{soul}
\usepackage{titlesec}
\usepackage{times}
\usepackage[colorlinks=true,citecolor=blue,linkcolor=magenta,hypertexnames=false]{hyperref}
\usepackage{enumitem}
\usepackage[normalem]{ulem} %for striking out text with \sout{} (useful for corrections)
\usepackage{cancel}
\usepackage[capitalise]{cleveref}
\usepackage[dvipsnames]{xcolor}
\usepackage[compat=1.1.0]{tikz-feynman}
\usepackage{tikz}
\usepackage{feynmp-auto}
\usepackage{tikz}
\usepackage{lipsum}
\usepackage{comment}

\usetikzlibrary{decorations.pathreplacing,decorations.markings,calc,intersections}
\newcommand {\beq} {\begin{equation}}
		\newcommand {\eeq} {\end{equation}}
	\newcommand {\bqa} {\begin{eqnarray}}
		\newcommand {\eqa} {\end{eqnarray}}
	\newcommand {\bseq} {\begin{subequations}}
		\newcommand {\eseq} {\end{subequations}}
	\newcommand {\baln} {\begin{align}}
		\newcommand {\ealn} {\end{align}}

	\newcommand{\tcav}{T_{\mathrm{cav}}}
        \newcommand{\tcry}{T_{\mathrm{cry}}}
        \newcommand{\tc}{T^{\mathrm{c}}_{\mathrm{cry}}}
        \newcommand{\tcz}{T^{\mathrm{c}0}_{\mathrm{cry}}}
        \newcommand{\tcu}{T^{\mathrm{u}}_{\mathrm{cry}}}
        \newcommand{\tcl}{T^{\mathrm{l}}_{\mathrm{cry}}}
        \newcommand{\tcm}{T^{\mathrm{m}}_{\mathrm{cry}}}

\begin{document}

%\title{ nonthermal steady states from open cavity fermion systems}
%\title{Non-thermal cavity control of order in electronic systems}
\title{Cavity-induced Eliashberg effect: superconductivity vs charge density wave}
%\title{Multi-stage Eliashberg effect for charge density waves in cavities}

\author{Md Mursalin Islam}\email{md.islam@uni-a.de}
\affiliation{Theoretical Physics III, Center for Electronic Correlations and Magnetism,
Institute of Physics, University of Augsburg, 86135 Augsburg, Germany}
\affiliation{Max Planck Institute for the Physics of Complex Systems, Nöthnitzer Stra{\ss}e 38, 01187 Dresden, Germany}
\author{Michele Pini}
\affiliation{Theoretical Physics III, Center for Electronic Correlations and Magnetism,
Institute of Physics, University of Augsburg, 86135 Augsburg, Germany}
\affiliation{Max Planck Institute for the Physics of Complex Systems, Nöthnitzer Stra{\ss}e 38, 01187 Dresden, Germany}
\author{R. Flores-Calderón}
\affiliation{Max Planck Institute for the Physics of Complex Systems, Nöthnitzer Stra{\ss}e 38, 01187 Dresden, Germany}
\affiliation{Technical University of Munich, TUM School of Natural Sciences, Physics Department, 85748 Garching, Germany}
\affiliation{Munich Center for Quantum Science and Technology (MCQST), Schellingstr. 4, 80799 M{\"u}nchen, Germany}
\author{Francesco Piazza}
\affiliation{Theoretical Physics III, Center for Electronic Correlations and Magnetism,
Institute of Physics, University of Augsburg, 86135 Augsburg, Germany}
\affiliation{Max Planck Institute for the Physics of Complex Systems, Nöthnitzer Stra{\ss}e 38, 01187 Dresden, Germany}

\begin{abstract}
Recent experiments have shown that non-equilibrium effects can play a key role in cavity-based control of material phases, notably in systems with charge-density-wave order. Motivated by this, we extend the theory of the Eliashberg effect, originally developed for superconducting phases, to charge-density-wave phases. Starting from a minimal electronic model where superconductivity and charge-density-wave order are equivalent at equilibrium, we introduce coupling to cavity photons, which are in turn coupled to an environment at a temperature different from the one of the electronic environment. This drives the system into a non-thermal steady state, which breaks the equivalence between superconductivity and charge-density-wave order. In the superconducting case, we recover the known behavior: a shift from continuous to discontinuous phase transitions with bistability. In contrast, the charge-density-wave case displays richer behavior: tuning the cavity frequency induces both continuous and discontinuous transitions, two distinct ordered phases, and a bistable regime ending at a critical point. These findings demonstrate that the scope of cavity-based non-thermal control of quantum materials is broader than at thermal equilibrium, and strongly depends on the targeted phases.
\end{abstract}
\maketitle

%\fpc{General comment : I suggest ti replace "Normal phase" with "Metallic phase" everywhere.}\mic{Done.}

%\fpc{General comment 2: when referring to further details of the calculations, let us try to cite the EndMatter, which in turn should cite the SupplementalMaterial. The latter should be cited direcltly from the main text only if unavoidable.}\mic{Done.}

\begin{figure}[t]
    \centering
    \includegraphics[width=\textwidth]{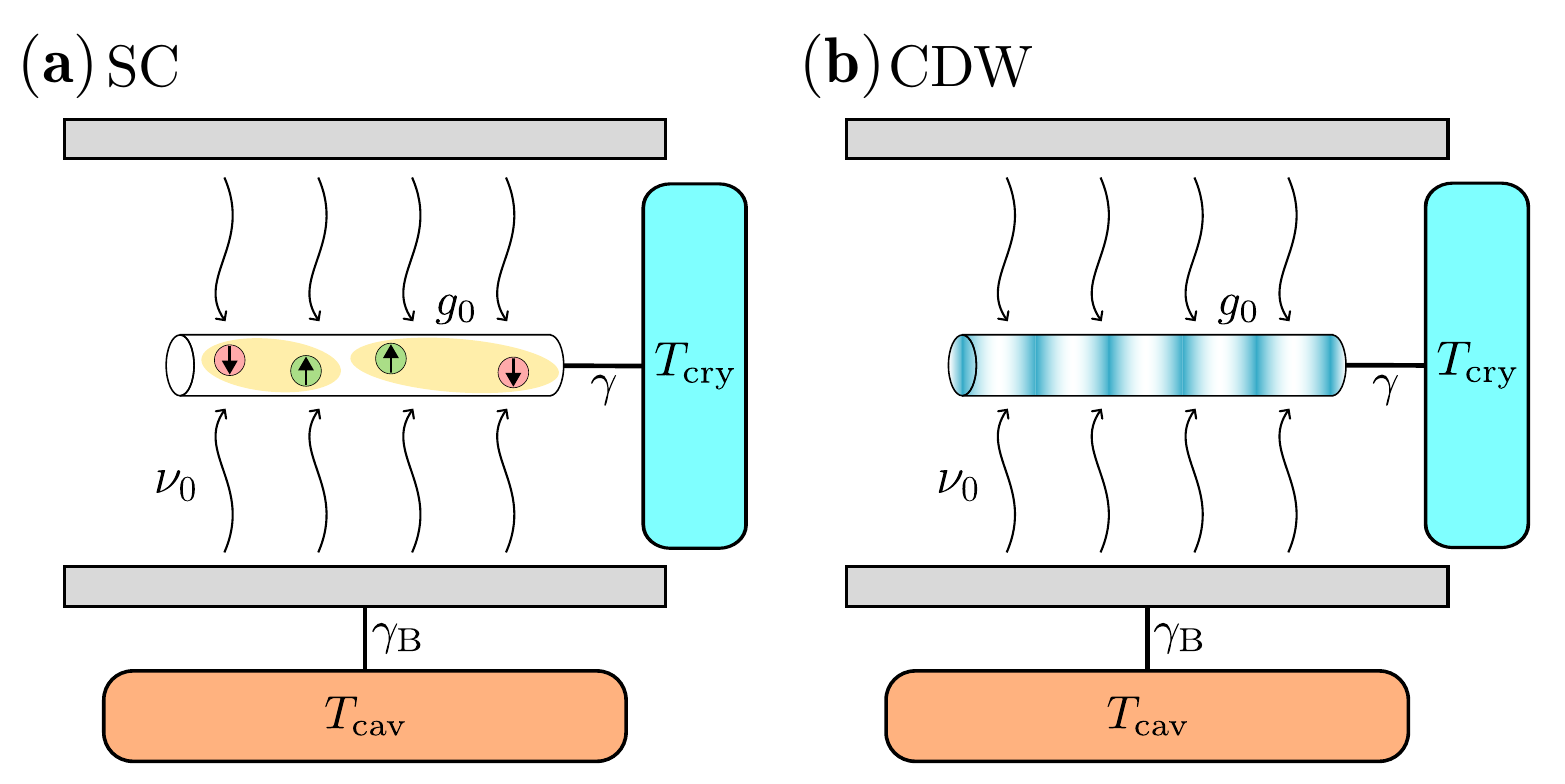}
    \caption{Schematics of the setup: a quasi-1D system of electrons featuring (a) superconducting (SC) or (b) 
 charge-density-wave (CDW) order
is placed inside a Fabry-Pérot cavity, along a direction parallel to the cavity mirrors. The cavity has a fundamental frequency of $\nu_0$. The electron current couples to the cavity photons with strength $g_0$.  Furthermore, the electrons are coupled to a cryostat with temperature $\tcry$, which induces a finite electron damping $\gamma$. The cavity photons are also coupled via the mirrors to the electromagnetic (EM) continuum outside, the latter is set at a temperature $\tcav$ and also induces a finite photon damping $\gamma_B$.}
    \label{fig:setup}
\end{figure}

\begin{figure}[t]
    \centering
    \includegraphics[width=\textwidth]{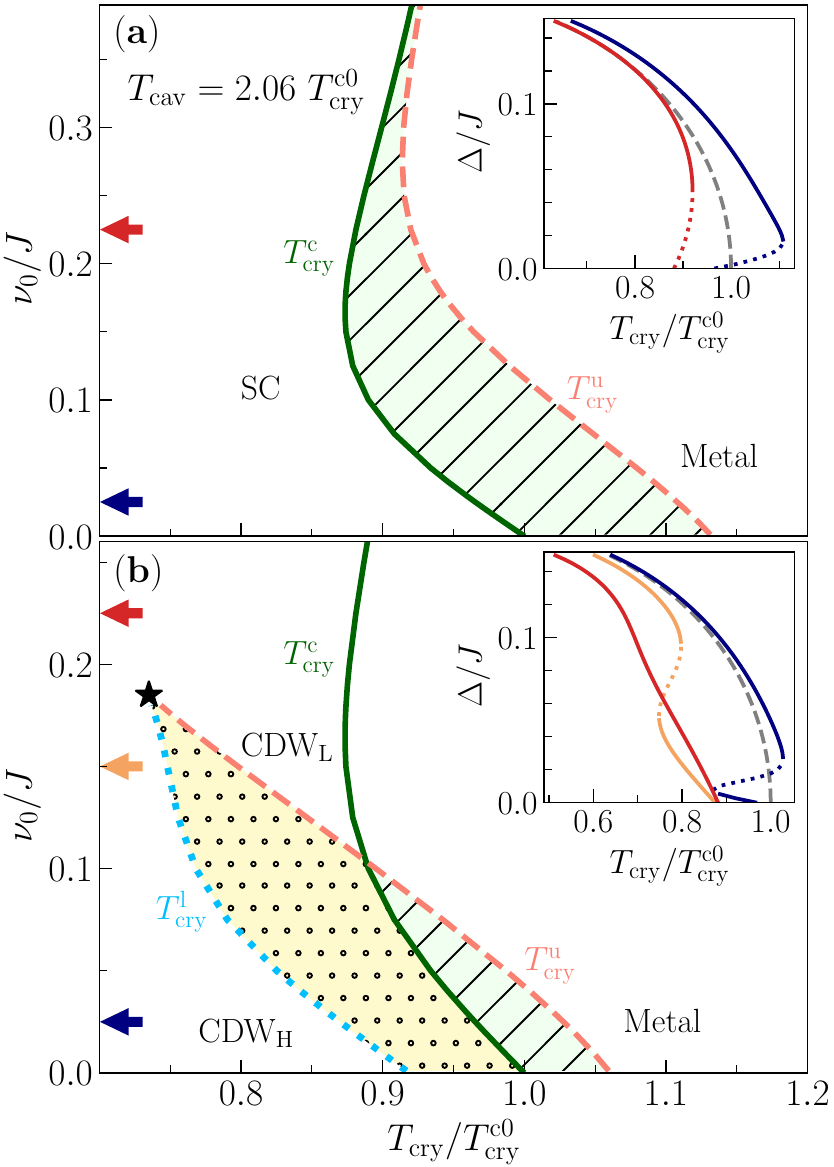}
    \caption{NESS phase diagram in the $\nu_0$ - $\tcry$ plane for the SC (a) and the CDW (b) case. 
    %\fps{ inside cavity on} 
    We set $\tcav=2.06\,\tcz$, $\kappa=0.5\,J$, $g_0=0.1\,J$, $\gamma=0.01\,J$ and $\gamma_B=0.5\,\nu_0$. 
    The shaded areas indicate phase-coexistence regions and the black star a critical point.
    %\fps{(a) SC: On the right of the dashed light-red line ($\tcry>\tcu$), we have normal metallic phase (Normal) with $\Delta=0$. On the left of the solid green line ($\tcry<\tc$), there is ordered superconducting phase (SC) with finite $\Delta$. In between these two lines in the striped light-green region ($\tc<\tcry<\tcu$), SC and Normal phases coexist and it indicates a first-order transition. (b) CDW: On the right-side blank region [$\tcry>\text{max}(\tc,\tcu)$], we have Normal phase with $\Delta=0$. On the left of the dotted light-blue line ($\tcry<\tcl$) we have order phase with higher value of $\Delta$ ($\mathrm{CDW}_{\mathrm{H}}$). In the blank area between dashed light-red line and solid green line ($\tcu<\tcry<\tc$), we have another ordered phase with lower value of $\Delta$ ($\mathrm{CDW}_{\mathrm{L}}$). In the striped light-green region Normal phase and $\mathrm{CDW}_{\mathrm{H}}$ coexist. In the dotted yellow region $\mathrm{CDW}_{\mathrm{L}}$ and $\mathrm{CDW}_{\mathrm{H}}$ coexist. This regions ends in a critical point indicate by a black star.} 
   %\fpc{All explained in the text: in a letter, no repetition is necessary.}
    Insets: $\Delta$ vs $\tcry$ for different values of $\nu_0$ which are indicated by arrows of the corresponding colors in the main panel. 
    %\fps{The curves touch zero at $\tcry=\tc$, have a maxima in temperature (if exists) at $\tcry=\tcu$ and a minima in temperature (if exists) at $\tcry=\tcl$.} 
    The solid (dotted) part of the curves represents a stable (unstable) solution of the gap equation.
    The dashed gray equilibrium curve is added for reference. 
    }
    \label{fig:phasedia}
\end{figure}

%\section{Introduction}
%\fpm{
%Out of equilibrium behavior breaks the thermal constraints and allows for new physics.
%In the solid state physics in particular, 
\textit{Introduction---} Excitation of a material out of equilibrium via laser light has allowed to observe collective phenomena which would have been otherwise prohibited in thermal equilibrium at the given external conditions \cite{de2021colloquium}. 
A substantial portion of the work done so far has dealt with light-assisted superconductivity, also due to its technological relevance. This research direction is currently receiving major attention, due to recent experiments observing superconducting behavior close to ambient conditions after applying a laser pulse which excites selected phonon modes that mediate pairing between electrons \cite{doi:10.1126/science.1197294,PhysRevB.90.100503,Hu2014,Mitrano2016,Cantaluppi2018,Budden2021,Isoyama2021}.
%\fpc{@Mursalin: can you cite Refs. 9,10,11,12,13,14,15 from this paper: https://arxiv.org/pdf/2502.20276}. 
A different route to the control of superconductivity, implemented experimentally much earlier, is the one that goes under the name of Eliashberg effect (EE) \cite{Eliashberg1970,*Eliashbergrus,Ivlev1973, PhysRevLett.16.1166,PhysRev.155.419,Klapwijk1977}.
%\fpc{Please add also the experimental work}. 
The latter involves a microwave pulse which directly excites the electrons into a non-equilibrium distribution that favors a larger superconducting gap.
\\
A drawback of both the previous routes is that light pulses are used and thus the interesting non-equilibrium behavior is restricted to a transient. This issue can be solved by confining the light around the material using for instance the two mirrors of a cavity, so that the impact of a single photon is enhanced by reducing the volume of the electromagnetic (EM) modes, which become also separated in frequency \cite{Garcia-Vidal:2021, Mivehvar:2021, Schlawin:2022, Bloch:2022}. Experiments have demonstrated that the material's properties can be altered without laser driving but confining the light \cite{appugliese2022breakdown,Thomas:2021,Jarc2023,Thomas:2019,keren2025cavityalteredsuperconductivity}, so that only vacuum fluctuations and thermal excitations of photons are present. 
\\
Such cavity-material platforms also allow to use light to take the matter out of equilibrium and thus to escape thermal constraints. Since intense light pulses are not needed here, non-thermal stationary states instead of transients can be achieved for an extended cavity-based control of materials \cite{Rafa2025}. The EE mentioned above has also theoretically been revisited for cavity-based non-thermal control of superconductors in the stationary state \cite{Galitski2019}. 
\\
More recently, the first experimental evidence of the crucial role that non-equilibrium effects can play in the cavity-based control of a material's phase has appeared \cite{Jarc2023,fassioli2025controlling,chiriaco2024thermal}. This has however not been observed for a superconducting (SC) phase, but  for another type of ordered phase, where the system spontaneously breaks translation-invariance by developing a modulation of the electrons' density -- a so-called charge-density-wave (CDW) phase. This immediately calls for an extension of the theory of the EE from SC to CDW order.
\\
In this work, we undertake this step. We consider a simple model of electrons, for which SC and CDW order are equivalent at thermal equilibrium.
%, that is, the gap equation for the order parameter is the same in both cases
We then add the coupling between electrons and photons, whereby the latter are confined within a cavity and coupled to the EM modes outside via the cavity mirrors (see Fig.~\ref{fig:setup}). This EM environment has a temperature which is assumed to be different from the one of the cryogenic bath coupled to the electrons. The latter are thus in general found in a non-thermal steady state \cite{Rafa2025}. We find that this breaks the %\mpc{Not sure if fluctuations at thermal eq can break it. I would anyway avoid to specify "mean-field" here.}\fpc{I think we can leave the discussion as it is in the intro since the level is quite superficial and thus such addition would confuse more than anything else.} 
equivalence between SC and CDW order and leads to a different phenomenology of the EE. For SC, we find the standard EE phenomenology \cite{Eliashberg1970,*Eliashbergrus,Ivlev1973,Galitski2019} for any finite cavity resonance frequency, leading to a change from a continuous to a discontinuous transition featuring bistable behavior [see Fig.~\ref{fig:phasedia}(a)]. For CDW instead, the EE leads to a much richer 
phenomenology: By tuning the cavity resonance frequency (via mirror separation) one can switch between three qualitatively different behaviors as a function of temperature [see Fig.~\ref{fig:phasedia}(b)]. Both continuous and discontinuous transitions are found, and two distinct CDW phases appear, sharing their own additional bistable region which ends at a critical point.
\\
Our findings showcase how cavity-based non-thermal control of materials strongly depends on the targeted phase and can be much larger in scope than its equilibrium counterpart, enabling to switch between qualitatively different behaviors by tuning the cavity frequency.
%}

%\section{Model}

\begin{table}[b]
\centering
\begin{tabular}{ |c||c|c| } 
 \hline
 & SC & CDW \\ 
 \hline
 $\hat{c}_+(k)~/~\hat{c}_-(k)$ & $\hat{c}_{\uparrow}(k)~/~\hat{c}^{\dagger}_{\downarrow}(-k)$ & $\hat{c}(k+k_F)~/~\hat{c}(k-k_F)$ \\ 
 \hline
 $\xi_k$ & $-J\cos(ka)$ & $J\sin(ka)$ \\
 \hline
 BZ & $\left[-\pi/a,\pi/a\right)$ & $\left[-\pi/2a,\pi/2a\right)$ \\
 \hline
 $g^{\pm}_{k,k'}$ & $g_0\sin\left(\frac{ka+k'a}{2}\right)$ & $\pm g_0\cos\left(\frac{ka+k'a}{2}\right)$ \\
 \hline
 $\eta$ & 1 & -1 \\
 \hline
\end{tabular}
\caption{Mapping of the Hamiltonian \cref{eq:Ham} for SC and CDW. Rows indicate (from top to bottom) the electron operators, the free electron dispersion, corresponding Brillouin zone (BZ), the light-matter (LM) coupling, and the parameter $\eta$ that is used below to differentiate between SC and CDW.
}
\label{tab:scncdw}
\end{table}

\textit{Model and approach---} We consider the setup sketched in \cref{fig:setup}
%\fps{, where the system of fermions is placed between the parallel mirrors of a Fabry-Pérot cavity. Additionally, the fermions are connected to a cryostat and the photons interact with the environment outside through the cavity mirrors.}
described by the following Hamiltonian ($\hbar$ and $k_B$ are set to unity throughout the manuscript)

\beq 
\begin{split}
    \label{eq:Ham}
    &H=H_{\mathrm{el}}+H_{\mathrm{ph}}+H_{\mathrm{el-ph}}+H_{\mathrm{el-bath}}+H_{\mathrm{ph-bath}}\\
    &H_{\mathrm{el}}=\sum_{k\in\rm BZ}
    \begin{bmatrix}
        \hat{c}^{\dagger}_{+}(k) & \hat{c}^{\dagger}_{-}(k)
    \end{bmatrix}
    \begin{bmatrix}
        \xi_k & \Delta \\
        \Delta^* & -\xi_k
    \end{bmatrix}
    \begin{bmatrix}
        \hat{c}_{+}(k) \\ \hat{c}_{-}(k)
    \end{bmatrix}+\frac{|\Delta|^2}{\kappa}\\
    &H_{\mathrm{ph}}=\sum_{q}\nu_q\left(\hat{a}^{\dagger}_q\hat{a}_q+\frac{1}{2}\right)\\
    %&H_{\mathrm{el-ph}}=\sum_{k,k'}\sum_{\alpha=\{+,-\}}g^{\alpha}_{k,k'}\hat{c}^{\dagger}_{\alpha}(k)\hat{c}_{\alpha}(k')\hat{\phi}(k-k').\\
    &H_{\mathrm{el-ph}}=\sum_{k,k'}\sum_{\alpha=\{+,-\}}g^{\alpha}_{k,k'}\frac{\left(\hat a_{k-k'}+\hat a^{\dagger}_{k'-k}\right)}{\sqrt{2\pi\nu_{k-k'}/\nu_0}}\hat{c}^{\dagger}_{\alpha}(k)\hat{c}_{\alpha}(k').\\
\end{split}
\eeq
Here, $H_{\mathrm{el}}$ is a mean-field Hamiltonian for the electrons, allowing for spontaneous symmetry breaking with order parameter $\Delta$ 
and $\kappa$ is the strength of the pairing coupling.
The electrons are divided into two Nambu components ($\hat{c}_+$ and $\hat{c}_-$) and are coupled through $\Delta$ and $\Delta^*$. In the rest of the paper, we will choose the gauge where $\Delta$ is real without loss of generality.
%\fps{that appears as the off-diagonal terms in $H_{\mathrm{el}}$.}
This simple model can be used to study both superconducting (SC) and charge-density-wave (CDW) transitions with the mapping described in \cref{tab:scncdw}. We consider electrons  hopping with strength $J$ along a 1D lattice with lattice constant $a$, corresponding to the free dispersion $\xi_k$ 
%\mis{in the Brillouin zone $\left[-\frac{\pi}{a},\frac{\pi}{a}\right)$} 
%\mpc{BZ is actually half for CDW. Add a line in the table.} 
with bandwidth $2J$, and we assume the band to be half-filled. This can model electrons in quasi-1D nanotubes \cite{nanotube:Hümmer2016,nanotube:Jeantet2016}. 
Apart from its simplicity, the main reason to choose this model is that, at thermal equilibrium, the SC and CDW order parameters satisfy the same equation \eqref{eq:gapeq0}. This allows us to clearly identify the asymmetric out-of-equilibrium behavior of the EE between SC and CDW. 
%\mpc{Simple model that treats CDW and SC on the same footing.} 
%\fps{However, we expect most of the results to be qualitatively similar in higher dimensions} %\mic{is it okay to make this statement?}\rafc{If we think of the assumptions we use, mean field on the $\Delta$ term, this should be okay close and above the upper critical dimension. I guess the momentum structure doesn't change much above the mean field.}\\
%\mpc{We discussed this and I remember that that is was not trivial to generalize to higher dimensions (the problem was more about the momentum structure of the interaction than about the validity of mean-field). If a referee asks, do we have good arguments to convince them about this?}. We also considered the free-electron band to be half-filled, so that the Fermi momentum is $k_F=\pi/2a$. Note that with this choice, the CDW becomes commensurate with the underlying lattice in the ordered phase.
%\mpc{We should mention that this choice makes the CDW commensurate with the lattice} 

In order to drive the system out of equilibrium, coupling to cavity photons is added. The latter are described by the Hamiltonian
$H_{\mathrm{ph}}$, where $\hat{a}_q$ is the photon annihilation operator, and $\nu_q=\sqrt{\nu_0^2+c^2q^2}$ is the dispersion, where the momentum $q$ is along the 1D tube, $c$ is the speed of light, and $\nu_0=\pi c/L$ is the cavity fundamental frequency that can be tuned using the distance between the cavity-mirrors $L$. 
The cavity photons couple to the electron current 
via $H_{\mathrm{el-ph}}$ \cite{Schlawin2019,Rao2023}.
The light-matter (LM) coupling $g^{\alpha}_{k,k'}$ 
(with strength $g_0$) 
depends on the electronic momenta and Nambu components, and it differs between SC and CDW, as tabulated in \cref{tab:scncdw}. 
The cavity photons are connected to the EM environment with temperature $\tcav$ through the mirrors. This coupling  is indicated by
\beq\label{eq:phbath}
\begin{split}
    H_{\rm ph-bath}= &\sum_{s,q} \left[\nu_s(q)\left(\hat{b}^\dagger_{sq}\hat{b}_{sq}+\dfrac{1}{2}\right) \right.\\&\left.+\frac{t^{\rm ph}_s(q)}{2\sqrt{\nu_{q} \nu_s(q)}}(\hat{a}^\dagger_{-q}+\hat{a}_{q})(\hat{b}^\dagger_{s-q}+\hat{b}_{sq})\right],
\end{split}
\eeq
where we consider an extensive number of modes $\hat b_{sq}$ of the external EM environment couple to each cavity photon mode. The bath induces a spectral-width $\gamma_{\mathrm{B}}$ for the photons. We consider 
%$\gamma_B>>g_0$,
it to be large compared to the LM coupling
so that the feedback on photons by the electrons can be neglected and the photons are always at thermal equilibrium at temperature $\tcav$. %\fpm{(see the End Matter and \cite{Rafa2025}).}
%\mic{What we should show in End Matter for this?}
%\mpc{At the moment it does not fit in the End Matter, and it was used but not shown explicitly in \cite{Rafa2025}. It is shown in an appendix of the manuscript I am writing with Christian (not published yet).}
The electrons are connected to a cryostat with temperature $\tcry$ via the Hamiltonian 
\beq\label{eq:elbath}
\begin{split}
   H_{\mathrm{el-bath}} = &\sum_{\alpha,k,s} \left\{\epsilon_{s\alpha}(k)\hat{f}^\dagger_{s\alpha k}\hat{f}_{s\alpha k}\right.\\&\left.+t_{s\alpha }(k)\left[\hat{c}^\dagger_{\alpha}(k)\hat{f}_{s\alpha k}+\hat{f}^\dagger_{s\alpha k}\hat c_{\alpha}(k)\right]\right\}.
\end{split}
\eeq
Here we consider an extensive number of modes $\hat{f}_{s\alpha k}$ for the cryostat, labeled by $s$ and coupled to each electron.
%momentum $k$ and Nambu component $\alpha$.
The cryostat provides spectral width $\gamma$ to the electrons \cite{Rafa2025}.

Without coupling to the photons ($g_0=0$), the electrons are at thermal equilibrium, and the phase transition is described by the 
gap-equation (see the End Matter and Refs.~\cite{agd,altlandsimons,tinkham2004}) 
\beq\label{eq:gapeq0}
\int_{\Delta}^{\infty} \! \! dE~ \Lambda(E) \,  \frac{1-2n_0(E)}{E}=\frac{1}{\kappa},\\
\eeq
where $n_0(\omega)=\frac{1}{\exp(\omega/T_{\mathrm{cry}})+1}$ is the Fermi-Dirac distribution function and $\Lambda(E)=\frac{2}{\pi}\frac{E~\Theta(E-\Delta)\Theta(\sqrt{J^2+\Delta^2}-E)}{\sqrt{E^2-\Delta^2}\sqrt{J^2+\Delta^2-E^2}}$ is the density of states (DOS) on a lattice for gaped quasi-particles. The gap equation \eqref{eq:gapeq0} is valid for both SC and CDW, which thus behave identically in this thermal equilibrium case, due to the particle-hole symmetry (see the End Matter). %\fpc{How to see this? Explain briefly or refer to the EndMatter}
%In both cases, below a certain temperature $\tcz$, the system goes to an ordered state with a finite order parameter $\Delta$. 
%Using Landau-Ginzberg theory, one can show that near the transition $\Delta\sim\sqrt{\tcz-\tcry}$. 

%\mpc{Maybe mention that the only effect of LM coupling considered is in $\delta n$ (kinetic equation) and we neglect retarded self energy.}
For finite LM coupling, the system in general achieves a non-equilibrium steady state (NESS), due to the presence of two separate baths. So, to describe it, we resort to the Schwinger-Keldysh field theory (SKFT) \cite{keldysh,*keldyshrus,schwinger,kamenev_book,rammer_book,DiehlReview2016}. Under a weak-coupling approximation and for well-defined fermionic quasi-particles (for which we also neglect the light-induced modification of the spectrum)
%\mpc{We use also the $\gamma$ small approximation, right? If yes, I would add "and for well-defined fermionic quasiparticles"} 
we arrive at the following out-of-equilibrium gap equation (see the End Matter)
%for a sketch of the derivation)}
%\fpc{As suggested above, let's try to have reference to the EndMatter, which in turn refers to the SuppMat}
\beq\label{eq:gapeq}
\begin{split}
&\int_{\Delta}^{\infty} \! \! dE~ \Lambda(E) \,  \frac{1-2[n_0(E)+\delta n(E)]}{E}=\frac{1}{\kappa}.
\end{split}
\eeq
As in the standard description of the EE \cite{Eliashberg1970,Ivlev1973},
the above approximations lead to the fact that the coupling to light does not modify the structure of the gap equation, but only the electron energy distribution $\delta n$ entering the latter. 
%\mpc{Should we justify this? Like saying that "as they do not change the overall qualitative picture" or referring to the original paper by Eliashberg?}  \rafc{I think as long as we assume the phase is well described by the mean-field fix point Hamiltonians that should be enough.}
The non-equilibrium modification of the distribution reads
\beq\label{eq:deln}
\delta n(E)=\frac{2\,g_0^2}{\gamma}\int_{\Delta}^{\infty} \! \!  dE'~ \Lambda(E')\left[\Gamma^{\eta}_{\Delta}(E,E')+\Gamma^{\eta}_{\Delta}(E,-E')\right],
\eeq
where 
\beq\label{eq:scat}
\Gamma^{\eta}_{\Delta}(E,E')=\left(1+\eta\frac{\Delta^2}{EE'}\right)\mathrm{Im}D^R_0(E-E')H_0(E,E')
\eeq
describes the electron-scattering processes mediated by photons with spectral function
\beq\label{eq:phsf}
\mathrm{Im}D^R_0(\omega)=\frac{1}{4\pi}\left[\frac{\gamma_B}{(\omega+\nu_0)^2+\gamma_B^2}-\frac{\gamma_B}{(\omega-\nu_0)^2+\gamma_B^2}\right].
\eeq
The scattering processes depend on both the electron and photon distributions, which are contained in the factor
\beq\label{eq:dist}
\begin{split}
   & H_0(\omega,\omega')\\&=\left[N_{T_{\mathrm{cav}}}(\omega-\omega')-N_{T_{\mathrm{cry}}}(\omega-\omega')\right]\left[n_0(\omega)-n_0(\omega')\right]
\end{split}
\eeq
that keeps track of the difference in temperature between the two baths. Here, $N_T(\omega)=\frac{1}{\exp(\omega/T)-1}$ is the Bose-Einstein distribution function at the temperature $T$.

For the photon spectral function [see \cref{eq:phsf}]
we have considered only zero momentum photons as their effective mass is much less than that of the electrons and hence finite momenta contributions are negligible (forward scattering approximation).
%The distribution factor \mpc{define $B_T(\omega)$? $N_T(\omega)$?} encodes the information about the distributions of the thermal baths. Note that $H_0$ and consequently $\delta n(E)$ vanishes when $\tcry=\tcav$. This is not surprising as in that case the system reaches thermal equilibrium. 
This would make the light-induced scattering negligible unless one includes the effect of elastic scattering on static impurities that are generally present in real materials. This is indeed always done in the description of the EE and allows for scattering processes where momentum is not conserved. In the simplest modeling, one can replace the momentum conserving delta function by a constant $(\Lambda_F/\tau_{\mathrm{el}})^{-1}$ where $\Lambda_F=2/\pi J$ is the DOS at the Fermi surface and $\tau_{\mathrm{el}}$ is the elastic scattering rate \cite{MattisBardeen1958,Galitski2019}. We further absorb the factor $(\Lambda_F/\tau_{\mathrm{el}})$ in $\gamma$. 
%For explicitly momentum dependent quantities \mim{(e.g.~the LM coupling)} one can use Fermi surface averaged values and the processes can be described in terms of energy transfers only \cite{Galitski2019}.
%Further, one uses Fermi surface averaged LM coupling and hence the processes can be described in terms of energy transfers only 
Moreover, one can average the LM coupling on the Fermi Surface and describe the processes in terms of energy transfers only \cite{Galitski2019}.

%\mpc{Move discussion after you defined all the elements of the equation.}
Note that within our description the whole difference between SC and CDW is condensed in a single sign $\eta=\pm 1$ (see  \cref{tab:scncdw}), which enters the non-equilibrium correction to the electron distribution via the scattering function $\Gamma_\Delta^\eta$.
The relevant scattering processes are best understood in the Bogoliubov rotated basis where the free electron Hamiltonian becomes diagonal and we have two bands separated by the gap. In this construction, $\Gamma^{\eta}_{\Delta}(E,E')$ corresponds to intraband processes, while $\Gamma^{\eta}_{\Delta}(E,-E')$ to interband processes. Note that for SC (CDW), the intraband (interband) processes have larger amplitudes than the interband (intraband) processes. The intraband processes enhance the gap, while interband processes always suppress the gap. We thus generically expect the EE to enhance the SC gap more than the CDW gap. Furthermore, for $\nu_0<\Delta$, the interband processes are heavily suppressed, while for $\nu_0>\Delta$ (but $\nu_0<2J$) they become stronger than intraband processes. Hence, the gap enhancement is expected for smaller cavity frequencies. 
%\mpc{Mention that depending on $\nu_0/\Delta$ some processes are activated/suppressed.}
%At this point it is instructive to describe the system in a Bogoliubov rotated basis 
%Here, the density of states for electrons on a lattice
%\beq\label{eq:dos}
%\Lambda(E)=\frac{2}{\pi}\frac{E~\Theta(E-\Delta)\Theta(\sqrt{J^2+\Delta^2}-E)}{\sqrt{E^2-\Delta^2}\sqrt{J^2+\Delta^2-E^2}},
%\eeq

 %\mpc{Mention the factor $(\Lambda_F/\tau_{\mathrm{el}})^{-1}$ in some way.$\to$ Reabsorbed in $\gamma$.}

%In our derivation, we got rid of the momentum dependence and described everything in terms of energy transfer. \mpc{Formulate better} This is justified by the fact that in real materials there is always some disorder present and the momentum conservation in the scattering processes need not be strictly obeyed as shown in Ref.~\cite{Galitski2019}. In the next section, we compare the non-thermal phase diagrams for the SC and CDW systems.   

%\section{Results}

\textit{Results---} At thermal equilibrium without coupling to light, the electronic system shows two phases as temperature is varied. Using \cref{eq:gapeq0}, one can show that at high temperatures the system stays in a metallic phase with $\Delta=0$. Below a certain temperature $\tcz$ (that depends on $\kappa$), the system undergoes a continuous phase transition to an ordered state with a finite gap $\Delta$. This behavior is depicted by the dashed line in the insets of \cref{fig:phasedia} and is exactly the same for SC and CDW.

On the other hand, the NESS behavior of the system at $g_0\neq 0$ is described by \cref{eq:gapeq}, where more parameters appear. 
We study the $\tcry$ dependence of the order parameter as $\nu_0$ is varied, 
since these are easily controllable parameters in experiments \cite{Jarc2023}.
In \cref{fig:phasedia}, we present the phase diagram in the $\nu_0-\tcry$ plane for SC [panel (a)] and CDW [panel (b)]. 
%Here, the cavity temperature $\tcav$ is kept higher than the equilibrium transition temperature $\tcz$,
Here, we keep $\tcav>\tcz$,
which is the case in typical experiments~\cite{Jarc2023}.  
%The phase diagram is constructed from the dependence of $\Delta$ on $\tcry$ as $\nu_0$ is varied. 
We show a few representative $\Delta$-vs-$\tcry$ curves in the inset, where the $\nu_0$ values are indicated in the main panel by arrows of corresponding colors. In the insets, the solid (dotted) lines represent stable (unstable) finite-$\Delta$ solutions of the  NESS gap equation \cref{eq:gapeq}. 

In the SC case, as apparent from the inset of \cref{fig:phasedia}(a), the finite-$\Delta$ solution is stable below $\tcry=\tcu$ (where the curves have a vertical tangent), while the $\Delta=0$ solution is stable above $\tcry=\tc$ (where the dotted lines touch zero).  
For $\tc<\tcry<\tcu$, both the finite-$\Delta$ and $\Delta=0$ solutions coexist. So, instead of the thermal continuous transition, in the NESS we have a first-order phase transition between the SC and metallic phase, and in the shaded-green region the two coexist. %\fps{in contrast to a continuous transition in thermal equilibrium (shown by the dashed gray curve in the inset).}
Furthermore, for small $\nu_0/J$ one has $\tcu>\tcz$, indicating an increase of the SC transition temperature. This cavity-induced EE behavior for SC order is consistent with the predictions of \cite{Galitski2019}, and qualitatively the same as found for the standard EE induced by driving with coherent light \cite{Eliashberg1970,*Eliashbergrus,Ivlev1973}. 

%at a low cavity frequency the gap is enhanced (blue curve) with respect to equilibrium (dashed gray curve). \mpc{Maybe refer to $T_c$ upper and lower? Then mention that gap enhancement is a necessary condition for $\tcry^{\mathrm{c, max}}>\tcz$.} We also note that by increasing the cavity frequency (red curve), the gap is mostly de-enhanced \cite{Galitski2019}. 
         
The CDW phase diagram is much richer, as shown in \cref{fig:phasedia}(b).
We find three different phases: a metallic phase with $\Delta=0$, a CDW with a smaller gap $\Delta$  ($\mathrm{CDW}_{\mathrm{L}}$) and a CDW with a larger $\Delta$ ($\mathrm{CDW}_{\mathrm{H}}$). 
%\fps{The Normal phase exists on the region on the right of the dashed light-red line (at lower $\nu_0$) and the solid green line (at higher $\nu_0$).  The $\mathrm{CDW}_{\mathrm{L}}$ phase lies on the region between the dashed light-red line and the solid green line. The $\mathrm{CDW}_{\mathrm{H}}$ phase appears on the left of the dotted light-blue line.}\fpc{This is all clear by looking at the phase diagram.} 
In the shaded-green region, metallic and $\mathrm{CDW}_{\mathrm{H}}$ phases coexist, while in the shaded-yellow region $\mathrm{CDW}_{\mathrm{L}}$ and $\mathrm{CDW}_{\mathrm{H}}$ coexist. Note that the latter region ends in a critical point indicated by a black star. 
%\fps{The system can go from one phase to another smoothly without any transition around this point. The phase coexistence indicates first-order transitions like in the SC case. However, depending on the cavity frequency, multi-staged transitions can be observed for CDW.} \fpc{It is explained later.}
The way in which the system transitions between these phases is correspondingly more complex.
In the inset of \cref{fig:phasedia}(b) we show $\Delta$ vs $\tcry$ for three representative $\nu_0$ values 
corresponding to qualitatively different types of transitions. 
%as temperature is varied.  
Note that, depending on $\nu_0$, the curves have a right inflection point at $\tcry=\tcu$ (corresponding to the dashed light-red line in the main panel) or a left inflection point at $\tcry=\tcl$ (corresponding to the dotted light-blue line in the main panel). We find stable finite-$\Delta$ solutions for $\tcry<\tcu$ (corresponding to 
 $\mathrm{CDW}_{\mathrm{H}}$) and for $\tcl<\tcry<\tc$ (corresponding to $\mathrm{CDW}_{\mathrm{L}}$).
For instance, for the blue curve in the inset (small $\nu_0/J$), starting from the metallic phase, as $\tcry$ is reduced the system first enters a coexistence region with metallic and $\mathrm{CDW}_{\mathrm{H}}$, then a coexistence region with $\mathrm{CDW}_{\mathrm{L}}$ and $\mathrm{CDW}_{\mathrm{H}}$, and finally the $\mathrm{CDW}_{\mathrm{H}}$ at low $\tcry$. That is, two first-order transitions are encountered in this case. For the orange curve instead (intermediate $\nu_0/J$), the system enters the $\mathrm{CDW}_{\mathrm{L}}$ smoothly, then a coexistence region with $\mathrm{CDW}_{\mathrm{L}}$ and $\mathrm{CDW}_{\mathrm{H}}$, and finally the $\mathrm{CDW}_{\mathrm{H}}$ at low $\tcry$. That is, first a continuous and then a first-order transition. For the red curve (large $\nu_0/J$, which lies above the critical point), we only observe a continuous transition to the a CDW phase, as $\mathrm{CDW}_{\mathrm{H}}$ and $\mathrm{CDW}_{\mathrm{L}}$ are not anymore separated by a transition but rather a smooth crossover.
%These different kinds of transitions give rise to the phase diagram in \cref{fig:phasedia}(b).\\ 
Moreover, also for CDW we find $\tcu>\tcz$ for small $\nu_0/J$ values, indicating an increase of the transition temperature. As expected, the enhancement is smaller than for SC and persists up-to a smaller value of $\nu_0/J$. 

\begin{figure}[t]
    \centering
    \includegraphics[width=\textwidth]{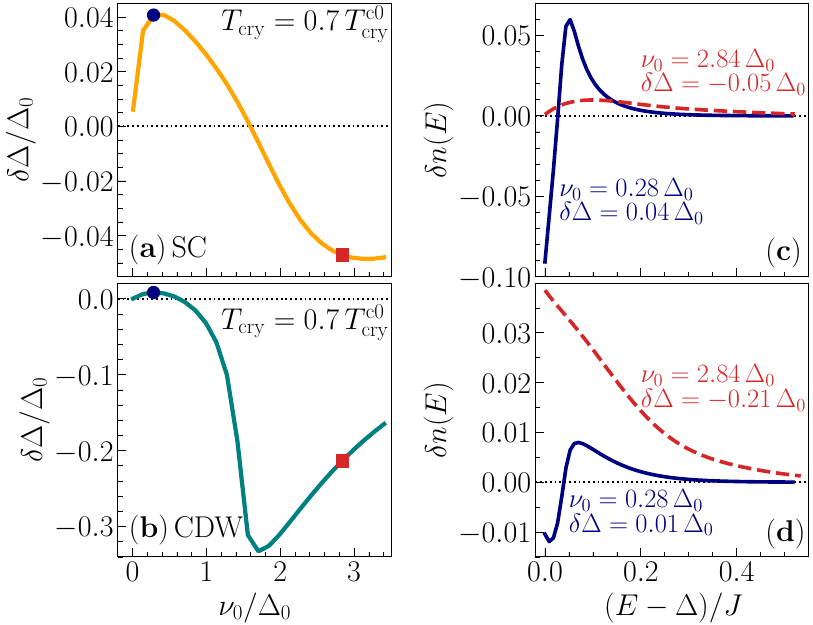}
    \caption{
    %Modification of the gap ($\delta \Delta=\Delta-\Delta_0$) and occupation ($\delta n$) in a NESS with respect to thermal equilibrium. We set $\tcry=0.7\,\tcz$ and other parameters are the same as in \cref{fig:phasedia}. For  
    %(a) SC and (b) CDW, the gap is enhanced ($\delta \Delta>0$) at small $\nu_0/\Delta_0$ and de-enhanced at larger $\nu_0/\Delta_0$. The de-enhancement is stronger for CDW compared to SC. 
    %$\delta n$ vs $E$ are shown for SC (c) and CDW (d) at two representative $\nu_0$ values marked in (a) and (b). 
    %In the case of gap enhancement (solid blue curves), the occupation is depleted near $E=\Delta$ making the effective gap larger.
    %For the de-enhancement of the gap (dashed red curves), the occupation near  $E=\Delta$ is not depleted. %\mps{Also, the curves are non-thermal in nature, not just an overall thermal heating or cooling.}  
    %\mpc{Caption can be shortened, remove repetition wrt text.}}
    Non-thermal modification of the gap $\delta \Delta=\Delta-\Delta_0$ vs $\nu_0$ is shown for SC (a) and CDW (b). $\Delta_0$ is the equilibrium gap value.  
    Modification of the distribution $\delta n$ vs $E$ is shown for SC (c) and CDW (d) at two representative $\nu_0$ values marked in (a) and (b).
    We set $\tcry=0.7\,\tcz$ and other parameters are the same as in \cref{fig:phasedia}.
    }
    \label{fig:delgapn}
\end{figure}

 A necessary (but not sufficient) condition for the increase of the transition temperature is the enhancement of the gap. In \cref{fig:delgapn}(a) and (b), we show how the gap $\Delta$ is modified from its equilibrium value $\Delta_0$ as $\nu_0$ is varied deep in the ordered phase for SC and CDW, respectively. The gap is enhanced for both at small $\nu_0/\Delta_0$.
 %, where $\Delta_0$ is the value of the gap at thermal equilibrium. 
 Consistently with the phase diagram of \cref{fig:phasedia}, for CDW the enhancement is smaller and persists up to a smaller value of $\nu_0/\Delta_0$. 
 Furthermore, the enhancement/suppression of the gap is related to the sign of $\delta n$ at energies close to the gap \cite{Eliashberg1970,*Eliashbergrus,Ivlev1973,Galitski2019}. In \cref{fig:delgapn}(c) and (d) we show $\delta n$ as a function of energy for SC and CDW, respectively, at specific cavity frequencies marked in \cref{fig:delgapn}(a) and (b) with corresponding colors. For the solid blue curves at small $\nu_0/\Delta_0$, $\delta n<0$ near $E=\Delta$ indicates that the quasiparticles are pushed away from the edge of the band, effectively enhancing the gap. On the other hand, for the dashed red curves at larger $\nu_0/\Delta_0$, the redistribution is not favorable for the gap. 
 %Note that the \fps{changing sign of the} \fpm{non-monotonous behavior of $\delta n$ as a function of energy is incompatible with a thermal state.}  \fps{curves indicates that the system does not thermalize to an effective temperature between $\tcry$ and $\tcav$, rather it reaches a NESS.} 
% \fpc{Add one sentence explaining where one sees it.}\mic{This is not true that $\delta n$ has to be monotonous for a thermal redistribution. However, it must be either positive or negative and cannot can sign for positive energies. $n=n_0+\delta n$ is monotonous for a thermal redistribution.}.

At thermal equilibrium, Ginzburg-Landau (GL) theory is a tool for understanding phase transitions based on the properties of the free energy as a function of the order parameter \cite{GLeng,*GLrus}. In the present model at $g_0=0$, an expansion of the free energy in powers of $\Delta$ shows that, while the equilibrium free energy contains terms quadratic and quartic in $\Delta$, the cubic term is absent. %\mps{owing to the symmetry of the thermal  electron distribution reflecting detailed balance}.
%\fpc{@Mursalin: please check and specify}
%\mic{The necessary ingredient for the absence of the cubic term is that $F(x)\to x$ as $x\to 0$ which is true for the thermal distribution function $F_0(x)=\tanh(x)$. I think it is fine to keep the above statement.}
%\rafc{Actually I think the cubic term is assumed to not be present because it goes like $F\propto\psi^*\psi \sqrt{\psi \psi^*}$ which is not analytic for small $\psi$}. 
The transition is continuous, with a gap that initially grows as $\Delta\sim\sqrt{\tcz-\tcry}$ by lowering the temperature below its critical value \cite{altlandsimons,tinkham2004}. 

%On the other hand, for $g_0>0$, the same expansion in powers of the gap in NESS yields the non-equilibrium free energy (see Supplemental Materials) \cite{Schmid1977}
For $g_0>0$, a similar expansion yields the NESS free energy (see Supplemental Materials) \cite{Schmid1977}
%\beq\label{eq:neqf}
\begin{align}
 \mathcal{F}_{\mathrm{neq}}[\Delta]&= \frac{\tcry-\tc}{2\tc}\Delta^2+\eta \frac{g_{0}^{2}[N_{\tcav}(\nu_0)-N_{\tc}(\nu_0)]}{3\pi \gamma J \tc}  \Delta^3\nonumber\\&
 + \frac{7\zeta(3)}{32\pi^{2}}\,
   \frac{1}{(\tc)^{2}}\Delta^4+\cdots, \label{eq:neqf}
\end{align}
%\eeq
where $\zeta(3)\approx1.2$ is the Riemann zeta function. In order to obtain an analytical expression, 
we assumed that $\gamma_B\ll\nu_0,\tcav$.
We see that  escaping thermal equilibrium allows for a cubic term in the NESS free energy. Such a term has also been found in the standard EE case of a SC driven by coherent light \cite{Ivlev1973,Schmid1977}. Here we extended this to the present two-bath cavity-electron setup, both for the SC and CDW case. We see that the cubic term alone encodes the information about the temperature difference between the two baths (through the boson-occupation difference), as well as the information about the difference between SC and CDW (through the sign-factor $\eta$).
%The presence of the cubic term in \cref{eq:neqf}, which differs between SC and CDW by the factor $\eta$, is the reason behind the existence of first-order transitions in the NESS. \mim{A similar cubic term was also noticed for the EE with classical light \cite{Ivlev1973}}.
%\mpc{Connect to \cite{Ivlev1973}, saying that they also see a similar cubic term.} 
Furthermore, the cubic term implies a linear-in-temperature growth of the order parameter solution as opposed to the square-root behavior at thermal equilibrium:
\beq\label{eq:gapLG}
\Delta[\tcry]\approx\frac{\pi\gamma J}{g_0^2}\frac{\eta}{N_{\tcav}(\nu_0)-N_{\tc}(\nu_0)}(\tcry-\tc).
\eeq
%i.e. the corresponding solution grows linearly with 
When $\tcav>\tc$, the 
%\mps{corresponding} 
solution for SC (CDW) is unstable (stable). The situation is reversed when $\tcav<\tc$ (See the End Matter for a more detailed analysis of the latter situation).  
\textit{Conclusions---} Motivated by recent experiments demonstrating cavity-based control of quantum materials in non-equilibrium steady states, we extended the theory of the Eliashberg effect from superconductivity to charge-density-wave order. The latter shows a much broader scope for the control of the ordering: by tuning the cavity resonance frequency, both continuous and discontinuous transitions, as well as multiple ordered phases merging at a critical point are selectively accessible.
In the future, our analysis shall be extended to include the non-equilibrium fluctuations of the order parameter, which are expected to lead to non-thermal distributions within the Ginzburg-Landau free-energy landscape. Moreover, analog extensions of the EE shall be applied to other types of order, including magnetic phases.
%In this work, we studied the SC and CDW transitions in optical cavity. The system achieves a NESS due to the presence of two baths with different temperatures. The NESS phase diagram shows richer phenomenology compared to thermal equilibrium. While in equilibrium SC and CDW are equivalent due to particle-hole symmetry, the NESS phase diagram for CDW is more feature-rich than SC. CDW shows multi-staged transitions depending on cavity frequency. Our findings add to the very interesting field of cavity control of quantum materials.  
%We find that by changing the cavity fundamental frequency i.e. the distance between the cavity mirrors, the nature of the transitions can be qualitatively modified.

\textit{Acknowledgements---} We thank Martin Eckstein, Daniele Fausti, Denis Golez, Zala Lenarcic and Vadim Plastovets for fruitful discussions. 
\bibliography{cavitytrans.bib}

\clearpage
\appendix*
\setcounter{equation}{0}
\section{End Matter}\label{sec:EM}
%\fpc{This end matter seems at an early stage of development. It needs a brief sketch of how the gap equation and the kinetic equation are derived, to be then completed by the supplemental material. Please let me know when it is ready to be looked into.}\\

\textit{NESS gap equation---}
\setcounter{equation}{0}
\renewcommand{\theequation}{A\arabic{equation}}
The NESS gap equation is derived using SKFT \cite{keldysh,*keldyshrus,schwinger,kamenev_book,rammer_book,DiehlReview2016}. We start by writing the Keldysh action for the Hamiltonian in \cref{eq:Ham} following the notations in Ref.~\cite{kamenev_book}. Note that one needs to treat $\Delta$ as a field to arrive at the correct Keldysh action.
We perform a Bogoliubov rotation on the Keldysh action so that the free part of the action corresponding to $H_{\rm el}$ becomes diagonal. The dispersions of the two energy bands are given by $\pm E_k$ where $E_k=\sqrt{\Delta^2+\xi_k^2}$. Under weak coupling approximation for $g_0\ll J,\gamma,\gamma_B$,   
the effect of the LM coupling is included perturbatively through its self-energy. Then, we integrate out the electronic degrees of freedom to obtain an effective Keldysh action in terms of the gap field $\Delta$.  Finally, we take the classical saddle point of the Keldysh action to get the NESS gap equation. 
Assuming $\gamma\ll \Delta, \gamma_B, \tcry,\tcav$ we perform a quasi-particle approximation to arrive at
\beq
\sum_{k}\frac{1-2[n_0(E_k)+\delta n(k,E_k)]}{E_k}=\frac{1}{\kappa},
\eeq
where the change in the distribution function is obtained through the quantum kinetic equation
\begin{widetext}
\beq
\begin{split}
&\delta n(k,E_k)=\frac{2}{\gamma}\sum_{k'}\sum_{q}\int d\Omega~(g^{+}_{k,k'})^2\left[\left(1+\frac{\xi_{k}\xi_{k'}+\eta\Delta^2}{E_{k}E_{k'}}\right)\mathrm{Im}D^R_0(q,\Omega) H_0(E_k,E_{k'})\delta(\Omega-(E_{k}-E_{k'}))\right.\\
  &~~~~~~~~~~~~~~~~~~~~~~~~~~~~~~~~~~~~~~~~~+\left.\left(1-\frac{\xi_{k}\xi_{k'}+\eta\Delta^2}{E_{k}E_{k'}}\right)\mathrm{Im}D^R_0(q,\Omega) H_0(E_k,-E_{k'})\delta(\Omega-(E_{k}+E_{k'}))\right]\delta(q-(k'-k)).\\
\end{split}
\eeq

See the Supplemental Material for a detailed derivation. In the above expression, the first term describes scattering processes within the same energy band and the second term between the bands. Note the energy and momenta conservations are described by the delta functions. 
Furthermore, we make a forward scattering approximation $\mathrm{Im}D^R_0(q,\omega)\approx\mathrm{Im}D^R_0(q=0,\omega)\equiv\mathrm{Im}D^R_0(\omega)$ and consider the effect of static impurities in real materials (see Refs.~\cite{Galitski2019, MattisBardeen1958}) to replace $\delta(q-(k-k'))$ with a constant, as discussed in the main text. We also average the LM coupling on the Fermi surface \cite{Galitski2019}. Finally, we convert the momentum sums to energy integration with the help of the density of states for the gaped quasi-particles $\Lambda(E)$ to arrive at \cref{eq:gapeq}.
%\mis{ However, in real materials, the presence of static impurities causes the momentum to be not conserved. Hence,  the corresponding delta function can be replaced by a constant $(\Lambda_F/\tau_{\rm el})^{-1}$, where $\Lambda_F=\frac{2}{\pi J}$ is the density of states at the Fermi surface and $\tau_{\rm el}$ is the elastic scattering rate (see Refs.~\cite{Galitski2019, MattisBardeen1958}). 
%We absorb the factor $(\Lambda_F/\tau_{\rm el})$ in $\gamma$.
%Furthermore, we convert the momentum sums to energy integration with the help of the density of states for the gaped quasi-particles $\Lambda(E)$. We further average the LM coupling on the Fermi surface \cite{Galitski2019}. Also note that the speed of light is much larger than the Fermi velocity, hence it suffices to consider only the $q=0$ photons. We define the shorthand notation $\mathrm{Im}D^R_0(\omega)=\mathrm{Im}D^R_0(0,\omega)$. Incorporating all these approximations, we arrive at \cref{eq:gapeq}.} 
Note that by putting $g_0=0$, we get $\delta n=0$ and recover the equilibrium gap equation in \cref{eq:gapeq0}.
\end{widetext}

\textit{Equivalence of SC and CDW gap equation at equilibrium---}
\setcounter{equation}{0}
\renewcommand{\theequation}{B\arabic{equation}}
At thermal equilibrium, the mapping between SC and CDW cases is ensured by the symmetry of the free electron dispersion around the Fermi surface, given by
$\epsilon_{k+k_F}=-\epsilon_{k-k_F}$. This is guaranteed since we consider a half-filled band ($k_F=\pi/2a$) with free-electron dispersion $\epsilon_k=-J\cos(ka)$. Note that this is a consequence of the particle-hole symmetry (PHS) of $H_{\rm el}$ under the transformation $\hat{c}_\sigma(k\pm k_F)\to \hat{c}^{\dagger}_\sigma(-k\pm k_F)$. 
%On the other hand, the LM interaction breaks particle-hole symmetry as $H_{\rm el-ph}$ is not invariant under this transformation.
However, $H_{\rm el-ph}$ is not invariant under this transformation and breaks PHS (see the Supplemental Material).
Hence, out-of-equilibrium the order parameters for the SC and CDW cases behave differently. 
%See the Supplemental Material for a detailed discussion.
%Particle-hole symmetry is given by $H\to H'=-H$  under the transformation $c(k)\to c^{\dagger}(-k)$ for free electrons \cite{deGennes1999}. For the electrons in the Nambu structure in \cref{eq:Ham}, this transformation is given by $c_{\alpha}(k)\to c^{\dagger}_{\alpha}(-k)$ for SC, and $c_{\alpha}(k)\to c^{\dagger}_{-\alpha}(-k)$ for CDW.
%\begin{align}
%    &c_{\alpha}(k)\to c^{\dagger}_{\alpha}(-k)~~\text{for SC, and}\notag\\
%    &c_{\alpha}(k)\to c^{\dagger}_{-\alpha}(-k)~~\text{for CDW.}
%\end{align}
%For both cases $H_{\rm el}\to H'_{\rm el}= -H_{\rm el}$. Presence of this particle-hole symmetry allows us to map one case to the other and we get identical gap equation at thermal equilibrium. On the other hand, the LM interaction breaks particle-hole symmetry as $H_{\rm el-ph}\to H'_{\rm el-ph}\ne -H_{\rm el-ph}$. Hence, out-of-equilibrium SC and CDW behave differently. See the Supplemental Material for a detailed discussion.

\textit{Additional results for cooler photon-bath---}
\setcounter{equation}{0}
\renewcommand{\theequation}{C\arabic{equation}}
In the main text, we discuss the phase diagram for $\tcav>\tcz$, which is generally the case in experiments \cite{Jarc2023}. However, for completeness, it is also interesting to explore the opposite case where $\tcav<\tcz$. From \cref{eq:gapLG} within GL theory, we can see that in this case the linear growth of the gap solution with $|\tcry-\tc|$ will flip sign for both SC and CDW with respect to the situation described in the main text. The stability of the solution will also change accordingly.  

\begin{figure}[htbp]
    \centering   \includegraphics[width=\textwidth]{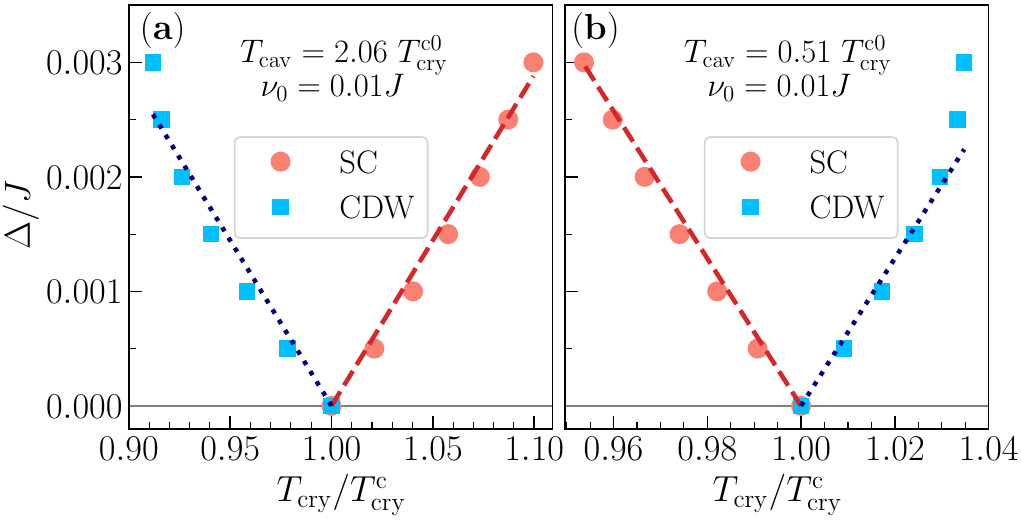}
    \caption{$\Delta$ vs $\tcry$ for (a) hotter photon-bath ($\tcav=2.06\,\tcz$) and (b) cooler photon-bath ($\tcav=0.51\,\tcz$) for small $\Delta$. We set $\nu_0=0.01\,J$. Other parameters are the same as in \cref{fig:phasedia}. The markers represent the numerical solution of the gap equation and the lines are obtained using the analytical expression in \cref{eq:gapLG}. 
    %\mps{They are in excellent agreement.}
    }
    \label{fig:gapsmall}
\end{figure}

In \cref{fig:gapsmall}, we show the solution of the gap equation for a hotter [\cref{fig:gapsmall}(a)] and a cooler [\cref{fig:gapsmall}(b)] photon-bath. The numerical solution is in excellent agreement with the analytical expression of \cref{eq:gapLG} when the gap is small. Further we notice an emergent symmetry between SC and CDW in this case. From \cref{eq:gapLG} we infer that the symmetry operation is given by $\tcav\rightarrow\tcav'$ such that $N_{\tcav'}(\nu_0)=2N_{\tc}(\nu_0)-N_{\tcav}(\nu_0)$ and SC $\rightarrow$ CDW i.e. $\eta\rightarrow \eta'=-\eta$. Note that this emergent symmetry is valid only in the small-$\Delta$ region where the GL expansion is valid and not for the general NESS gap equation in \cref{eq:gapeq}. 

\begin{figure}[htbp]
    \centering
    \includegraphics[width=\textwidth]{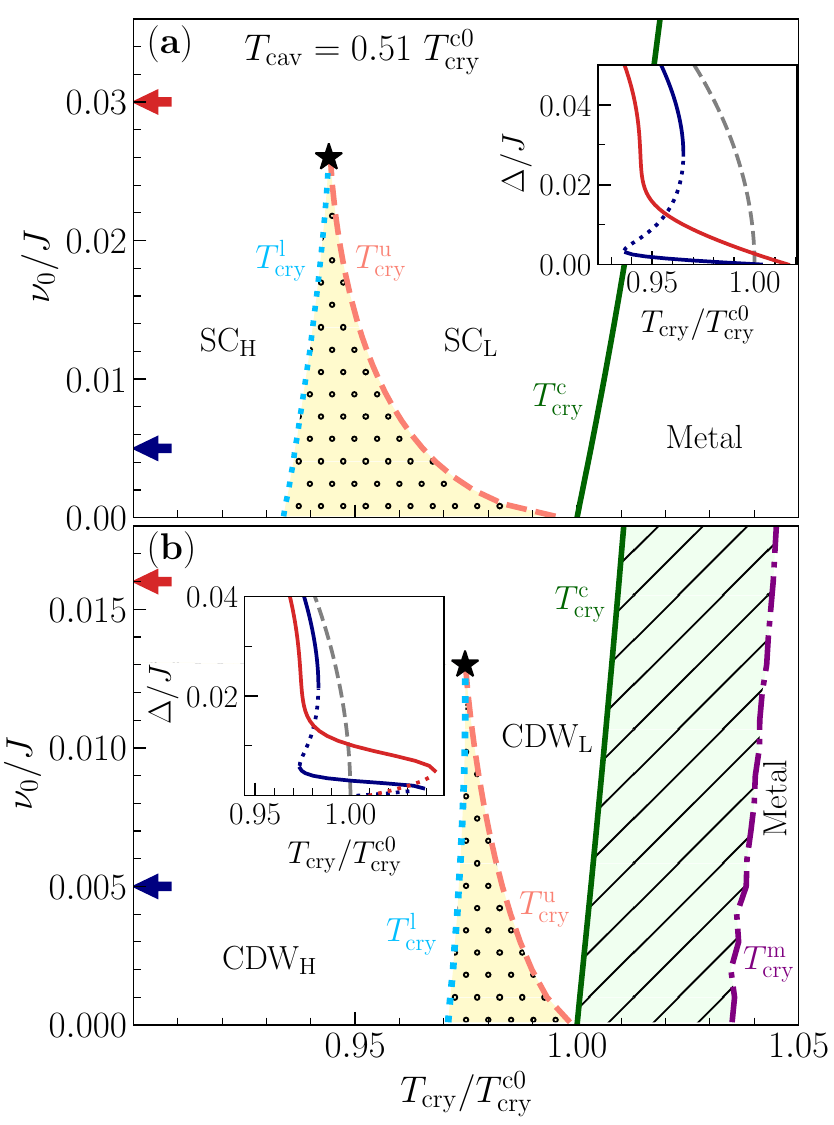}
\caption{NESS phase diagram in the $\nu_0$ - $\tcry$ plane for the SC (a) and the CDW (b) case for a cooler photon-bath.  
    We set $\tcav=0.51\,\tcz$, $\kappa=0.5\,J$, $g_0=0.1\,J$, $\gamma=0.01\,J$ and $\gamma_B=0.5\,\nu_0$. 
    The shaded areas indicate phase-coexistence regions and the black stars critical points.
    Insets: $\Delta$ vs $\tcry$ for different values of $\nu_0$ which are indicated by arrows of the corresponding colors in the main panel. 
    The solid (dotted) part of the curves represents a stable (unstable) solution of the gap equation.
    The dashed gray equilibrium curve is added for reference. 
    }
    \label{fig:phasedia2}
\end{figure}

In \cref{fig:phasedia2}, we show the phase diagram in the $\nu_0-\tcry$ plane for a cooler photon-bath i.e. $\tcav<\tcz$. For the SC case in \cref{fig:phasedia2}(a), we see that at high temperatures we have  a metallic phase with $\Delta=0$ that exists on the right of the solid green line. On the left of the solid green line, we have ordered phases with finite $\Delta$. 
We see two kinds of ordered phases: $\mathrm{SC}_{\mathrm{L}}$, with a smaller value of $\Delta$, which exists between the dashed light-red line and the solid green line, and $\mathrm{SC}_{\mathrm{H}}$, with a larger value of $\Delta$, which exists on the left of the dotted light-blue line. 
In the shaded-yellow region 
$\mathrm{SC}_{\mathrm{H}}$ and $\mathrm{SC}_{\mathrm{L}}$ can coexist, similarly to the behavior of CDW in the dotted yellow region of \cref{fig:phasedia}(b) in the main text. Note that also here the phase-coexistence region ends in a critical point marked by a black star.
The phase diagram is constructed from the dependence of $\Delta$ on $\tcry$ as $\nu_0$ is varied. We show a few representative $\Delta$ vs $\tcry$ curves in the inset, where the $\nu_0$ values are indicated in the main plot by arrows of the corresponding colors. Here, the solid (dotted) lines represent stable (unstable) finite-$\Delta$ solutions of the  NESS gap equation in \cref{eq:gapeq}. Note that the finite-$\Delta$ solution is stable below $\tcry=\tc$ (represented by the solid green line in the main plot) where the curves touch zero, while the $\Delta=0$ solution is stable above $\tcry=\tc$, indicating a second-order continuous transition. For the blue curve, we see two stable branches connected by a dotted line. The branch with smaller $\Delta$ exists for $\tcl<\tcry<\tc$ while the branch with larger $\Delta$ exists for $\tcry<\tcu$. Hence for $\tcl<\tcry<\tcu$, both the finite-$\Delta$ solutions exist and this indicates a first-order transition between $\mathrm{SC}_{\mathrm{H}}$ and $\mathrm{SC}_{\mathrm{L}}$. Note that in the main plot the dashed light-red line represents $\tcu$ and the dotted light-blue line stands for $\tcl$, which constitute the boundaries of the phase-coexistence region in the phase diagram. So, depending on $\nu_0$, we see only a continuous transition or a continuous transition accompanied by a first-order one as $\tcry$ is lowered.

On the other hand, for the CDW case in \cref{fig:phasedia2}(b), we always see at least one first-order transition. In this phase diagram, on the right of the dashed-dotted purple line ($\tcry>\tcm$), there is the metallic phase.
We see again two kinds of ordered phases: $\mathrm{CDW}_{\mathrm{L}}$, with a smaller value of $\Delta$, which exists between the dashed light-red line and the solid green line ($\tcu<\tcry<\tc$), and $\mathrm{CDW}_{\mathrm{H}}$, with a larger value of $\Delta$, which exists on the left of the dotted light-blue line ($\tcry<\tcl$).
%\mps{Between the dashed light-red line and the solid green line ($\tcu<\tcry<\tc$), \mpm{the} $\mathrm{CDW}_{\mathrm{L}}$ phase (an ordered phase with a smaller value of $\Delta$) exists and on the left of the dotted light-blue line ($\tcry<\tcl$), $\mathrm{CDW}_{\mathrm{H}}$ (an ordered phase with a smaller value of $\Delta$) exists.}
In this case, however, there are two phase-coexistence regions: in the shaded-green region ($\tc<\tcry<\tcm$), metallic phase and $\mathrm{CDW}_{\mathrm{L}}$ coexist, while  in the shaded-yellow region ($\tcl<\tcry<\tcu$), $\mathrm{CDW}_{\mathrm{L}}$ and $\mathrm{CDW}_{\mathrm{H}}$ coexist. The latter region ends in a critical point marked by a black star. So, at small $\nu_0$, we can observe two first-order transitions as $\tcry$ is varied, while at larger $\nu_0$ we observe only one first-order transition. In the inset we show a few representative $\Delta$ vs $\tcry$, from which the phase diagram is constructed. For the red curve, a finite-$\Delta$ solution and a $\Delta=0$ solution coexist between the temperature where the dotted part of the curve touches zero and the rightmost inflection point ($\tc<\tcry<\tcm$). This indicates a first-order transition. For $\tcry<\tc$, $\Delta$ grows smoothly without any further transition. On the other hand for the blue curve, in addition to the $\Delta=0$ and finite $\Delta$ coexistence we see another range of temperature between the left inflection point and a right (not the rightmost) inflection point where two finite $\Delta$ solutions coexist. This indicates an additional first-order transition.     

%\mpc{I typed this quickly so feel free to modify. I would anyway summarize what is new here for a cooler photon-bath.}
In conclusion, for a cooler photon-bath the behavior of the gap solution close to $\Delta=0$ is inverted with respect to the hotter photon bath case due to an emergent symmetry for small $\Delta$. However, for the whole phase diagram which is determined by finite values of $\Delta$, this symmetry is broken and a new structure of phase coexistence regions emerges both for the SC and the CDW cases.

%\mpc{This conclusion seems disconnected from the rest.}
%\mps{In conclusion, the difference of the phase diagrams for SC and CDW is a result of the broken equivalence between them out-of-equilibrium. However, an emergent symmetry emerges for small  $\Delta$ in the GL limit.}   

\end{document}

%\title{ nonthermal steady states from open cavity fermion systems}
\title{Supplemental material for ``Cavity-induced Eliashberg effect: superconductivity vs charge density wave''}

\author{Md Mursalin Islam}\email{md.islam@uni-a.de}
\affiliation{Theoretical Physics III, Center for Electronic Correlations and Magnetism,
Institute of Physics, University of Augsburg, 86135 Augsburg, Germany}
\affiliation{Max Planck Institute for the Physics of Complex Systems, Nöthnitzer Stra{\ss}e 38, 01187 Dresden, Germany}
\author{Michele Pini}
\affiliation{Theoretical Physics III, Center for Electronic Correlations and Magnetism,
Institute of Physics, University of Augsburg, 86135 Augsburg, Germany}
\affiliation{Max Planck Institute for the Physics of Complex Systems, Nöthnitzer Stra{\ss}e 38, 01187 Dresden, Germany}
\author{R. Flores-Calderón}
\affiliation{Max Planck Institute for the Physics of Complex Systems, Nöthnitzer Stra{\ss}e 38, 01187 Dresden, Germany}
\affiliation{Technical University of Munich, TUM School of Natural Sciences, Physics Department, 85748 Garching, Germany}
\affiliation{Munich Center for Quantum Science and Technology (MCQST), Schellingstr. 4, 80799 M{\"u}nchen, Germany}
\author{Francesco Piazza}
\affiliation{Theoretical Physics III, Center for Electronic Correlations and Magnetism,
Institute of Physics, University of Augsburg, 86135 Augsburg, Germany}
\affiliation{Max Planck Institute for the Physics of Complex Systems, Nöthnitzer Stra{\ss}e 38, 01187 Dresden, Germany}

\begin{abstract}
\centering
The Supplemental Material contains details on the calculations described in the main text.
\end{abstract}
\maketitle

\tableofcontents
\section{The non-thermal gap equation: a Keldysh field theory derivation}
\subsection{The Model}
We consider the system sketched in Fig.~1 in the main text. A system of fermions in a quasi-1D lattice is placed between the parallel mirrors of a Fabry-Perot cavity. Additionally, the fermions are connected to a cryostat and the photons interact with the environment outside through the cavity mirrors. This setup is described by the following Hamiltonian ($\hbar$ and $k_B$ are set to unity throughout the letter)
\beq\label{seq:hamtot}
\begin{split}
    &H=(H_{\mathrm{el}}+H_{\mathrm{el-bath}})+(H_{\mathrm{ph}}+H_{\mathrm{ph-bath}})+H_{\mathrm{el-ph}}\\
\end{split}
\eeq
We divided the Hamiltonian in three parts: (i) the matter part $(H_{\mathrm{el}}+H_{\mathrm{el-bath}})$,  
(ii) the light part $(H_{\mathrm{ph}}+H_{\mathrm{ph-bath}})$ and the (iii) the light-matter coupling $H_{\mathrm{el-ph}}$. We describe them separately below.
\subsubsection*{Matter}
We consider a system of fermions in Nambu basis described by the following mean-field Hamiltonian
\beq\label{seq:ham0}
H_{\mathrm{el}}=\sum_{k}
    \begin{bmatrix}
        \hat c^{\dagger}_{+}(k) & \hat c^{\dagger}_{-}(k)
    \end{bmatrix}
    \begin{bmatrix}
        \xi_k & \Delta \\
        \Delta^* & -\xi_k
    \end{bmatrix}
    \begin{bmatrix}
        \hat c_{+}(k) \\ \hat c_{-}(k)
    \end{bmatrix}+\frac{|\Delta|^2}{\kappa},
\eeq
where $\kappa$ is the strength of pairing coupling. In this context it can be electron-phonon coupling or two-particle electron-electron interaction strength. Due to the presence of the off-diagonal terms ($\Delta$), this model can describe a spontaneous symmetry-breaking transition where $\Delta$ acts as an order parameter as well as the energy gap.   
We can describe both the superconducting (SC) and charge-density-wave (CDW) transitions with the following definition.\\
For CDW (particle-hole channel) :
\beq
\begin{split}
    \hat c_+(k)&=\hat c(k+k_F)~~~\mathrm{and}~~~
    \hat c_-(k)=\hat c(k-k_F).\\
    %\epsilon_+(\bk)&=\epsilon(\bk+\bk_F)~~~\mathrm{and}~~~\epsilon_-(\bk)=\epsilon(\bk-\bk_F)\\
    %g^{\pm}_{\bk,\bk'}&=g_0\sin\left(\frac{k+k'}{2}\pm k_{F}\right)\\
\end{split}
\eeq
and for SC (particle-particle channel) :
\beq
\begin{split}
    \hat c_+(k)&=\hat c_{\uparrow}(k)~~~\mathrm{and}~~~
    \hat c_-(k)=\hat c^{\dagger}_{\downarrow}(-k)\\
    %\epsilon_+(\bk)&=\epsilon(\bk)~~~~\mathrm{and}~~~~\epsilon_-(\bk)=-\epsilon(\bk)\\
   % g^{\pm}_{\bk,\bk'}&=g_0\sin\left(\frac{k+k'}{2}\right)\\
\end{split}
\eeq
We consider that the free fermions are hopping on a lattice with lattice constant $a$. This results in a dispersion $\epsilon(k)=-J\cos(ka)$ in the Brillouin zone $k\in \left[-\pi/a, \pi/a\right)$. Here, $J$ determines the bandwidth for the fermions and the energy ranges from $-J$ to $+J$. We also consider a half-filled band so that $\epsilon_F=0$ and $k_F=\frac{\pi}{2a}$.
So we have
\beq
\xi_k=
\begin{cases}
    \epsilon(k+k_F)=-J\cos(ka+\pi/2)=J\sin(ka)~~\text{where}~~k\in\left[-\pi/2a, \pi/2a\right)~~~\mathrm{for ~~CDW}\\
    \epsilon(k)=-J\cos(ka)~~\text{where}~~k\in\left[-\pi/a, \pi/a\right)~~~\mathrm{for~~SC}.\\
\end{cases}
\eeq
So, in \cref{seq:ham0}, the momentum sum represents
\beq
    \sum_{k}\to 
\begin{cases}
    \int_{-\pi/a}^{\pi/a}\frac{dk}{2\pi}~~\text{for SC, and } \\
    \int_{-\pi/2a}^{\pi/2a}\frac{dk}{\pi}~~\text{for CDW. }
\end{cases}
\eeq
We have also considered that the fermions are connected to a cryostat with temperature $\tcry$ that is described by %$H_{\mathrm{el-bath}}$. 
\beq\label{eq:elbath}
\begin{split}
   H_{\mathrm{el-bath}} = &\sum_{\alpha,k,s} \left\{\epsilon_{s\alpha}(k)\hat f^\dagger_{s\alpha k}\hat f_{s\alpha k}+t_{s\alpha }(k)\left[\hat{c}^\dagger_{\alpha}(k)\hat f_{s\alpha k}+\hat f^\dagger_{s\alpha k}\hat c_{\alpha}(k)\right]\right\}.
\end{split}
\eeq
Here we consider an extensive number of degrees of freedom $\hat f_{s\alpha k}$ for the cryostat, labeled by $s$ and coupled to each electron momentum $k$ and Nambu component $\alpha$.
Th cryostat provides the electrons with a spectral width $\gamma$ \cite{Rafa2025}. Note that $\tcry$ is a controllable parameter in an experiment, and we will study the behavior of the system as $\tcry$ is varied.
\subsubsection*{Light}
The cavity photons are described by 
\beq
H_{\mathrm{ph}}=\sum_{q}\nu_q\left(\hat a^{\dagger}_q\hat a_q+\frac{1}{2}\right)
\eeq

The bosonic anihilation and creation operators $\hat a_q, \hat a^{\dagger}_q$ describe the photons where $\left[\hat a_q,\hat  a^{\dagger}_{q'}\right]=i\delta(q-q')$. The photons dispersion is given by $\nu_q=\sqrt{\nu_0^2+c^2q^2}$,
where $c$ is the speed of light and $\nu_0$ is the cavity fundamental frequency given in terms of the distance between the cavity mirrors as $\nu_0=\pi c/L$. 
%In the light- matter coupling $H_{\mathrm{el-ph}}$, we have $\hat\phi(q)=\sqrt{\frac{\nu_0}{\nu_q}}\frac{\left(\hat a_q+\hat a^{\dagger}_{-q}\right)}{\sqrt{2\pi}}$.

Cavity photons are also connected to the electromagnetic continuum of the external environment through the cavity mirrors. This coupling is described by 
\beq\label{eq:phbath}
\begin{split}
    H_{\rm ph-bath}= &\sum_{s,q} \left[\nu_s(q)\left(\hat b^\dagger_{sq}\hat b_{sq}+\dfrac{1}{2}\right) +\frac{t^{\rm ph}_s(q)}{2\sqrt{\nu_{q} \nu_s(q)}}(\hat a^\dagger_{-q}+\hat a_{q})(\hat b^\dagger_{s-q}+\hat b_{sq})\right],
\end{split}
\eeq
where we consider an extensive number of modes $\hat b_{sq}$ of the external EM environment couple to each cavity photon mode.
This keeps the photons in thermal equilibrium at temperature $\tcav$. This coupling also provide the photons with a spectral width $\gamma_B$ \cite{Rafa2025}. 

\subsubsection*{Light-matter coupling}
The fermion current couples to the cavity photons under paramagnetic approximation \cite{Schlawin2019,Rao2023}. In terms of free fermion operators it is given by
\beq
H_{\mathrm{el-ph}}=\sum_{k,k'}\sum_{\sigma=\{\uparrow,\downarrow\}}g_0\sin\left(\frac{ka+k'a}{2}\right)\hat{c}^{\dagger}_{\sigma}(k)\hat{c}_{\sigma}(k')\hat\phi(k-k'),
\eeq
where $\hat\phi(q)=\sqrt{\frac{\nu_0}{\nu_q}}\frac{\left(\hat a_q+\hat a^{\dagger}_{-q}\right)}{\sqrt{2\pi}}$ and $g_0$ is the light-matter coupling strength. 
Now we need to re-write this in terms of the Nambu fermions. For SC,
\beq
\begin{split}
H_{\mathrm{el-ph}}&=\sum_{k,k'}g_0\sin\left(\frac{ka+k'a}{2}\right)\left[\hat{c}^{\dagger}_{\uparrow}(k)\hat{c}_{\uparrow}(k')\hat\phi(k-k')+\hat{c}^{\dagger}_{\downarrow}(k)\hat{c}_{\downarrow}(k')\hat\phi(k-k')\right]\\
&=\sum_{k,k'}g_0\sin\left(\frac{ka+k'a}{2}\right)\left[\hat{c}^{\dagger}_{\uparrow}(k)\hat{c}_{\uparrow}(k')\hat\phi(k-k')-\hat{c}^{\dagger}_{\downarrow}(-k)\hat{c}_{\downarrow}(-k')\hat\phi(k'-k)\right]\\
&=\sum_{k,k'}g_0\sin\left(\frac{ka+k'a}{2}\right)\left[\hat{c}^{\dagger}_{+}(k)\hat{c}_{+}(k')\hat\phi(k-k')-\hat{c}_{-}(k)\hat{c}^{\dagger}_{-}(k')\hat\phi(k'-k)\right]\\
&=\sum_{k,k'}g_0\sin\left(\frac{ka+k'a}{2}\right)\left[\hat{c}^{\dagger}_{+}(k)\hat{c}_{+}(k')\hat\phi(k-k')+\hat{c}^{\dagger}_{-}(k)\hat{c}_{-}(k')\hat\phi(k-k')\right]\\
\end{split}
\eeq
For CDW we can ignore the spin as the phenomenology is independent of spin. We have
\beq
\begin{split}
H_{\mathrm{el-ph}}&=\sum_{k,k'}g_0\left[\sin\left(\frac{ka+k'a}{2}+\frac{\pi}{2}\right)\hat{c}^{\dagger}(k+k_F)\hat{c}(k'+k_F)+\sin\left(\frac{ka+k'a}{2}-\frac{\pi}{2}\right)\hat{c}^{\dagger}(k-k_F)\hat{c}(k'-k_F)\right]\hat\phi(k-k')\\
&+\sum_{k,k'}g_0\sin\left(\frac{ka+k'a}{2}\right)\left[\hat{c}^{\dagger}(k+k_F)\hat{c}(k'-k_F)\hat\phi(k-k'+2k_F)-\hat{c}^{\dagger}(k-k_F)\hat{c}(k'+k_F)\hat\phi(k-k'-2k_F)\right]\\
&=\sum_{k,k'}g_0\cos\left(\frac{ka+k'a}{2}\right)\left[\hat{c}^{\dagger}_{+}(k)\hat{c}_{+}(k')\hat\phi(k-k')-\hat{c}^{\dagger}_{-}(k)\hat{c}_{-}(k')\hat\phi(k-k')\right]\\
&+\sum_{k,k'}g_0\sin\left(\frac{ka+k'a}{2}\right)\left[\hat{c}^{\dagger}_+(k)\hat{c}_-(k')\hat\phi(k-k'+2k_F)-\hat{c}_{-}^{\dagger}(k)\hat{c}_{+}(k')\hat\phi(k-k'-2k_F)\right]\\
&\approx\sum_{k,k'}g_0\cos\left(\frac{ka+k'a}{2}\right)\left[\hat{c}^{\dagger}_{+}(k)\hat{c}_{+}(k')\hat\phi(k-k')-\hat{c}^{\dagger}_{-}(k)\hat{c}_{-}(k')\hat\phi(k-k')\right]\\
\end{split}
\eeq
Note that we can neglect the terms that couple $\hat c_+$ and $\hat c_-$ as the photon has to carry large momenta while we will see later that photons can carry only negligible momentum. 

We can summarize both cases in the compact notation
\beq
H_{\mathrm{el-ph}}=\sum_{k,k'}\sum_{\alpha=\{+,-\}}g^{\alpha}_{k,k'}\hat{c}^{\dagger}_{\alpha}(k)\hat{c}_{\alpha}(k')\hat{\phi}(k-k').
\eeq
Here the light-matter couplings are given by
\beq
g^{+}_{k,k'}=\eta g^{-}_{k,k'}=g_{k,k'}=
\begin{cases}
g_0\sin\left(\frac{ka+k'a}{2}+ k_{F}a\right)= g_0\cos\left(\frac{ka+k'a}{2}\right)~~~\mathrm{for ~~CDW}\\
g_0\sin\left(\frac{ka+k'a}{2}\right)~~~\mathrm{for~~SC,}
\end{cases}
\eeq
where 
\beq
\eta=
\begin{cases}
-1~~~\mathrm{for ~~CDW}\\
+1~~~\mathrm{for~~SC}.
\end{cases}
\eeq

For zero light-matter coupling ($g_0=0$), fermions at thermal equilibrium with the cryostat and have Fermi-Dirac distribution $n_0(\omega)=\frac{1}{\exp(\omega/\tcry)+1}$. We define the distribution factor $F_0(\omega)=1-2n_0(\omega)=\tanh\left(\frac{\omega}{2\tcry}\right)$. Similarly the photons are at thermal equilibrium with the environment and have distribution $N_{\tcav}$, where we define   Bose-Einstein distribution $N_T=\frac{1}{\exp(\omega/T)-1}$. We also define $B_T(\omega)=1+2N_T(\omega)=\coth\left(\frac{\omega}{2T}\right)$ for future purpose. 
%\subsubsection*{Particle-hole symmetry}

\subsection{Keldysh action and the gap equation}
We follow the standard procedure described in Ref.~\cite{kamenev_book} to write 
the Keldysh action for the Hamiltonian in Eq.~\eqref{seq:ham0} and it is given by (in Nambu basis for electrons)
\beq\label{seq:actionN}
\begin{split}
S_N=&\int d\omega \sum_{k}
\begin{bmatrix}
	\bar{c}_{+1}(k,\omega) && 	\bar{c}_{-1}(k,\omega) && 	\bar{c}_{+2}(k,\omega) && 	\bar{c}_{-2}(k,\omega)
\end{bmatrix}
\begin{bmatrix}
	G^{-1}_{N0}(\delc,\delq)
\end{bmatrix}
\begin{bmatrix}
	c_{+1}(k,\omega) \\ c_{-1}(k,\omega) \\  c_{+2}(k,\omega) \\	c_{-2}(k,\omega)
\end{bmatrix}\\
&-\frac{1}{\kappa}\left[\delc^*\delq+\delq^*\delc\right]+S_{L}+S_{LM},~~\text{where}\\
S_L&=\int \frac{d\Omega}{2\pi}\sum_{q}L
\begin{bmatrix}
    \phi_{\rm cl}(q,\Omega) &  \phi_{\rm q}(q,\Omega)
\end{bmatrix}
\begin{bmatrix}
    0 &  (\Omega-i\gamma_B)^2-\nu_q^2\\
    (\Omega+i\gamma_B)^2-\nu_q^2 & 2i\gamma_B\Omega B_{\tcav}(\Omega)
\end{bmatrix}
\begin{bmatrix}
    \phi_{\rm cl}(q,\Omega) \\  \phi_{\rm q}(q,\Omega)
\end{bmatrix}~~\text{and}
\\
S_{LM}&=\int \frac{d\omega}{2\pi}\frac{d\omega'}{2\pi}\sum_{\alpha=\{+,-\}}\sum_{k,k'}g^{\alpha}_{k,k'}\left\{\phi_{\rm cl}(k-k',\omega-\omega')\left[\bar{c}_{\alpha 1}(k,\omega)c_{\alpha 1}(k',\omega')+\bar{c}_{\alpha 2}(k,\omega)c_{\alpha 2}(k',\omega')\right]\right.\\
&~~~~~~~~~~~~~~~~~~~~~~~~~~~~~~~~~~~~~~~~~~~~~~~\left.+\phi_{\rm q}(k-k',\omega-\omega')\left[\bar{c}_{\alpha 1}(k,\omega)c_{\alpha 2}(k',\omega')+\bar{c}_{\alpha 2}(k,\omega)c_{\alpha 1}(k',\omega')\right]\right\}.\\
\end{split}
\eeq
Here, $\{+,-\}$ are the Nambu indices and $\{1,2\}$ are the Larkin-Ovchinikov indices.
The ``free" fermionic inverse Green's function $G^{-1}$ is given by
\beq\label{seq:Ginv}
\begin{bmatrix}
	G^{-1}_{N0}(\delc,\delq)
\end{bmatrix}
=
\begin{bmatrix}
	\omega-\xi_k+i\gamma && -\delc && [G^{-1}_0]^{K}_{++}  && -\delq+[G^{-1}_0]^{K}_{+-}\\
	-\delc^* && \omega+\xi_k+i\gamma && -\delq^*+[G^{-1}_0]^{K}_{-+} && [G^{-1}_0]^{K}_{--}\\
	0 && -\delq && \omega-\xi_k-i\gamma && -\delc\\
	-\delq^* && 0 && -\delc^* && \omega+\xi_k-i\gamma. 
\end{bmatrix}
\eeq
%and where we have considered that $\gamma$ is the quasiparticle damping (due to a bath) and 
Note that we have assumed the static limit (zero frequency for the bosonic $\Delta$ fields). The distribution function that enters $[G^{-1}]^K_{\alpha\beta}$  and they can be non-thermal in general.
It is useful to write this $4\times4$ matrix in terms of $2\times2$ blocks as
\beq
G^{-1}_{N0}(\delc,\delq)=
\begin{bmatrix}
	G^{R^{-1}}_{N0} && G^{K^{-1}}_{N0}+G^{q^{-1}}_{N0}\\
 G^{q^{-1}}_{N0} && G^{A^{-1}}_{N0}
\end{bmatrix}
\eeq
where we have in the Nambu basis
\beq
\begin{split}
&G^{R^{-1}}_{N0}=
\begin{bmatrix}
\omega-\xi_k+i\gamma && -\delc\\
-\delc^* && \omega+\xi_k+i\gamma
\end{bmatrix}~~~~~~~
G^{K^{-1}}_{N0}=
\begin{bmatrix}
[G^{-1}_0]^{K}_{++}  && [G^{-1}_0]^{K}_{+-}\\
[G^{-1}_0]^{K}_{-+}  && [G^{-1}_0]^{K}_{--}
\end{bmatrix}\\
&G^{q^{-1}}_{N0}=
\begin{bmatrix}
0 && -\delq\\
-\delq^* && 0
\end{bmatrix}~~~~~~~~~~~~~~~~~~~~~~~
G^{A^{-1}}_{N0}=
\begin{bmatrix}
\omega-\xi_k-i\gamma && -\delc\\
-\delc^* && \omega+\xi_k-i\gamma
\end{bmatrix}
\end{split}
\eeq
At this point we do the Bogoliubov transformation given by (generalized for the additional Keldysh indices) 
\beq
\begin{bmatrix}
	\psi_{p1}(k,\omega) \\ \psi_{m1}(k,\omega) \\  \psi_{p2}(k,\omega) \\	\psi_{m2}(k,\omega)
\end{bmatrix}
=\mathcal{U}(\delc,k)
\begin{bmatrix}
	c_{+1}(k,\omega) \\ c_{-1}(k,\omega) \\  c_{+2}(k,\omega) \\	c_{-2}(k,\omega)
\end{bmatrix}
\eeq
where $\{p,m\}$ are Bogoliubov indices.
We have 
\beq
\mathcal{U}=
\begin{bmatrix}
    U && 0_{2\times2}\\
    0_{2\times2 } && U
\end{bmatrix}
~~~\mathrm{and}~~~
U=
\begin{bmatrix}
    u_k && v_k\\
    -v^*_k && u^*_k
\end{bmatrix}
=
\begin{bmatrix}
    \sqrt{\frac{1}{2}\left(1+\frac{\xi_k}{E_k}\right)} && \frac{\delc}{|\delc|}\sqrt{\frac{1}{2}\left(1-\frac{\xi_k}{E_k}\right)}\\
    -\frac{\delc^*}{|\delc|}\sqrt{\frac{1}{2}\left(1-\frac{\xi_k}{E_k}\right)} && \sqrt{\frac{1}{2}\left(1+\frac{\xi_k}{E_k}\right)}
\end{bmatrix}
\eeq
where
\beq
E_k=\sqrt{\xi_k^2+|\delc|^2}.
\eeq

Under this transformation the action becomes
\beq\label{seq:actionBG}
\begin{split}
S_B=&\int d\omega \sum_{k}
\begin{bmatrix}
	\bar{\psi}_{p1}(k,\omega) && 	\bar{\psi}_{m1}(k,\omega) && 	\bar{\psi}_{p2}(k,\omega) && 	\bar{\psi}_{m2}(k,\omega)
\end{bmatrix}
\begin{bmatrix}
	G^{-1}_0(\delc,\delq)
\end{bmatrix}
\begin{bmatrix}
	\psi_{p1}(k,\omega) \\ \psi_{m1}(k,\omega) \\  \psi_{p2}(k,\omega) \\	\psi_{m2}(k,\omega)
\end{bmatrix}\\
&-\frac{1}{\kappa}\left[\delc^*\delq+\delq^*\delc\right]+S_{L}+S_{LM},
\end{split}
\eeq
where 
\beq
G^{-1}_{0}(\delc,\delq)=
\begin{bmatrix}
	G^{R^{-1}}_{0} && G^{K^{-1}}_{0}+G^{q^{-1}}_{0}\\
 G^{q^{-1}}_{0} && G^{A^{-1}}_{0}
\end{bmatrix}
=\mathcal{U}G^{-1}_{N0}\mathcal{U}^{\dagger}.
\eeq
The $2\times2$  blocks
\beq
G^{R^{-1}}_{0}=UG^{R^{-1}}_{N0}U^{\dagger}=
\begin{bmatrix}
    \omega-E_k+i\gamma && 0\\
    0 && \omega+E_k+i\gamma
\end{bmatrix}
=[G^{A^{-1}}_{0}]^{\dagger},
\eeq
\beq
G^{K^{-1}}_{0}=UG^{K^{-1}}_{N0}U^{\dagger}=
\begin{bmatrix}
    2i\gamma F_0(\omega) && 0\\
    0 && 2i\gamma F_0(\omega)
\end{bmatrix}
\eeq
and
\beq\label{seq:ginvq}
G^{q^{-1}}_{0}=UG^{q^{-1}}_{N0}U^{\dagger}=
\begin{bmatrix}
    -u_kv_k^*\delq-u_k^*v_k\delq^* && -u_k^2\delq+v_k^2\delq^*\\
    +(v_k^*)^2\delq-(u_k^*)^2\delq^* && u_kv_k^*\delq+u_k^*v_k\delq^*
\end{bmatrix}.
\eeq
The light-matter coupling in this rotated basis
\beq
\begin{split}
   S_{LM}&=\int \frac{d\omega}{2\pi}\frac{d\omega'}{2\pi}\sum_{\alpha,\beta=\{p,m\}}\sum_{k,k'}\tilde{g}^{\alpha\beta}_{k,k'}\left\{\phi_{\rm cl}(k-k',\omega-\omega')\left[\bar{\psi}_{\alpha 1}(k,\omega)\psi_{\beta 1}(k',\omega')+\bar{\psi}_{\alpha 2}(k,\omega)\psi_{\beta 2}(k',\omega')\right]\right.\\
&~~~~~~~~~~~~~~~~~~~~~~~~~~~~~~~~~~~~~~~~~~~~~~~\left.+\phi_{\rm q}(k-k',\omega-\omega')\left[\bar{\psi}_{\alpha 1}(k,\omega)\psi_{\beta 2}(k',\omega')+\bar{\psi}_{\alpha 2}(k,\omega)\psi_{\beta 1}(k',\omega')\right]\right\}\\
\end{split}
\eeq
where
\beq\label{seq:LMcoupling_BG}
\begin{split}
\tilde{g}^{pp}_{k,k'}&=g^+_{k,k'}u_ku_{k'}^*+g^-_{k,k'}v_kv_{k'}^*=\frac{g_{k,k'}}{2}\left[\sqrt{\left(1+\frac{\xi_k}{E_k}\right)\left(1+\frac{\xi_{k'}}{E_{k'}}\right)}+\eta\sqrt{\left(1-\frac{\xi_k}{E_k}\right)\left(1-\frac{\xi_{k'}}{E_{k'}}\right)}\right]\\
\tilde{g}^{pm}_{k,k'}&=-g^+_{k,k'}u_kv_{k'}+g^-_{k,k'}v_ku_{k'}=-\frac{g_{k,k'}}{2}\frac{\delc}{|\delc|}\left[\sqrt{\left(1+\frac{\xi_k}{E_k}\right)\left(1-\frac{\xi_{k'}}{E_{k'}}\right)}-\eta\sqrt{\left(1-\frac{\xi_k}{E_k}\right)\left(1+\frac{\xi_{k'}}{E_{k'}}\right)}\right]\\
\tilde{g}^{mp}_{k,k'}&=-g^+_{k,k'}v_k^*u_{k'}^*+g^-_{k,k'}u_k^*v_{k'}^*=-\frac{g_{k,k'}}{2}\frac{\delc^*}{|\delc|}\left[\sqrt{\left(1-\frac{\xi_k}{E_k}\right)\left(1+\frac{\xi_{k'}}{E_{k'}}\right)}-\eta\sqrt{\left(1+\frac{\xi_k}{E_k}\right)\left(1-\frac{\xi_{k'}}{E_{k'}}\right)}\right]\\
\tilde{g}^{mm}_{k,k'}&=g^+_{k,k'}v_k^*v_{k'}+g^-_{k,k'}u_k^*u_{k'}=\frac{g_{k,k'}}{2}\left[\sqrt{\left(1-\frac{\xi_k}{E_k}\right)\left(1-\frac{\xi_{k'}}{E_{k'}}\right)}+\eta\sqrt{\left(1+\frac{\xi_k}{E_k}\right)\left(1+\frac{\xi_{k'}}{E_{k'}}\right)}\right].\\
\end{split}
\eeq
Now, we will treat the light-matter coupling in a mean-field way. This will induce self energy to the fermions of the form
\beq
\Sigma=
\begin{bmatrix}
    \Sigma^R && \Sigma^K\\
    \Sigma^q && \Sigma^A
\end{bmatrix}
\eeq
where $\Sigma^{R/A/K/q}$ is $2\times2$ matrix in the Bogoliubov basis. So, the action becomes
\beq\label{seq:actionBGSE}
\begin{split}
S_B\approx&\int d\omega \sum_{k}
\begin{bmatrix}
	\bar{\psi}_{p1}(k,\omega) && 	\bar{\psi}_{m1}(k,\omega) && 	\bar{\psi}_{p2}(k,\omega) && 	\bar{\psi}_{m2}(k,\omega)
\end{bmatrix}
\begin{bmatrix}
	G^{-1}(\delc,\delq)
\end{bmatrix}
\begin{bmatrix}
	\psi_{p1}(k,\omega) \\ \psi_{m1}(k,\omega) \\  \psi_{p2}(k,\omega) \\	\psi_{m2}(k,\omega)
\end{bmatrix}\\
&-\frac{1}{\kappa}\left[\delc^*\delq+\delq^*\delc\right]+S_{L},
\end{split}
\eeq
where 
\beq
G^{-1}(\delc,\delq)=G^{-1}_0(\delc,\delq)-\Sigma
\eeq

Integrating out the fermions one gets the effective action
\beq\label{seq:actioneffective}
\begin{split}
	\tilde{S}_B=&-i\text{Tr}\left[\ln\left\{-iG^{-1}(\delc,\delq)\right\}\right]-\frac{1}{\kappa}\left[\delc^*\delq+\delq^*\delc\right]+S_{L}\\
\end{split}
\eeq

Classical saddle point equations for this action can be found out by setting 
\beq
\left.\frac{\delta\tilde{S}_B}{\delta \delq}\right|_{\delq=0}=0~~~~\text{and}~~~~\left.\frac{\delta\tilde{S}_B}{\delta \delq^*}\right|_{\delq=0}=0.
\eeq
%where $X_q=0$ is a shorthand notation for $X_q(\bQ,\nu)=X_q(-\bQ,\nu)=0$.

So, we get
\beq\label{seq:sadleq}
%\beq\begin{split}
-i\text{Tr}\left[iG(\delc,\delq=0)\left.\frac{\delta\left\{-iG^{-1}(\delc,\delq)\right\}}{\delta \delq}\right|_{\delq=0}\right]-\frac{1}{\kappa}\delc^*=0.
%	&-i\text{Tr}\left[iG(\delc,\delq=0)\left.\frac{\delta\left\{-iG^{-1}(\delc,\delq)\right\}}{\delta \delq^*}\right|_{\delq=0}\right]-\frac{1}{\kappa}\delc=0~~\text{and}\label{sadleq1}
%\end{split}\eeq
\eeq
Note that it is okay to consider only one of the equations as they are Hermitian conjugate of each other and at the end produce the same equation.
To evaluate these equations we need $G(\delc,\delq=0)$, which can be found out by inverting 
%(we also use $\Delta=\delc$ )
\beq\label{seq:Ginv0}
%G^{-1}(\Delta)=
	G^{-1}(\delc,\delq=0)
=
\begin{bmatrix}
G^{R^{-1}} && [G^{-1}]^K\\
 0 && G^{A^{-1}}  
\end{bmatrix}
=
\begin{bmatrix}
G^{R^{-1}}_0-\Sigma^R(\delc,0) && [G^{-1}_0]^K-\Sigma^K(\delc,0)\\
 0 && G^{A^{-1}}_0-\Sigma^A(\delc,0)  
\end{bmatrix},
\eeq
where $G^{R^{-1}}_0=\begin{bmatrix}
	\omega-E_k+i\gamma && 0\\
	0 &&  	\omega+E_k+i\gamma
\end{bmatrix}=[G^{A^{-1}}_0(\delc)]^{\dagger}$.

By inverting the free part we get,
\beq
G_0(\delc)=
\begin{bmatrix}
	G_0(\delc,\delq=0)
\end{bmatrix}
=
\begin{bmatrix}
G^{R}_0(\delc) && G^K_0(\delc)\\
 0 && G^{A}_0(\delc)  
\end{bmatrix}
\eeq
where
\beq\label{seq:grgap}
G^R_0=[G^A_0]^{\dagger}=
\begin{bmatrix}
    G^R_{0pp} && G^R_{0pm}\\
    G^R_{0mp} && G^R_{0mm}
\end{bmatrix}=
\begin{bmatrix}
	\frac{1}{\omega-E_k+i\gamma} && 0\\
	0 &&  	\frac{1}{\omega+E_k+i\gamma}
\end{bmatrix},
\eeq
and 
\beq\label{seq:gkgap}
G^K_0=[G^R_0-G^A_0]F_0=
\begin{bmatrix}
    G^K_{0pp} && G^K_{0pm}\\
    G^K_{0mp} && G^K_{0mm}
\end{bmatrix}=
\begin{bmatrix}
	2i\mathrm{Im}G^R_{0pp}F_0 && 0\\
	0 &&	2i\mathrm{Im}G^R_{0mm}F_0
\end{bmatrix}.
\eeq

 In general we will have 
 \beq\label{seq:GF}
%G^{-1}(\Delta)=
	G(\delc,\delq=0)
=
\begin{bmatrix}
G^{R}(\delc) && G^K(\delc)\\
 0 && G^{A}(\delc)  
\end{bmatrix}
=
\begin{bmatrix}
G^{R}_0+G^R_0\Sigma^RG^R && G^K\\
 0 && G^{A}_0+G^A_0\Sigma^AG^A
\end{bmatrix},
\eeq
 For the Keldysh Green's function we use the standard steady-state ansatz  
 \beq\label{seq:gkparam}
 \begin{split}
     &G^K=\begin{bmatrix}
    G^K_{pp} && G^K_{pm}\\
    G^K_{mp} && G^K_{mm}
\end{bmatrix}=G^RF-FG^A=
\begin{bmatrix}
    G^R_{pp} && G^R_{pm}\\
    G^R_{mp} && G^R_{mm}
\end{bmatrix}
\begin{bmatrix}
    F_{pp} && F_{pm}\\
    F_{mp} && F_{mm}
\end{bmatrix}
-
\begin{bmatrix}
    F_{pp} && F_{pm}\\
    F_{mp} && F_{mm}
\end{bmatrix}
\begin{bmatrix}
    G^A_{pp} && G^A_{pm}\\
    G^A_{mp} && G^A_{mm}
\end{bmatrix}\\    
&=\begin{bmatrix}
    (G^R_{pp}-G^A_{pp})F_{pp}+G^R_{pm}F_{mp}-F_{pm}G^A_{mp} && G^R_{pm}F_{mm}-F_{pp}G^A_{pm}+(G^R_{pp}-G^A_{mm})F_{pm}\\
    G^R_{mp}F_{pp}-F_{mm}G^A_{mp}+(G^R_{mm}-G^A_{pp})F_{mp} &&    (G^R_{mm}-G^A_{mm})F_{mm}+G^R_{mp}F_{pm}-F_{mp}G^A_{pm}
\end{bmatrix}
\end{split}
\eeq

Now using the similar notation of breaking up the $4\times 4 $ matrix into $2\times2$ blocks, we get
\beq
\left.\frac{\delta\left\{-iG^{-1}(\delc,\delq)\right\}}{\delta \delq}\right|_{\delq=0}=-i
\begin{bmatrix}
	0 && \frac{\delta G^{q^{-1}}_0}{\delta \delq}|_{\delq=0}\\
	\frac{\delta G^{q^{-1}}_0}{\delta \delq}|_{\delq=0} && 0
\end{bmatrix}
+i\begin{bmatrix}
	\frac{\delta\Sigma^R}{\delta \delq}|_{\delq=0} && \frac{\delta\Sigma^K}{\delta \delq}|_{\delq=0}\\
	\frac{\delta\Sigma^q}{\delta \delq}|_{\delq=0} && \frac{\delta\Sigma^A}{\delta \delq}|_{\delq=0}
\end{bmatrix}
\eeq
where 
\beq
\begin{split}
\frac{\delta G^{q^{-1}}_0}{\delta \delq}|_{\delq=0}&=
\begin{bmatrix}
    -u_kv_k^* && -u_k^2\\
    (v_k^*)^2 && u_kv_k^*
\end{bmatrix}=
\begin{bmatrix}
    -\frac{\delc^*}{2E_k} && -\frac{1}{2}\left(1+\frac{\xi_k}{E_k}\right)\\
   \frac{(\delc^*)^2}{2|\delc|^2}\left(1-\frac{\xi_k}{E_k}\right) && \frac{\delc^*}{2E_k}
   \end{bmatrix}~~~\mathrm{and}\\
\frac{\delta G^{q^{-1}}_0}{\delta \delq^*}|_{\delq=0}&=
\begin{bmatrix}
   -u_k^*v_k && v_k^2\\
    -(u_k^*)^2 && u_k^*v_k
\end{bmatrix}=
\begin{bmatrix}
    -\frac{\delc}{2E_k} && \frac{(\delc)^2}{2|\delc|^2}\left(1-\frac{\xi_k}{E_k}\right)\\
   -\frac{1}{2}\left(1+\frac{\xi_k}{E_k}\right) && \frac{\delc}{2E_k}
   \end{bmatrix}
\end{split}
\eeq
Note that we have also considered a $\Sigma^q$ that is absent in the standard Keldysh structure and vanishes when $\delq=0$.

So, Eq.~\eqref{seq:sadleq} produces
\beq
	-i\text{Tr}\left[G^K\frac{\delta G^{q^{-1}}_0}{\delta \delq}|_{\delq=0}\right]+i\text{Tr}\left[G^R\bar{\Sigma}^R+G^A\bar{\Sigma}^A+G^K\bar{\Sigma}^q\right]=\frac{\delc^*}{\kappa}.
\eeq
Here, we have used the short-hand notation $\bar{\Sigma}^{(R/A/K/q)}=\left.\frac{\delta\Sigma^{(R/A/K/q)}}{\delta \delq}\right|_{\delq=0}$ and $\bar{\Sigma}^{*(R/A/K/q)}=\left.\frac{\delta\Sigma^{(R/A/K/q)}}{\delta \Delta^*_q}\right|_{\delq=0}$.

Writing the components of $G^K$ explicitly for the first part we get 
\beq
i\sum_{k}\int \frac{d\omega}{2\pi}~\left[u_kv_k^*(G^K_{pp}-G^K_{mm})-(v_k^*)^2G^K_{pm}+u_k^2G^K_{mp}\right]+i\text{Tr}\left[G^R\bar{\Sigma}^R+G^A\bar{\Sigma}^A+G^K\bar{\Sigma}^q\right]=\frac{\delc^*}{\kappa}.
\eeq

Using the parametrization of $G^K$ in Eq.~\eqref{seq:gkparam}, we get 
\beq\begin{split}\label{seq:gapeqgen}
    &i\sum_{k}\int \frac{d\omega}{2\pi}~\left\{\left[u_kv_k^*(G^R_{pp}-G^A_{pp})+(v_k^*)^2G^A_{pm}+u_k^2G^R_{mp}\right]F_{pp}\right.+\left[-u_kv_k^*(G^R_{mm}-G^A_{mm})-(v_k^*)^2G^R_{pm}-u_k^2G^A_{mp}\right]F_{mm}\\
    &~~~~~~~~~~~~~~~~~\left.+\left[-u_kv_k^*(G^R_{mp}+G^A_{mp})-(v_k^*)^2(G^R_{pp}-G^A_{mm})\right]F_{pm}+\left[u_kv_k^*(G^R_{pm}+G^A_{pm})+(u_k)^2(G^R_{mm}-G^A_{pp})\right]F_{mp}\right\}\\
 &+i\sum_{k}\int \frac{d\omega}{2\pi}\sum_{\alpha,\beta=\{p,m\}}\left[G^R_{\alpha\beta}\bar{\Sigma}^R_{\beta\alpha}+G^A_{\alpha\beta}\bar{\Sigma}^A_{\beta\alpha}+\sum_{\lambda=\{p,m\}}\left(G^R_{\alpha\lambda}F_{\lambda\beta}-F_{\alpha\lambda}G^A_{\lambda\beta}\right)\bar{\Sigma}^q_{\beta\alpha}\right]=\frac{\delc^*}{\kappa}.
\end{split}\eeq	  

%More generally, if the electrons are interacting or connected to the cavity, we will have
%\beq
%G^{-1}(\delc,\delq)=G^{-1}_0(\delc,\delq)-\Sigma(\delc,\delq)
%\eeq
 This is a general gap equation that does not assume anything about the fermion dispersion (or the Green's functions) and the distribution other than the steady state ansatz.

\subsection{Equilibrium gap equation: equivalence of SC and CDW}
In thermal equilibrium (decoupled from the cavity: $\Sigma=0$, $G^{R/A}_{pm}=G^{R/A}_{mp}=0$) we have $F_{pm}=F_{mp}=0$ and $F_{pp}=F_{mm}=F_0=\tanh\left(\frac{\omega}{2T}\right)$. %If we consider the first equation (both of them produce the same equation eventually), 
So, from \cref{seq:gapeqgen} we get
\beq\begin{split}
	&i\sum_{\bk}\int \frac{d\omega}{2\pi}u_kv_k^*[2i\mathrm{Im}G^R_{0pp}-2i\mathrm{Im}G^R_{0mm}]F_0=\frac{\delc^*}{\kappa}
\end{split}\eeq
or, explicitly,
\beq
%\beq\begin{split}
	-2\sum_{\bk}\int \frac{d\omega}{2\pi}~\frac{\delc^*}{2E_k}\left[-\frac{\gamma}{(\omega-E_k)^2+\gamma^2}+\frac{\gamma}{(\omega+E_k)^2+\gamma^2}\right]F_0(\omega)=\frac{\delc^*}{\kappa}.
%\end{split}\eeq
\eeq
In the $\gamma\rightarrow0$ limit we get 
\beq
\sum_{\bk}\frac{F_0(E_k)}{E_k}=\frac{1}{\kappa}.
\eeq
This is the standard gap equation at thermal equilibrium.

Let us now show that this gap equation is identical for SC and CDW. We define the notation $\Delta=\delc$ as this is the physical gap.
We have $E_k=\sqrt{\xi^2_k+|\Delta|^2}=\sqrt{J^2\cos^2(ka)+|\Delta|^2}$ for SC and $E_k=\sqrt{\xi^2_k+|\Delta|^2}=\sqrt{J^2\sin^2(ka)+|\Delta|^2}$ for CDW. Hence the gap equations
\beq\label{seq:gapeqboth}
\int_{-\pi/a}^{\pi/a}\frac{dk}{2\pi}\frac{F_0\left(\sqrt{J^2\cos^2(ka)+|\Delta|^2}\right)}{\sqrt{J^2\cos^2(ka)+|\Delta|^2}}=\frac{1}{\kappa}~\text{for SC,} \ \ \text{and} \ \ \int_{-\pi/2a}^{\pi/2a}\frac{dk}{\pi}\frac{F_0\left(\sqrt{J^2\sin^2(ka)+|\Delta|^2}\right)}{\sqrt{J^2\sin^2(ka)+|\Delta|^2}}=\frac{1}{\kappa}~\text{for CDW.}
\eeq
The SC gap equation can be easily mapped to the CDW gap equation by a change of variables $k\to k\pm k_F$ where $k_F=\pi/2a$. Note that we can do this because we considered a half-filled free electron band $\epsilon_k=-J\cos(ka)$ that results in the symmetry $\epsilon_{k+k_F}=-\epsilon_{-k+k_F}$. This is a consequence of the symmetry of $H_{\rm el}$ under the particle-hole transformation $\hat c_\sigma(k\pm k_F)\to \hat c^{\dagger}_\sigma(-k\pm k_F)$. In the Nambu structure this transformation translates to $\hat c_{\alpha}(k\pm k_F)\to \hat c^{\dagger}_{\alpha}(-k\pm k_F)$ for SC and $\hat c_{\alpha}(k)\to \hat c^{\dagger}_{\alpha}(-k)$ for CDW. 
Let us show that $H_{\rm{el}}$ is invariant under this transformation. For SC, we have
\beq
\begin{split}
    H_{\rm el}&=\int_{-\pi/a}^{\pi/a}\frac{dk}{2\pi}\left[\xi_k\hat{c}^{\dagger}_+(k)\hat{c}_+(k)-\xi_k\hat{c}^{\dagger}_-(k)\hat{c}_-(k)+\Delta\hat{c}^{\dagger}_+(k)\hat{c}_-(k)+\Delta^*\hat{c}^{\dagger}_-(k)\hat{c}_+(k)\right]\\
    &=\int_{-\pi/2a}^{\pi/2a}\frac{dk}{2\pi}\left[\xi_{k+k_F}\hat{c}^{\dagger}_+(k+k_F)\hat{c}_+(k+k_F)-\xi_{k+k_F}\hat{c}^{\dagger}_-(k+k_F)\hat{c}_-(k+k_F)\right.\\&~~~~~~~~~~~~~~~~~~~~~~~\left.+\Delta\hat{c}^{\dagger}_+(k+k_F)\hat{c}_-(k+k_F)+\Delta^*\hat{c}^{\dagger}_-(k+k_F)\hat{c}_+(k+k_F)\right]\\
    &+\int_{-\pi/2a}^{\pi/2a}\frac{dk}{2\pi}\left[\xi_{k-k_F}\hat{c}^{\dagger}_+(k-k_F)\hat{c}_+(k-k_F)-\xi_{k-k_F}\hat{c}^{\dagger}_-(k-k_F)\hat{c}_-(k-k_F)\right.\\&~~~~~~~~~~~~~~~~~~~~~~~\left.+\Delta\hat{c}^{\dagger}_+(k-k_F)\hat{c}_-(k-k_F)+\Delta^*\hat{c}^{\dagger}_-(k-k_F)\hat{c}_+(k-k_F)\right]\\
    &\to \tilde{H}_{\rm el}=\int_{-\pi/2a}^{\pi/2a}\frac{dk}{2\pi}\left[\xi_{k+k_F}\hat{c}_+(-k+k_F)\hat{c}^{\dagger}_+(-k+k_F)-\xi_{k+k_F}\hat{c}_-(-k+k_F)\hat{c}^{\dagger}_-(-k+k_F)\right.\\&~~~~~~~~~~~~~~~~~~~~~~~\left.+\Delta^*\hat{c}_+(-k+k_F)\hat{c}^{\dagger}_-(-k+k_F)+\Delta\hat{c}_-(-k+k_F)\hat{c}^{\dagger}_+(-k+k_F)\right]\\
    &+\int_{-\pi/2a}^{\pi/2a}\frac{dk}{2\pi}\left[\xi_{k-k_F}\hat{c}_+(-k-k_F)\hat{c}^{\dagger}_+(-k-k_F)-\xi_{k-k_F}\hat{c}_-(-k-k_F)\hat{c}^{\dagger}_-(-k-k_F)\right.\\&~~~~~~~~~~~~~~~~~~~~~~~\left.+\Delta^*\hat{c}_+(-k-k_F)\hat{c}^{\dagger}_-(-k-k_F)+\Delta\hat{c}_-(-k-k_F)\hat{c}^{\dagger}_+(-k-k_F)\right]\\
    &=\int_{-\pi/2a}^{\pi/2a}\frac{dk}{2\pi}\left[\xi_{k+k_F}\hat{c}^{\dagger}_+(k+k_F)\hat{c}_+(k+k_F)-\xi_{k+k_F}\hat{c}^{\dagger}_-(k+k_F)\hat{c}_-(k+k_F)\right.\\&~~~~~~~~~~~~~~~~~~~~~~~\left.-\Delta\hat{c}^{\dagger}_+(k+k_F)\hat{c}_-(k+k_F)-\Delta^*\hat{c}^{\dagger}_-(k+k_F)\hat{c}_+(k+k_F)\right]\\
    &+\int_{-\pi/2a}^{\pi/2a}\frac{dk}{2\pi}\left[\xi_{k-k_F}\hat{c}^{\dagger}_+(k-k_F)\hat{c}_+(k-k_F)-\xi_{k-k_F}\hat{c}^{\dagger}_-(k-k_F)\hat{c}_-(k-k_F)\right.\\&~~~~~~~~~~~~~~~~~~~~~~~\left.-\Delta\hat{c}^{\dagger}_+(k-k_F)\hat{c}_-(k-k_F)-\Delta^*\hat{c}^{\dagger}_-(k-k_F)\hat{c}_+(k-k_F)\right]\\&=\int_{-\pi/2a}^{\pi/2a}\frac{dk}{2\pi}\left[\xi_{k+k_F}\hat{c}^{\dagger}_+(k+k_F)\hat{c}_+(k+k_F)-\xi_{k+k_F}\hat{c}^{\dagger}_-(k+k_F)\hat{c}_-(k+k_F)\right.\\&~~~~~~~~~~~~~~~~~~~~~~~\left.+\Delta\hat{c}^{\dagger}_+(k+k_F)\hat{c}_-(k+k_F)+\Delta^*\hat{c}^{\dagger}_-(k+k_F)\hat{c}_+(k+k_F)\right]\\
    &+\int_{-\pi/2a}^{\pi/2a}\frac{dk}{2\pi}\left[\xi_{k-k_F}\hat{c}^{\dagger}_+(k-k_F)\hat{c}_+(k-k_F)-\xi_{k-k_F}\hat{c}^{\dagger}_-(k-k_F)\hat{c}_-(k-k_F)\right.\\&~~~~~~~~~~~~~~~~~~~~~~~\left.+\Delta\hat{c}^{\dagger}_+(k-k_F)\hat{c}_-(k-k_F)+\Delta^*\hat{c}^{\dagger}_-(k-k_F)\hat{c}_+(k-k_F)\right]=H_{\rm el}\\
\end{split}
\eeq
Note that we used the fact that $\xi_{-k\pm k_F}=-\xi_{k\pm k_F}$. We also absorbed a minus sign in $\Delta$ as it has a gauge symmetry and the gap equation only depends on $|\Delta|^2$.
For CDW we have instead
\beq
\begin{split}
    H_{\rm el}&=\int_{-\pi/2a}^{\pi/2a}\frac{dk}{\pi}\left[\xi_k\hat{c}^{\dagger}_+(k)\hat{c}_+(k)-\xi_k\hat{c}^{\dagger}_-(k)\hat{c}_-(k)+\Delta\hat{c}^{\dagger}_+(k)\hat{c}_-(k)+\Delta^*\hat{c}^{\dagger}_-(k)\hat{c}_+(k)\right]\\
    &\to \tilde{H}_{\rm el}=\int_{-\pi/2a}^{\pi/2a}\frac{dk}{\pi}\left[\xi_k\hat{c}_+(-k)\hat{c}^{\dagger}_+(-k)-\xi_k\hat{c}_-(-k)\hat{c}^{\dagger}_-(-k)+\Delta^*\hat{c}_+(-k)\hat{c}^{\dagger}_-(-k)+\Delta\hat{c}_-(-k)\hat{c}^{\dagger}_+(-k)\right]\\
    &=\int_{-\pi/2a}^{\pi/2a}\frac{dk}{\pi}\left[\xi_k\hat{c}^{\dagger}_+(k)\hat{c}_+(k)-\xi_k\hat{c}^{\dagger}_-(k)\hat{c}_-(k)-\Delta\hat{c}^{\dagger}_+(k)\hat{c}_-(k)-\Delta^*\hat{c}^{\dagger}_-(k)\hat{c}_+(k)\right]\\
    &=\int_{-\pi/2a}^{\pi/2a}\frac{dk}{\pi}\left[\xi_k\hat{c}^{\dagger}_+(k)\hat{c}_+(k)-\xi_k\hat{c}^{\dagger}_-(k)\hat{c}_-(k)+\Delta\hat{c}^{\dagger}_+(k)\hat{c}_-(k)+\Delta^*\hat{c}^{\dagger}_-(k)\hat{c}_+(k)\right]=H_{\rm el}\\
\end{split}
\eeq
Here we used the fact that $\xi_{-k}=-\xi_k$.
So, under this transformation $H_{\rm el}$ stays invariant.

Now let us see how $H_{\rm el-ph}$ behaves under this transformation.
For SC, we have
\beq
\begin{split}
    &H_{\mathrm{el-ph}}=\sum_{\alpha=\{+,-\}}\int _{-\pi/a}^{\pi/a}\frac{dk}{2\pi}\int _{-\pi/a}^{\pi/a}\frac{dk'}{2\pi}~g^{\alpha}_{k,k'}\hat{c}^{\dagger}_{\alpha}(k)\hat{c}_{\alpha}(k')\hat{\phi}(k-k')\\
    &=\sum_{\alpha=\{+,-\}}\int _{-\pi/2a}^{\pi/2a}\frac{dk}{2\pi}\int _{-\pi/2a}^{\pi/2a}\frac{dk'}{2\pi}~\left\{\left[g^{\alpha}_{k+k_F,k'+k_F}\hat{c}^{\dagger}_{\alpha}(k+k_F)\hat{c}_{\alpha}(k'+k_F)+g^{\alpha}_{k-k_F,k'-k_F}\hat{c}^{\dagger}_{\alpha}(k-k_F)\hat{c}_{\alpha}(k'-k_F)\right]\hat{\phi}(k-k')\right.\\
    &~~~~~~~~~~~~~~~~~~~~~~\left.+g^{\alpha}_{k+k_F,k'-k_F}\hat{c}^{\dagger}_{\alpha}(k+k_F)\hat{c}_{\alpha}(k'-k_F)\hat{\phi}(k-k'+2k_F)+g^{\alpha}_{k-k_F,k'+k_F}\hat{c}^{\dagger}_{\alpha}(k-k_F)\hat{c}_{\alpha}(k'+k_F)\hat{\phi}(k-k'-2k_F)\right\}\\
    &\to  
    \tilde{H}_{\rm el-ph}=\sum_{\alpha=\{+,-\}}\int _{-\pi/2a}^{\pi/2a}\frac{dk}{2\pi}\int _{-\pi/2a}^{\pi/2a}\frac{dk'}{2\pi}~\left\{g^{\alpha}_{k+k_F,k'+k_F}\hat{c}_{\alpha}(-k+k_F)\hat{c}^{\dagger}_{\alpha}(-k'+k_F)\hat{\phi}(k-k')\right.\\
&~~~~~~~~~~~~~~~~~~~~~~~~~~~~~~~~~~~~~~~~~~~~~~~~~~~~~~~~~~~~~~~~~~~~~~~~~~~~+g^{\alpha}_{k-k_F,k'-k_F}\hat{c}_{\alpha}(-k-k_F)\hat{c}^{\dagger}_{\alpha}(-k'-k_F)\hat{\phi}(k-k')\\
&~~~~~~~~~~~~~~~~~~~~~~~~~~~~~~~~~~~~~~~~~~~~~~~~~~~~~~~~~~~~~+g^{\alpha}_{k+k_F,k'-k_F}\hat{c}_{\alpha}(-k+k_F)\hat{c}^{\dagger}_{\alpha}(-k'-k_F)\hat{\phi}(k-k'+2k_F)\\
&~~~~~~~~~~~~~~~~~~~~~~~~~~~~~~~~~~~~~~~~~~~~~~~~~~~~~~~~~~~~~\left.+g^{\alpha}_{k-k_F,k'+k_F}\hat{c}_{\alpha}(-k-k_F)\hat{c}^{\dagger}_{\alpha}(-k'+k_F)\hat{\phi}(k-k'-2k_F)\right\}\\
    &=\sum_{\alpha=\{+,-\}}\int _{-\pi/2a}^{\pi/2a}\frac{dk}{2\pi}\int _{-\pi/2a}^{\pi/2a}\frac{dk'}{2\pi}~\left\{-\left[g^{\alpha}_{-k'+k_F,-k+k_F}\hat{c}^{\dagger}_{\alpha}(k+k_F)\hat{c}_{\alpha}(k'+k_F)+g^{\alpha}_{-k'-k_F,-k-k_F}\hat{c}^{\dagger}_{\alpha}(k-k_F)\hat{c}_{\alpha}(k'-k_F)\right]\hat{\phi}(k-k')\right.\\
    &~~~~~~~~~~~~~~~~~~~~~~\left.-g^{\alpha}_{-k'+k_F,-k-k_F}\hat{c}^{\dagger}_{\alpha}(k+k_F)\hat{c}_{\alpha}(k'-k_F)\hat{\phi}(k-k'-2k_F)-g^{\alpha}_{-k'-k_F,-k+k_F}\hat{c}^{\dagger}_{\alpha}(k-k_F)\hat{c}_{\alpha}(k'+k_F)\hat{\phi}(k-k'+2k_F)\right\}\\
    &=\sum_{\alpha=\{+,-\}}\int _{-\pi/2a}^{\pi/2a}\frac{dk}{2\pi}\int _{-\pi/2a}^{\pi/2a}\frac{dk'}{2\pi}~\left\{-\left[g^{\alpha}_{k+k_F,k'+k_F}\hat{c}^{\dagger}_{\alpha}(k+k_F)\hat{c}_{\alpha}(k'+k_F)+g^{\alpha}_{k'-k_F,k-k_F}\hat{c}^{\dagger}_{\alpha}(k-k_F)\hat{c}_{\alpha}(k'-k_F)\right]\hat{\phi}(k-k')\right.\\
    &~~~~~~~~~~~~~~~~~~~~~~\left.+g^{\alpha}_{k+k_F,k'-k_F}\hat{c}^{\dagger}_{\alpha}(k+k_F)\hat{c}_{\alpha}(k'-k_F)\hat{\phi}(k-k'-2k_F)-g^{\alpha}_{k-k_F,k'+k_F}\hat{c}^{\dagger}_{\alpha}(k-k_F)\hat{c}_{\alpha}(k'+k_F)\hat{\phi}(k-k'+2k_F)\right\}\\&\ne H_{\rm el-ph}.\\
\end{split}
\eeq
Here we have used the fact that $g^{\alpha}_{-k+k_F,-k'+k_F}=g^{\alpha}_{k+k_F,k'+k_F}$, $g^{\alpha}_{-kk_F,-k'-k_F}=g^{\alpha}_{k-k_F,k'-k_F}$, $g^{\alpha}_{-k+k_F,-k'-k_F}=-g^{\alpha}_{k+k_F,k'-k_F}$ and $g^{\alpha}_{-k-k_F,-k'+k_F}=-g^{\alpha}_{k-k_F,k'+k_F}$.

For CDW we have 
\beq
\begin{split}
    &H_{\mathrm{el-ph}}=\sum_{\alpha=\{+,-\}}\int _{-\pi/2a}^{\pi/2a}\frac{dk}{\pi}\int _{-\pi/2a}^{\pi/2a}\frac{dk'}{\pi}~g^{\alpha}_{k,k'}\hat{c}^{\dagger}_{\alpha}(k)\hat{c}_{\alpha}(k')\hat{\phi}(k-k')\\
    &\to  \tilde{H}_{\rm el-ph}=\sum_{\alpha=\{+,-\}}\int _{-\pi/2a}^{\pi/2a}\frac{dk}{\pi}\int _{-\pi/2a}^{\pi/2a}\frac{dk'}{\pi}~g^{\alpha}_{k,k'}\hat{c}_{\alpha}(-k)\hat{c}^{\dagger}_{\alpha}(-k')\hat{\phi}(k-k')\\
    &~~~~~~~~~~~~~~~~~=\sum_{\alpha=\{+,-\}}\int _{-\pi/2a}^{\pi/2a}\frac{dk}{\pi}\int _{-\pi/2a}^{\pi/2a}\frac{dk'}{\pi}~(-1)g^{\alpha}_{-k',-k}\hat{c}^{\dagger}_{\alpha}(k)\hat{c}_{\alpha}(k')\hat{\phi}(k-k')\\
    &~~~~~~~~~~~~~~~~~=\sum_{\alpha=\{+,-\}}\int _{-\pi/2a}^{\pi/2a}\frac{dk}{\pi}\int _{-\pi/2a}^{\pi/2a}\frac{dk'}{\pi}~(-1)g^{\alpha}_{k,k'}\hat{c}^{\dagger}_{\alpha}(k)\hat{c}_{\alpha}(k')\hat{\phi}(k-k')\\
&\ne H_{\rm el-ph}.\\
\end{split}
\eeq
So, $H_{\rm el-ph}$ is not invariant under the particle-hole transformation. So, SC and CDW cannot be mapped into each other when they are coupled to the cavity photons. Hence, we expect to observe different out-of-equilibrium behavior for them.

\section{Material in optical cavity}
The gap equation in Eq.~\eqref{seq:gapeqgen} depends on the specifics of the distribution function. We put the material in an optical cavity and use the resultant non-thermal distribution function in the gap equation. In the weak light-matter coupling strength ($g_0$) we do a perturbative calculation of $\mathcal{O}(g_0^2)$. So, we derive the self-energies of the electrons and the non-thermal distribution using quantum kinetic equations (QKE) due to coupling to the cavity before deriving the non-thermal gap equation. 
\subsection{Quantum Kinetic Equations}
From the gap equation in Eq.~\eqref{seq:gapeqgen}, it is clear that we need to calculate the self-energies of the electrons due to the light-matter coupling. In this section we do that before deriving the non-thermal gap equation.

Let's re-derive QKE assuming a full $F$ matrix in the Bogoliubov basis.
We have \cite{kamenev_book,Rafa2025}
\beq
[G^{-1}_0]^{K}-\left(G^{R^{-1}}_0\cdot F- F\cdot G^{A^{-1}}_0\right)=\Sigma^K-\left(\Sigma^R\cdot F- F\cdot \Sigma^A\right)
\eeq
where
\beq
\begin{split}
\mathrm{L.H.S.}&=[G^{-1}_0]^{K}-\left(G^{R^{-1}}_0\cdot F- F\cdot G^{A^{-1}}_0\right)\\
&=
\begin{bmatrix}
2i\gamma F_0 && 0\\
0 && 2i\gamma F_0
\end{bmatrix}
-\left(
\begin{bmatrix}
\omega-E_k+i\gamma  && 0  \\
 0  &&  \omega+E_k+i\gamma
\end{bmatrix}
\cdot 
\begin{bmatrix}
F_{pp}  && F_{pm}  \\
F_{mp}   && F_{mm}
\end{bmatrix}
-h.c.
\right)\\
&=
\begin{bmatrix}
2i\gamma(F_0-F_{pp})  && 2(E_k-i\gamma)F_{pm}  \\
-2(E_k+i\gamma)F_{mp}  && 2i\gamma(F_0-F_{mm})
\end{bmatrix}
\end{split}
\eeq
and 
\beq
\begin{split}
&\mathrm{R.H.S.}\\
&=\Sigma^{K}-\left(\Sigma^{R}\cdot F- F\cdot \Sigma^{A}\right)\\
&=
\begin{bmatrix}
\Sigma^K_{pp} && \Sigma^K_{pm}\\
\Sigma^K_{mp} && \Sigma^K_{mm}
\end{bmatrix}
-\left(
\begin{bmatrix}
\Sigma^R_{pp}  && \Sigma^R_{pm}  \\
 \Sigma^R_{mp}  &&  \Sigma^R_{mm}
\end{bmatrix}
\cdot 
\begin{bmatrix}
F_{pp}   && F_{pm}  \\
F_{mp}   && F_{mm}
\end{bmatrix}
-
\begin{bmatrix}
F_{pp}   && F_{pm}  \\
F_{mp}   && F_{mm}
\end{bmatrix}
\cdot
\begin{bmatrix}
\Sigma^A_{pp}  && \Sigma^A_{pm}  \\
 \Sigma^A_{mp}  &&  \Sigma^A_{mm}
\end{bmatrix}
\right)\\
&=
\begin{bmatrix}
\Sigma^K_{pp}-(\Sigma^R_{pp}-\Sigma^A_{pp})F_{pp} + (F_{pm}\Sigma^A_{mp}-\Sigma^R_{pm}F_{mp})   && \Sigma^K_{pm}+(F_{pp}\Sigma^A_{pm}-\Sigma^R_{pm}F_{mm})-(\Sigma^R_{pp}-\Sigma^A_{mm})F_{pm} \\
\Sigma^K_{mp}+(F_{mm}\Sigma^A_{mp}-\Sigma^R_{mp}F_{pp})-(\Sigma^R_{mm}-\Sigma^A_{pp})F_{mp}  && \Sigma^K_{mm} - (\Sigma^R_{mm} - \Sigma^A_{mm})F_{mm} + 
(F_{mp}\Sigma^A_{pm}-\Sigma^R_{mp}F_{pm})
\end{bmatrix}
\end{split}
\eeq
We will do a perturbative calculation in the light-matter coupling $g$ and keep terms upto $\mathcal{O}(g^2)$.
Equating off-diagonal terms of L.H.S. and R.H.S. we get 
\beq\label{Fpm}
\begin{split}
    F_{pm}(k,\omega)&\approx\frac{\Sigma^K_{pm}-(\Sigma^R_{pm}-\Sigma^A_{pm})F_0}{2(E_k-i\gamma)}
    =\frac{\Sigma_{pm}(k,\omega)}{2(E_k-i\gamma)}\\
    F_{mp}(k,\omega)&\approx-\frac{\Sigma^K_{mp}-(\Sigma^R_{mp}-\Sigma^A_{mp})F_0}{2(E_k+i\gamma)}
    =-\frac{\Sigma_{mp}(k,\omega)}{2(E_k+i\gamma)}\\
\end{split}
\eeq
From the diagonal terms we get 
\beq
\begin{split}
    2i\gamma(F_0-F_{pp})&\approx\Sigma^K_{pp}-(\Sigma^R_{pp}-\Sigma^A_{pp})F_0 =\Sigma_{pp}~~~\mathrm{and}\\
       2i\gamma(F_0-F_{mm})&\approx\Sigma^K_{mm}-(\Sigma^R_{mm}-\Sigma^A_{mm})F_0 =\Sigma_{mm}
\end{split}
\eeq
So,
\beq\label{Fdiag}
\begin{split}
    F_{pp}(k,\omega)&=F_0(\omega)-\frac{\Sigma_{pp}(k,\omega)}{2i\gamma}~~~\mathrm{and}\\
       F_{mm}(k,\omega)&=F_0(\omega)-\frac{\Sigma_{mm}(k,\omega)}{2i\gamma}
\end{split}
\eeq
Note that we have defined
\beq\label{seq:sigfactor}
\Sigma_{\alpha\beta}(k,\omega)=\Sigma^K_{\alpha\beta}(k,\omega)-[\Sigma^R_{\alpha\beta}(k,\omega)-\Sigma^A_{\alpha\beta}(k,\omega)]F_0(\omega).
\eeq
\subsection{Self-energies and non-thermal distribution}
%In the non-thermal gap equation in \cref{gapeqgen}, we need self-energies of the fermions due to the coupling to the photons for $\delq=0$ and linear in $\delq$. 
We need to derive the self-energies to the fermions due to the light-matter coupling. Note that we will derive them at the classical saddle point i.e. mean field approximation. In that case we set $\delq=0$ and consider only $\delc$.
Before deriving them using diagrammatic techniques, not that the photon Green's functions
\beq
\begin{split}
&D^R_{0}(q,\omega)=[D^A_{0}(q,\omega)]^*=\frac{\nu_0}{2\pi}\frac{1}{(\omega+i\gamma_B)^2-\nu_q^2}=\frac{\nu_0}{4\pi\nu_q}\left[\frac{1}{\omega+i\gamma_B-\nu_q}-\frac{1}{\omega+i\gamma_B+\nu_q}\right],\\
&\mathrm{Im}D^R_{0}(q,\omega)=-\frac{\nu_0}{4\pi\nu_q}\left[\frac{\gamma_B}{(\omega-\nu_q)^2+\gamma_B^2}-\frac{\gamma_B}{(\omega+\nu_q)^2+\gamma_B^2}\right],~~~\mathrm{and}\\
&D^K(q,\omega)=2i\mathrm{Im}D^R_{0}(q,\omega)B_{T_{\mathrm{cav}}}(\omega)\\
\end{split}
\eeq

\begin{figure}[htbp]
    \centering
    \includegraphics[width=\textwidth]{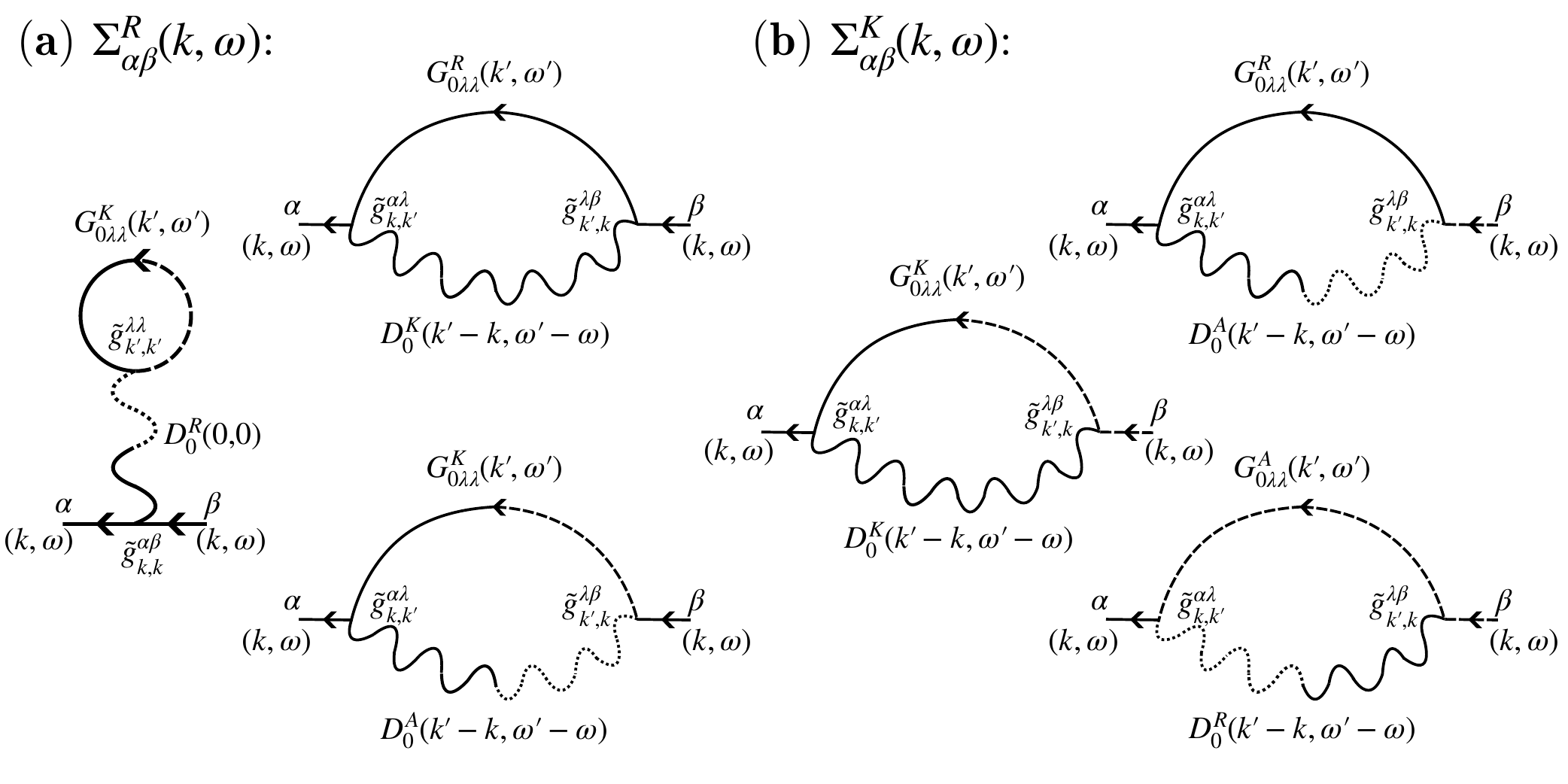}
    \caption{Self-energy diagrams of electron due to light-matter coupling for $\delq=0$. Solid lines represent $\psi_{\alpha1}(k,\omega)$ and dashed lines represent $\psi_{\alpha2}(k,\omega)$. Solid curly lines represent $\phi_{\rm cl}(q,\Omega)$ and dotted curly lines represent $\phi_{\rm q}(q,\Omega)$. All the diagrams are of $\mathcal{O}(g_0^2)$. (a) Retarded self-energy diagrams are listed. The left-most one is Hartree diagram while the other two are Fock diagrams. (b) Keldysh self-energy diagrams are listed. All of them are Fock diagrams.}
    \label{sfig:selfenergy}
\end{figure}

In \cref{sfig:selfenergy} we sketch the self-energies which are lowest order in the light-matter coupling [$\mathcal{O}(g_0^2)$]. For these diagrams we followed the notation in Ref.~\onlinecite{kamenev_book}, i.e. solid lines represent, $\psi_1/\bar{\psi}_1$, dashed lines represent $\psi_1/\bar{\psi}_1$, solid wavy lines represent $\phi_{\rm cl}$ and dotted wavy lines represent $\phi_{\rm q}$.
From \cref{sfig:selfenergy}(a), we get the retarded self-energy
\beq
\begin{split}
    &\Sigma^R_{\alpha\beta}(k,\omega)=[\Sigma^A_{\beta\alpha}(k,\omega)]^*\\
    &=i\sum_{\lambda=\{p,m\}}\sum_{k'}\int\frac{d\omega'}{2\pi}(-1)(i\tilde{g}^{\alpha\beta}_{k,k})(i\tilde{g}^{\lambda\lambda}_{k',k'})iG^K_{0\lambda\lambda}(k',\omega')iD^R_0(0,0)\\
    &+i\sum_{\lambda=\{p,m\}}\sum_{k'}\int\frac{d\omega'}{2\pi}(i\tilde{g}^{\alpha\lambda}_{k,k'})(i\tilde{g}^{\lambda\beta}_{k',k})\left[iG^R_{0\lambda\lambda}(k',\omega')iD^K_0(k'-k,\omega'-\omega)+iG^K_{0\lambda\lambda}(k',\omega')iD^A_0(k'-k,\omega'-\omega)\right]\\
    &=-\sum_{\lambda=\{p,m\}}\sum_{k'}\int\frac{d\omega'}{\pi}\tilde{g}^{\alpha\lambda}_{k,k'}\tilde{g}^{\lambda\beta}_{k',k}[G^R_{0\lambda\lambda}(k',\omega')\mathrm{Im}D^R_0(k'-k,\omega'-\omega)B_{\tcav}(\omega'-\omega)+\mathrm{Im}G^R_{0\lambda\lambda}(k',\omega')D^A_0(k'-k,\omega'-\omega)F_0(\omega')].\\
\end{split}
\eeq
Note that the Hartree term vanishes.
From \cref{sfig:selfenergy}(b), we get the Keldysh self-energy
\beq
\begin{split}
    &\Sigma^K_{\alpha\beta}(k,\omega)\\
    &=i\sum_{\lambda=\{p,m\}}\sum_{k'}\int\frac{d\omega'}{2\pi}(i\tilde{g}^{\alpha\lambda}_{k,k'})(i\tilde{g}^{\lambda\beta}_{k',k})\left[iG^R_{0\lambda\lambda}(k',\omega')iD^A_0(k'-k,\omega'-\omega)+iG^A_{0\lambda\lambda}(k',\omega')iD^R_0(k'-k,\omega'-\omega)\right]\\
    &+i\sum_{\lambda=\{p,m\}}\sum_{k'}\int\frac{d\omega'}{2\pi}(i\tilde{g}^{\alpha\lambda}_{k,k'})(i\tilde{g}^{\lambda\beta}_{k',k})\left[iG^K_{0\lambda\lambda}(k',\omega')iD^K_0(k'-k,\omega'-\omega)\right]\\
    &=-2\sum_{\lambda=\{p,m\}}\sum_{k'}\int\frac{d\omega'}{\pi}\tilde{g}^{\alpha\lambda}_{k,k'}\tilde{g}^{\lambda\beta}_{k',k}\left\{\im G^R_{0\lambda\lambda}(k',\omega')\im D^R_0(k'-k,\omega'-\omega)[F_0(\omega')B_{\tcav}(\omega'-\omega)-1]\right\}.\\
\end{split}
\eeq
Hence we get (noting $\tilde{g}^{\alpha\beta}_{k,k'}=[\tilde{g}^{\beta\alpha}_{k',k}]^*$),
\beq
\begin{split}
    &\Sigma_{\alpha\beta}(k,\omega)=\Sigma^K_{\alpha\beta}(k,\omega)-[\Sigma^R_{\alpha\beta}(k,\omega)-\Sigma^A_{\alpha\beta}(k,\omega)]F_0(\omega)\\
  &=-i2\sum_{\lambda=\{p,m\}}\sum_{k'}\int\frac{d\omega'}{\pi}\tilde{g}^{\alpha\lambda}_{k,k'}\tilde{g}^{\lambda\beta}_{k',k}\mathrm{Im}G^R_{0\lambda\lambda}(k',\omega')\mathrm{Im}D^R_0(k'-k,\omega'-\omega)\tilde H_0(\omega,\omega')\\
\end{split}
\eeq
where the distribution factor
\beq
\begin{split}
\tilde H_0(\omega,\omega')&=B_{\tcav}(\omega'-\omega)\left[F_0(\omega')-F_0(\omega)\right]+F_0(\omega')F_0(\omega)-1=\left[B_{\tcav}(\omega-\omega')-B_{\tcry}(\omega-\omega')\right]\left[F_0(\omega)-F_0(\omega')\right].
\end{split}
\eeq
So for the distribution functions we get
\beq\label{seq:nthFdia}
\begin{split}
    F_{pp}(k,\omega)=-F_{mm}(k,-\omega)&=F_0(\omega)-\frac{1}{2\gamma}\sum_{k'}(g_{k,k'})^2\left[\left(1+\frac{\xi_{k}\xi_{k'}+\eta|\delc|^2}{E_{k}E_{k'}}\right)\mathrm{Im}D^R_0(k'-k,E_{k'}-\omega)\tilde H_0(\omega,E_{k'})\right.\\
     &~~~~~~~~~~~~~~~~~~~~~~~~~~~~~~~~~~~~~~~~~~~\left.-\left(1-\frac{\xi_{k}\xi_{k'}+\eta|\delc|^2}{E_{k}E_{k'}}\right)\mathrm{Im}D^R_0(k'-k,E_{k'}+\omega)\tilde H_0(-\omega,E_{k'})\right].\\
\end{split}
\eeq
and
\beq\label{seq:nthFdiaoff}
\begin{split}
    &F_{pm}(k,\omega)=[F_{mp}(k,\omega)]^*\\
      &= i\frac{\delc}{2(E_k-i\gamma)}\sum_{k'}(g^{+}_{k,k'})^2\left(\frac{\eta\xi_{k}-\xi_{k'}}{E_{k}E_{k'}}\right)\left[\mathrm{Im}D^R_0(k'-k,E_{k'}-\omega)\tilde H_0(\omega,E_{k'})+\mathrm{Im}D^R_0(k'-k,E_{k'}+\omega)\tilde H_0(-\omega,E_{k'})\right].\\
\end{split}
\eeq

\begin{figure}[h]
    \centering
    \includegraphics[width=\textwidth]{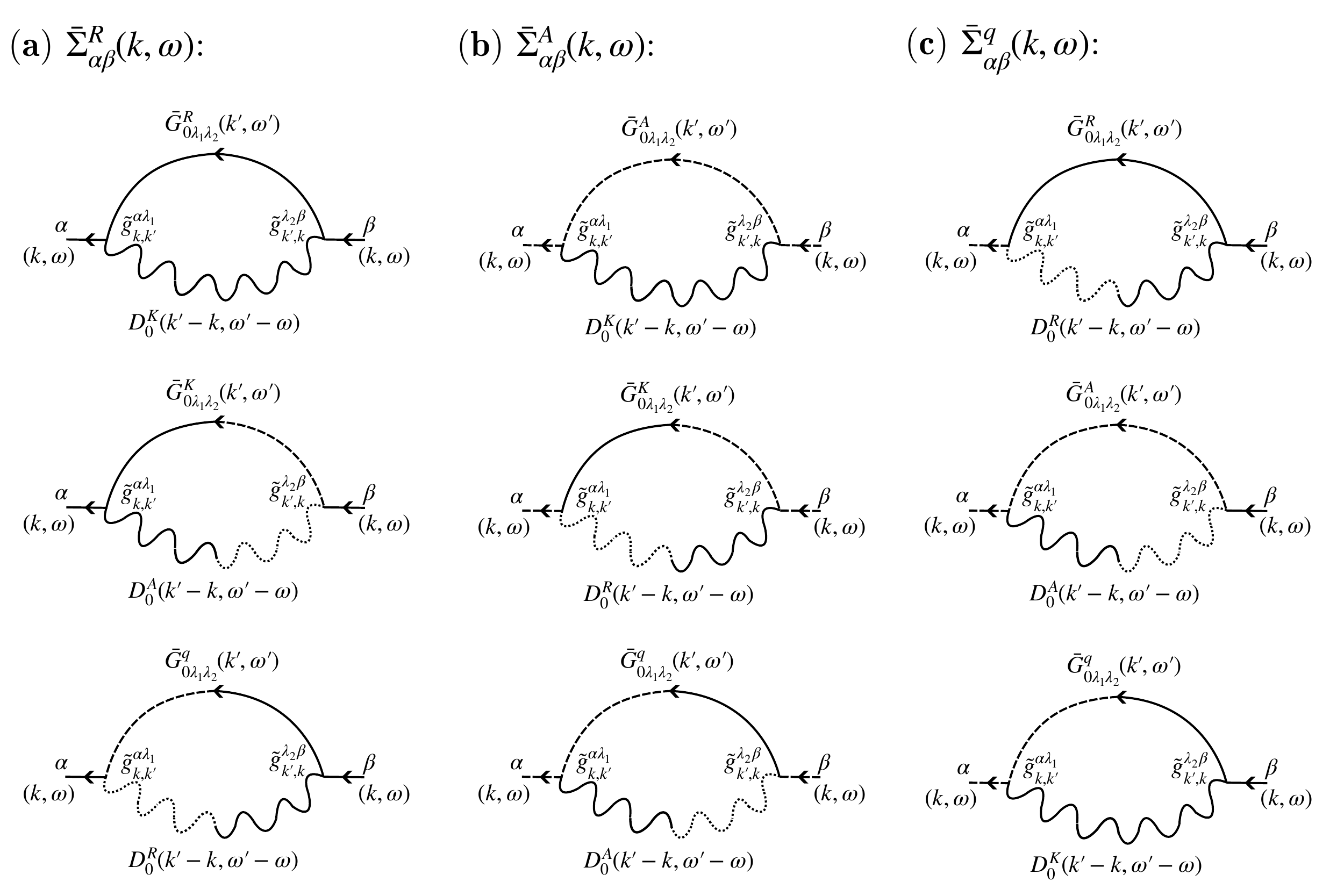}
    \caption{Diagrammatic representation of the derivative of the self-energies with respect to $\delq$ defined as $\bar{\Sigma}^{R/A/q}=\left.\frac{\partial\Sigma^{R/A/q}}{\partial \delq}\right|_{\delq=0}$. The derivative directly acts on the electronic Green's function in the diagrams and they are defined by $\bar{G}^{R/A/K/q}=\left.\frac{\partial G^{R/A/K/q}}{\partial \delq}\right|_{\delq=0}$. (a) Diagrams for derivative of the retarded self-energy. (b) Diagrams for derivative of the advanced self-energy. (c) Diagrams for derivative of the `quantum' self-energy. Note that while $\Sigma^{q}=0$, $\bar{\Sigma}^{q}$ is finite. }
    \label{sfig:selfenergyqdot}
\end{figure}
$\bullet$\textbf{Variation of self-energies:}\\

In the non-thermal gap equation in \cref{seq:gapeqgen}, in addition to the mean field self-energies, we need the variation of them w.r.t. $\delq$ and for that we need to derive self-energies linear in $\delq$. We sketch them in \cref{sfig:selfenergyqdot}.
In the self-energy diagrams we need to consider fermion Green's functions that depends on $\delq$. When we differentiate the self-energy w.r.t. $\delq$, that will apply on the fermion Green's function in the diagrams. So, let us first list them before evaluating the $\bar{\Sigma}$. 
Note that
\beq
\begin{split}
    G_0(\delq)G^{-1}_0(\delq)=\mathbb{I}_{4\times4}
    &\implies \frac{\delta G_0(\delq)}{\delta \delq}G^{-1}_0(\delq)+G_0(\delq)\frac{\delta G^{-1}_0(\delq)}{\delta \delq}=0\\
     &\implies\frac{\delta G_0(\delq)}{\delta \delq}|_{\delq=0}=-G_0(0)\frac{\delta G^{-1}_0(\delq)}{\delta \delq}|_{\delq=0}G_0(0)\\
\end{split}
\eeq
Now,
\beq
G_0=
\begin{bmatrix}
    G^R_0 && G^K_0\\
    0 && G^A_0
\end{bmatrix}
~~~\mathrm{and}~~~
\frac{\delta G_0^{-1}(\delq)}{\delta \delq}|_{\delq=0}
=\begin{bmatrix}
	0 && \frac{\delta G^{q^{-1}}_0}{\delta \delq}|_{\delq=0}\\
	\frac{\delta G^{q^{-1}}_0}{\delta \delq}|_{\delq=0} && 0
\end{bmatrix}
\eeq
So,
\beq
\begin{split}
    & \frac{\delta G_0(\delq)}{\delta \delq}|_{\delq=0}=   \begin{bmatrix}
	 \bar{G}^R_0 &&  \bar{G}^K_0 \\
	 \bar{G}^q_0  &&  \bar{G}^A_0 
\end{bmatrix} =
   \begin{bmatrix}
	\frac{\delta G^R_0(\delq)}{\delta \delq}|_{\delq=0}&& \frac{\delta G^K_0(\delq)}{\delta \delq}|_{\delq=0}\\
	\frac{\delta G^q_0(\delq)}{\delta \delq}|_{\delq=0} && \frac{\delta G^A_0(\delq)}{\delta \delq}|_{\delq=0}
\end{bmatrix} \\
&=
\begin{bmatrix}
	-G^K_0\frac{\delta G^{q^{-1}}_0}{\delta \delq}|_{\delq=0}G^R_0 && -G^R_0\frac{\delta G^{q^{-1}}_0}{\delta \delq}|_{\delq=0}G^A_0-G^K_0\frac{\delta G^{q^{-1}}_0}{\delta \delq}|_{\delq=0}G^K_0\\
	-G^A_0\frac{\delta G^{q^{-1}}_0}{\delta \delq}|_{\delq=0}G^R_0 && -G^A_0\frac{\delta G^{q^{-1}}_0}{\delta \delq}|_{\delq=0}G^K_0
\end{bmatrix}
\end{split}
\eeq
Hence
\beq
\begin{split}
    &\bar{G}^R_0=\frac{\delta G^R_0(\delq)}{\delta \delq}|_{\delq=0}=-G^K_0\frac{\delta G^{q^{-1}}_0}{\delta \delq}|_{\delq=0}G^R_0=-
     \begin{bmatrix}
    G^K_{0pp} && 0\\
    0 && G^K_{0mm}
    \end{bmatrix}
     \begin{bmatrix}
    -u_kv_k^* && -u_k^2\\
    (v_k^*)^2 && u_kv_k^*
    \end{bmatrix}
     \begin{bmatrix}
    G^R_{0pp} && 0\\
    0 && G^R_{0mm}
    \end{bmatrix}\\
    &=\begin{bmatrix}
  u_kv_k^*G^K_{0pp}G^R_{0pp} && u_k^2G^K_{0pp}G^R_{0mm} \\
    -(v_k^*)^2G^K_{0mm}G^R_{0pp}  && -u_kv_k^*G^K_{0mm}G^R_{0mm} 
    \end{bmatrix}\\
\end{split}
\eeq
\beq
\begin{split}
   &\bar{G}^A_0=\frac{\delta G^A_0(\delq)}{\delta \delq}|_{\delq=0}=-G^A_0\frac{\delta G^{q^{-1}}_0}{\delta \delq}|_{\delq=0}G^K_0=-
     \begin{bmatrix}
    G^A_{0pp} && 0\\
    0 && G^A_{0mm}
    \end{bmatrix}
     \begin{bmatrix}
    -u_kv_k^* && -u_k^2\\
    (v_k^*)^2 && u_kv_k^*
    \end{bmatrix}
     \begin{bmatrix}
    G^K_{0pp} && 0\\
    0 && G^K_{0mm}
    \end{bmatrix}\\
    &=\begin{bmatrix}
  u_kv_k^*G^A_{0pp}G^K_{0pp} && u_k^2G^A_{0pp}G^K_{0mm} \\
    -(v_k^*)^2G^A_{0mm}G^K_{0pp}  && -u_kv_k^*G^A_{0mm}G^K_{0mm} 
    \end{bmatrix}\\
\end{split}
\eeq

\beq
\begin{split}
    &\bar{G}^q_0=\frac{\delta G^q_0(\delq)}{\delta \delq}|_{\delq=0}=-G^A_0\frac{\delta G^{q^{-1}}_0}{\delta \delq}|_{\delq=0}G^R_0=-
     \begin{bmatrix}
    G^A_{0pp} && 0\\
    0 && G^A_{0mm}
    \end{bmatrix}
     \begin{bmatrix}
    -u_kv_k^* && -u_k^2\\
    (v_k^*)^2 && u_kv_k^*
    \end{bmatrix}
     \begin{bmatrix}
    G^R_{0pp} && 0\\
    0 && G^R_{0mm}
    \end{bmatrix}\\
    &=\begin{bmatrix}
  u_kv_k^*G^A_{0pp}G^R_{0pp} && u_k^2G^A_{0pp}G^R_{0mm} \\
    -(v_k^*)^2G^A_{0mm}G^R_{0pp}  && -u_kv_k^*G^A_{0mm}G^R_{0mm} 
    \end{bmatrix}\\
\end{split}
\eeq

\beq
\begin{split}
    &\bar{G}^K_0=\frac{\delta G^K_0(\delq)}{\delta \delq}|_{\delq=0}=-G^R_0\frac{\delta G^{q^{-1}}_0}{\delta \delq}|_{\delq=0}G^A_0-G^K_0\frac{\delta G^{q^{-1}}_0}{\delta \delq}|_{\delq=0}G^K_0\\
    &=-
     \begin{bmatrix}
    G^R_{0pp} && 0\\
    0 && G^R_{0mm}
    \end{bmatrix}
     \begin{bmatrix}
    -u_kv_k^* && -u_k^2\\
    (v_k^*)^2 && u_kv_k^*
    \end{bmatrix}
     \begin{bmatrix}
    G^A_{0pp} && 0\\
    0 && G^A_{0mm}
    \end{bmatrix}
    -
     \begin{bmatrix}
    G^K_{0pp} && 0\\
    0 && G^K_{0mm}
    \end{bmatrix}
     \begin{bmatrix}
    -u_kv_k^* && -u_k^2\\
    (v_k^*)^2 && u_kv_k^*
    \end{bmatrix}
     \begin{bmatrix}
    G^K_{0pp} && 0\\
    0 && G^K_{0mm}
    \end{bmatrix}\\
    &=\begin{bmatrix}
  u_kv_k^*(G^R_{0pp}G^A_{0pp}+G^K_{0pp}G^K_{0pp}) && u_k^2(G^R_{0pp}G^A_{0mm}+G^K_{0pp}G^K_{0mm}) \\
    -(v_k^*)^2(G^R_{0mm}G^A_{0pp}+G^K_{0mm}G^K_{0pp})  && -u_kv_k^*(G^R_{0mm}G^A_{0mm}+G^K_{0mm}G^K_{0mm}) 
    \end{bmatrix}.\\
\end{split}
\eeq
Evaluating the diagrams in \cref{sfig:selfenergyqdot} we get
\beq
\begin{split}
    &\bar{\Sigma}^R_{\alpha\beta}(k,\omega)\\
    &=i\sum_{\lambda_1,\lambda_2=\{p,m\}}\sum_{k'}\int\frac{d\omega'}{2\pi}\tilde{g}^{\alpha\lambda_1}_{k,k'}\tilde{g}^{\lambda_2\beta}_{k',k}\left[\bar{G}^R_{0\lambda_1\lambda_2}(k',\omega')D^K_0(k'-k,\omega'-\omega)+\bar{G}^K_{0\lambda_1\lambda_2}(k',\omega')D^A_0(k'-k,\omega'-\omega)\right.\\
    &\left.~~~~~~~~~~~~~~~~~~~~~~~~~~~~~~~~~~~~~~~~~~~~~~~~~~~~~~~~~~~~~~~~+\bar{G}^q_{0\lambda_1\lambda_2}(k',\omega')D^R_0(k'-k,\omega'-\omega)\right]\\
    &=i\sum_{k'}\int\frac{d\omega'}{2\pi}\tilde{g}^{\alpha p}_{k,k'}\tilde{g}^{p\beta}_{k',k}\left[\bar{G}^R_{0pp}D^K_0+\bar{G}^K_{0pp}D^A_0+\bar{G}^q_{0pp}D^R_0\right]+i\sum_{k'}\int\frac{d\omega'}{2\pi}\tilde{g}^{\alpha p}_{k,k'}\tilde{g}^{m\beta}_{k',k}\left[\bar{G}^R_{0pm}D^K_0+\bar{G}^K_{0pm}D^A_0+\bar{G}^q_{0pm}D^R_0\right]\\
    &+i\sum_{k'}\int\frac{d\omega'}{2\pi}\tilde{g}^{\alpha m}_{k,k'}\tilde{g}^{p\beta}_{k',k}\left[\bar{G}^R_{0mp}D^K_0+\bar{G}^K_{0mp}D^A_0+\bar{G}^q_{0mp}D^R_0\right]+i\sum_{k'}\int\frac{d\omega'}{2\pi}\tilde{g}^{\alpha m}_{k,k'}\tilde{g}^{m\beta}_{k',k}\left[\bar{G}^R_{0mm}D^K_0+\bar{G}^K_{0mm}D^A_0+\bar{G}^q_{0mm}D^R_0\right]\\
      &=i\sum_{k'}\int\frac{d\omega'}{2\pi}\frac{\delc^*}{2E_{k'}}\tilde{g}^{\alpha p}_{k,k'}\tilde{g}^{p\beta}_{k',k}\left[G^K_{0pp}G^R_{0pp}D^K_0+(G^R_{0pp}G^A_{0pp}+G^K_{0pp}G^K_{0pp})D^A_0+G^A_{0pp}G^R_{0pp}D^R_0\right]\\  
      &+i\sum_{k'}\int\frac{d\omega'}{2\pi}\frac{1}{2}\left(1+\frac{\xi_{k'}}{E_{k'}}\right)\tilde{g}^{\alpha p}_{k,k'}\tilde{g}^{m\beta}_{k',k}\left[G^K_{0pp}G^R_{0mm}D^K_0+(G^R_{0pp}G^A_{0mm}+G^K_{0pp}G^K_{0mm})D^A_0+G^A_{0pp}G^R_{0mm}D^R_0\right]\\ 
      &-i\sum_{k'}\int\frac{d\omega'}{2\pi}\frac{\delc^*}{2\delc}\left(1-\frac{\xi_{k'}}{E_{k'}}\right)\tilde{g}^{\alpha m}_{k,k'}\tilde{g}^{p\beta}_{k',k}\left[G^K_{0mm}G^R_{0pp}D^K_0+(G^R_{0mm}G^A_{0pp}+G^K_{0mm}G^K_{0pp})D^A_0+G^A_{0mm}G^R_{0pp}D^R_0\right]\\ 
      &-i\sum_{k'}\int\frac{d\omega'}{2\pi}\frac{\delc^*}{2E_{k'}}\tilde{g}^{\alpha m}_{k,k'}\tilde{g}^{m\beta}_{k',k}\left[G^K_{0mm}G^R_{0mm}D^K_0+(G^R_{0mm}G^A_{0mm}+G^K_{0mm}G^K_{0mm})D^A_0+G^A_{0mm}G^R_{0mm}D^R_0\right].\\
\end{split}
\eeq
 Note that for brevity we have omitted the momentum and energy arguments of the Green's functions. It is easily understood that all the fermionic Green's functions have the argument $(k',\omega')$ and the photonic Green's function carry $(k'-k,\omega'-\omega)$ as the argument.
 Similarly we also get 
 \beq
\begin{split}
    &\bar{\Sigma}^A_{\alpha\beta}(k,\omega)\\
    &=i\sum_{\lambda_1,\lambda_2=\{p,m\}}\sum_{k'}\int\frac{d\omega'}{2\pi}\tilde{g}^{\alpha\lambda_1}_{k,k'}\tilde{g}^{\lambda_2\beta}_{k',k}\left[\bar{G}^A_{0\lambda_1\lambda_2}(k',\omega')D^K_0(k'-k,\omega'-\omega)+\bar{G}^K_{0\lambda_1\lambda_2}(k',\omega')D^R_0(k'-k,\omega'-\omega)\right.\\
    &\left.~~~~~~~~~~~~~~~~~~~~~~~~~~~~~~~~~~~~~~~~~~~~~~~~~~~~~~~~~~~~~~~~+\bar{G}^q_{0\lambda_1\lambda_2}(k',\omega')D^A_0(k'-k,\omega'-\omega)\right]\\
    &=i\sum_{k'}\int\frac{d\omega'}{2\pi}\tilde{g}^{\alpha p}_{k,k'}\tilde{g}^{p\beta}_{k',k}\left[\bar{G}^A_{0pp}D^K_0+\bar{G}^K_{0pp}D^R_0+\bar{G}^q_{0pp}D^A_0\right]+i\sum_{k'}\int\frac{d\omega'}{2\pi}\tilde{g}^{\alpha p}_{k,k'}\tilde{g}^{m\beta}_{k',k}\left[\bar{G}^A_{0pm}D^K_0+\bar{G}^K_{0pm}D^R_0+\bar{G}^q_{0pm}D^A_0\right]\\
    &+i\sum_{k'}\int\frac{d\omega'}{2\pi}\tilde{g}^{\alpha m}_{k,k'}\tilde{g}^{p\beta}_{k',k}\left[\bar{G}^A_{0mp}D^K_0+\bar{G}^K_{0mp}D^R_0+\bar{G}^q_{0mp}D^A_0\right]+i\sum_{k'}\int\frac{d\omega'}{2\pi}\tilde{g}^{\alpha m}_{k,k'}\tilde{g}^{m\beta}_{k',k}\left[\bar{G}^A_{0mm}D^K_0+\bar{G}^K_{0mm}D^R_0+\bar{G}^q_{0mm}D^A_0\right]\\
      &=i\sum_{k'}\int\frac{d\omega'}{2\pi}\frac{\delc^*}{2E_{k'}}\tilde{g}^{\alpha p}_{k,k'}\tilde{g}^{p\beta}_{k',k}\left[G^A_{0pp}G^K_{0pp}D^K_0+(G^R_{0pp}G^A_{0pp}+G^K_{0pp}G^K_{0pp})D^R_0+G^A_{0pp}G^R_{0pp}D^A_0\right]\\  
      &+i\sum_{k'}\int\frac{d\omega'}{2\pi}\frac{1}{2}\left(1+\frac{\xi_{k'}}{E_{k'}}\right)\tilde{g}^{\alpha p}_{k,k'}\tilde{g}^{m\beta}_{k',k}\left[G^A_{0pp}G^K_{0mm}D^K_0+(G^R_{0pp}G^A_{0mm}+G^K_{0pp}G^K_{0mm})D^R_0+G^A_{0pp}G^R_{0mm}D^A_0\right]\\ 
      &-i\sum_{k'}\int\frac{d\omega'}{2\pi}\frac{\delc^*}{2\delc}\left(1-\frac{\xi_{k'}}{E_{k'}}\right)\tilde{g}^{\alpha m}_{k,k'}\tilde{g}^{p\beta}_{k',k}\left[G^A_{0mm}G^K_{0pp}D^K_0+(G^R_{0mm}G^A_{0pp}+G^K_{0mm}G^K_{0pp})D^R_0+G^A_{0mm}G^R_{0pp}D^A_0\right]\\ 
      &-i\sum_{k'}\int\frac{d\omega'}{2\pi}\frac{\delc^*}{2E_{k'}}\tilde{g}^{\alpha m}_{k,k'}\tilde{g}^{m\beta}_{k',k}\left[G^A_{0mm}G^K_{0mm}D^K_0+(G^R_{0mm}G^A_{0mm}+G^K_{0mm}G^K_{0mm})D^R_0+G^A_{0mm}G^R_{0mm}D^A_0\right].\\
\end{split}
\eeq
and 
\beq
\begin{split}
    &\bar{\Sigma}^q_{\alpha\beta}(k,\omega)\\
    &=i\sum_{\lambda_1,\lambda_2=\{p,m\}}\sum_{k'}\int\frac{d\omega'}{2\pi}\tilde{g}^{\alpha\lambda_1}_{k,k'}\tilde{g}^{\lambda_2\beta}_{k',k}\\&\times\left[\bar{G}^q_{0\lambda_1\lambda_2}(k',\omega')D^K_0(k'-k,\omega'-\omega)+\bar{G}^R_{0\lambda_1\lambda_2}(k',\omega')D^R_0(k'-k,\omega'-\omega)+\bar{G}^A_{0\lambda_1\lambda_2}(k',\omega')D^A_0(k'-k,\omega'-\omega)\right]\\
 &=i\sum_{k'}\int\frac{d\omega'}{2\pi}\left\{\tilde{g}^{\alpha p}_{k,k'}\tilde{g}^{p\beta}_{k',k}\left[\bar{G}^q_{0pp}D^K_0+\bar{G}^R_{0pp}D^R_0+\bar{G}^A_{0pp}D^A_0\right]+\tilde{g}^{\alpha p}_{k,k'}\tilde{g}^{m\beta}_{k',k}\left[\bar{G}^q_{0pm}D^K_0+\bar{G}^R_{0pm}D^R_0+\bar{G}^A_{0pm}D^A_0\right]\right.\\
&~~~~~~~~~~~~~~+\left.\tilde{g}^{\alpha m}_{k,k'}\tilde{g}^{p\beta}_{k',k}\left[\bar{G}^q_{0mp}D^K_0+\bar{G}^R_{0mp}D^R_0+\bar{G}^A_{0mp}D^A_0\right]+\tilde{g}^{\alpha m}_{k,k'}\tilde{g}^{m\beta}_{k',k}\left[\bar{G}^q_{0mm}D^K_0+\bar{G}^R_{0mm}D^R_0+\bar{G}^A_{0mm}D^A_0\right]\right\}\\
   &=i\sum_{k'}\int\frac{d\omega'}{2\pi}\frac{\delc^*}{2E_{k'}}\tilde{g}^{\alpha p}_{k,k'}\tilde{g}^{p\beta}_{k',k}\left[G^A_{0pp}G^R_{0pp}D^K_0+G^K_{0pp}G^R_{0pp}D^R_0+G^A_{0pp}G^K_{0pp}D^A_0\right]\\
      &+i\sum_{k'}\int\frac{d\omega'}{2\pi}\frac{1}{2}\left(1+\frac{\xi_{k'}}{E_{k'}}\right)\tilde{g}^{\alpha p}_{k,k'}\tilde{g}^{m\beta}_{k',k}\left[G^A_{0pp}G^R_{0mm}D^K_0+G^K_{0pp}G^R_{0mm}D^R_0+G^A_{0pp}G^K_{0mm}D^A_0\right]\\
      &-i\sum_{k'}\int\frac{d\omega'}{2\pi}\frac{\delc^*}{2\delc}\left(1-\frac{\xi_{k'}}{E_{k'}}\right)\tilde{g}^{\alpha m}_{k,k'}\tilde{g}^{p\beta}_{k',k}\left[G^A_{0mm}G^R_{0pp}D^K_0+G^K_{0mm}G^R_{0pp}D^R_0+G^A_{0mm}G^K_{0pp}D^A_0\right]\\
      &-i\sum_{k'}\int\frac{d\omega'}{2\pi}\frac{\delc^*}{2E_{k'}}\tilde{g}^{\alpha m}_{k,k'}\tilde{g}^{m\beta}_{k',k}\left[G^A_{0mm}G^R_{0mm}D^K_0+G^K_{0mm}G^R_{0mm}D^R_0+G^A_{0mm}G^K_{0mm}D^A_0\right].\\
\end{split}
\eeq

\subsection{The non-thermal gap equation for material in cavity}
We want to calculate the perturbative correction to the gap equation in the small $g_0$ limit. In this limit, the gap equation in \cref{seq:gapeqgen} can be written as 
\beq\label{gapeqgen_per1}
\begin{split}
    \nonumber&i\sum_{k}\int \frac{d\omega}{2\pi}~\left\{u_kv_k^*\left[(G^R_{0pp}-G^A_{0pp})-(G^R_{0mm}-G^A_{0mm})\right]F_0\right.\\
    &~~~~~~~~~~~~~~~~~~~~+u_kv_k^*\left[(G^R_{0pp}-G^A_{0pp})\delta F_{pp}-(G^R_{0mm}-G^A_{0mm})\delta F_{mm}\right]\\
       \nonumber &~~~~~~~~~~~~~~~~~+\left[u_kv_k^*(\delta G^R_{pp}-\delta G^A_{pp})-u_kv_k^*(\delta G^R_{mm}-\delta G^A_{mm})-(v_k^*)^2(\delta G^R_{pm}-\delta G^A_{pm})+u_k^2(\delta G^R_{mp}-\delta G^A_{mp})\right]F_{0}\\
    \nonumber &~~~~~~~~~~~~~~~~~+\left.\left[-(v_k^*)^2(G^R_{0pp}-G^A_{0mm})F_{pm}+u_k^2(G^R_{0mm}-G^A_{0pp})F_{mp}\right]\right\}\\
 &+i\sum_{k}\int \frac{d\omega}{2\pi}\sum_{\alpha=\{p,m\}}\left[G^R_{0\alpha\alpha}\bar{\Sigma}^R_{\alpha\alpha}+G^A_{0\alpha\alpha}\bar{\Sigma}^A_{\alpha\alpha}+\left(G^R_{0\alpha\alpha}-G^A_{0\alpha\alpha}\right)F_0\bar{\Sigma}^q_{\alpha\alpha}\right]=\frac{\Delta^*_{cl}}{\kappa}
\end{split}
\eeq
where $\delta G^{R/A}_{\alpha\beta}=G^{R/A}_{0\alpha\alpha}\Sigma^{R/A}_{\alpha\beta}G^{R/A}_{0\beta\beta}$ is the perturbative correction to the Green's functions. 
Noting that 
\beq
u_k=\sqrt{\frac{1}{2}\left(1+\frac{\xi_k}{E_k}\right)} ~~~\mathrm{and}~~~ v_k=\frac{\delc}{|\delc|}\sqrt{\frac{1}{2}\left(1-\frac{\xi_k}{E_k}\right)}
\eeq
we get 
\beq\label{gapeqgen_per2}
\begin{split}
    \nonumber&\sum_{k}\left\{\frac{F_0(E_k)}{E_k}+\left[\frac{\delta F_{pp}(k,E_k)-\delta F_{mm}(k,-E_k)}{2E_k}\right]\right.\\
       \nonumber &~~~~~+i\int\frac{d\omega}{2\pi}\left.\left[-\frac{1}{2}\left(1-\frac{\xi_k}{E_k}\right)\frac{(G^R_{0pp}-G^A_{0mm})}{\delc}F_{pm}
       +\frac{1}{2}\left(1+\frac{\xi_k}{E_k}\right)\frac{(G^R_{0mm}-G^A_{0pp})}{\delc^*}F_{mp}\right]\right\}\\
       \nonumber &+i\int\frac{d\omega}{2\pi}\left[\frac{(\delta G^R_{pp}-\delta G^A_{pp})-(\delta G^R_{mm}-\delta G^A_{mm})}{2E_k}\right.\\
       &~~~~~~~~~~~~~~~~~~~~\left.-\frac{1}{2}\left(1-\frac{\xi_k}{E_k}\right)\frac{(\delta G^R_{pm}-\delta G^A_{pm})}{\delc}+\frac{1}{2}\left(1+\frac{\xi_k}{E_k}\right)\frac{(\delta G^R_{mp}-\delta G^A_{mp})}{\delc^*}\right]F_{0}\\
 &+i\sum_{k}\int \frac{d\omega}{2\pi}\sum_{\alpha=\{p,m\}}\frac{1}{\delc^*}\left[G^R_{0\alpha\alpha}\bar{\Sigma}^R_{\alpha\alpha}+G^A_{0\alpha\alpha}\bar{\Sigma}^A_{\alpha\alpha}+\left(G^R_{0\alpha\alpha}-G^A_{0\alpha\alpha}\right)F_0\bar{\Sigma}^q_{\alpha\alpha}\right]=\frac{1}{\kappa}
\end{split}
\eeq
From now on we write $\Delta=\delc$ as this is the physically relevant one. Furthermore, we choose the gauge where $\Delta$ is real without loss of generality.
We make a quasi-particle approximation i.e. $\gamma\ll\tcry,\tcav,\Delta$ and get for the gap equation
\beq
\begin{split}
&\sum_{k}\frac{F_0(E_k)+\delta F(k,E_k)}{E_k}=\frac{1}{\kappa},\\
\end{split}
\eeq
where
\beq
\begin{split}
&\delta F(k,E_k)=-\frac{1}{\gamma}\sum_{k'}\sum_{q}\int d\Omega~(g^{+}_{k,k'})^2\left[\left(1+\frac{\xi_{k}\xi_{k'}+\eta\Delta^2}{E_{k}E_{k'}}\right)\mathrm{Im}D^R_0(q,\Omega)\tilde H_0(E_k,E_{k}+\Omega)\delta(\Omega-(E_{k'}-E_k))\right.\\
  &~~~~~~~~~~~~~~~~~~~~~~~~~~~~~~~~~~~~~~~~~-\left.\left(1-\frac{\xi_{k}\xi_{k'}+\eta\Delta^2}{E_{k}E_{k'}}\right)\mathrm{Im}D^R_0(q,\Omega)\tilde H_0(-E_k,-E_{k}+\Omega)\delta(\Omega-(E_{k'}+E_k))\right]\delta(q-(k'-k)).\\
\end{split}
\eeq

In the above equation, the delta functions represent energy and momentum conservations. However, in real materials, the presence of static impurities causes the momentum to be not conserved. Hence, the corresponding delta function can be replaced by a constant $(\Lambda_F/\tau_{\rm el})^{-1}$, where $\Lambda_F=\frac{2}{\pi J}$ is the density of states at the Fermi surface and $\tau_{\rm el}$ is the elastic scattering rate \cite{Galitski2019, MattisBardeen1958}. 
Absorbing the factor $(\Lambda_F/\tau_{\rm el})$ in $\gamma$, we get 
\beq
\begin{split}
&\delta F(k,E_k)=-\frac{1}{\gamma}\sum_{k'}\sum_{q}(g^{+}_{k,k'})^2\left[\left(1+\frac{\xi_{k}\xi_{k'}+\eta\Delta^2}{E_{k}E_{k'}}\right)\mathrm{Im}D^R_0(q,E_{k'}-E_k)\tilde H_0(E_k,E_{k'})\right.\\
  &~~~~~~~~~~~~~~~~~~~~~~~~~~~~~~~~~~~~~~~~~-\left.\left(1-\frac{\xi_{k}\xi_{k'}+\eta\Delta^2}{E_{k}E_{k'}}\right)\mathrm{Im}D^R_0(q,E_{k'}+E_k)\tilde H_0(-E_k,E_{k'})\right].\\
\end{split}
\eeq
Furthermore we convert the momentum sums to energy integration with the help of a density of states for the gapped quasi-particles $\Lambda(E)=\frac{2}{\pi}\,
  \frac{E}{\sqrt{E^{2}-\Delta^{2}}\sqrt{J^{2}+\Delta^{2}-E^{2}}}$ and use Fermi surface averaged values for explicitly momentum dependent quantities \cite{Galitski2019}. Also note the speed of light is much larger than the Fermi velocity, hence it suffices to consider only the $q=0$ photons. We define the shorthand notation $\mathrm{Im}D^R_0(\omega)=\mathrm{Im}D^R_0(0,\omega)$.
  Incorporating all of these we arrive at 
\begin{equation}
\begin{split}
&\int_{\Delta}^{\sqrt{J^{2}+\Delta^{2}}} dE\,
  \Lambda(E)\,
  \frac{F_{0}(E)+\delta F(E)}{E}
  \;=\;
  \frac{1}{\kappa}~,~\text{where}\\
&\delta F(E)=-\frac{g_0^2}{\gamma}\int_{\Delta}^{\sqrt{J^{2}+\Delta^{2}}} dE'\Lambda(E')\left[\left(1+\frac{\eta\Delta^2}{EE'}\right)\mathrm{Im}D^R_0(E'-E)\tilde H_0(E,E')-\left(1-\frac{\eta\Delta^2}{EE'}\right)\mathrm{Im}D^R_0(E'+E)\tilde H_0(-E,E')\right].\\
\end{split}
\end{equation}
We can also re-write the gap equation in terms of the Fermi and Bose functions as 
\begin{equation}
\begin{split}
&\int_{\Delta}^{\sqrt{J^{2}+\Delta^{2}}} dE\,
  \Lambda(E)\,
  \frac{1-2[n_{0}(E)+\delta n(E)]}{E}
  \;=\;
  \frac{1}{\kappa}~,~\text{where}\\
&\delta n(E)=\frac{2g_0^2}{\gamma}\int_{\Delta}^{\sqrt{J^{2}+\Delta^{2}}} dE'\Lambda(E')\left[\left(1+\frac{\eta\Delta^2}{EE'}\right)\mathrm{Im}D^R_0(E-E')H_0(E,E')+\left(1-\frac{\eta\Delta^2}{EE'}\right)\mathrm{Im}D^R_0(E+E')H_0(E,-E')\right]~\text{and}\\
&H_0(\omega,\omega')=\left[N_{\tcav}(\omega-\omega')-N_{\tcry}(\omega-\omega')\right]\left[n_0(\omega)-n_0(\omega')\right].
\end{split}
\end{equation}

\section{Ginzburg-Landau theory of the phase transition}

We start with the full gap equation expressed in terms of the energy $E$
\begin{equation}
\int_{\Delta}^{\sqrt{J^{2}+\Delta^{2}}}
  dE\,
  \Lambda(E)\,
  \frac{F_{0}(E)+\delta F(E)}{E}
  \;=\;
  \frac{1}{\kappa}\, .
\end{equation}
with 
\begin{equation}
\Lambda(E)=\frac{2}{\pi}\,
  \frac{E}{\sqrt{E^{2}-\Delta^{2}}\sqrt{J^{2}+\Delta^{2}-E^{2}}}\, .
\end{equation}

Set $E^{2}=x^{2}+\Delta^{2}$.
Then $E=\Delta\!\Rightarrow\!x=0$,
$E=\sqrt{J^{2}+\Delta^{2}}\!\Rightarrow\!x=J$,
and $E\,dE=x\,dx$:
\begin{equation}
\frac{2}{\pi}\int_{0}^{J}
  \frac{dx}{\sqrt{J^{2}-x^{2}}\sqrt{\Delta^2+x^2}}\,
  \Bigl[
       F_{0}\!\bigl(\sqrt{x^{2}+\Delta^{2}}\bigr)+
       \delta F\!\bigl(\sqrt{x^{2}+\Delta^{2}}\bigr)
  \Bigr]
  \;=\;\frac{1}{\kappa}\, .
\end{equation}
We now approximate as in equilibrium the metallic density of states $\frac{2}{\pi} \frac{1}{\sqrt{J^2-x^2}}$ as a constant with main contribution from the fermi energy $\varepsilon\simeq\varepsilon_{F}$ or here $x=0$, we denote it
$D\!\bigl(\varepsilon_{F}\bigr)=D_F=\frac{2}{\pi J}$.

\beq\begin{split}
\delta F\!\bigl(\sqrt{x^{2}+\Delta^{2}}\bigr)
  &= -\frac{g_{0}^2}{\gamma}
     \int_{0}^{J} \!
       d\tilde x\;
       D_F
       \Bigl\{
         \Bigl(1+\eta\frac{\Delta^{2}}{E(x)\,E(\tilde x)}\Bigr)
           \im D_{0}^{R}\!\bigl(E(\tilde x)-E( x)\bigr)\,
           \tilde H_0\!\bigl(E(x),E(\tilde x)\bigr)
         \\[2pt]
         &\hspace{2.5cm}
         -\Bigl(1-\eta\frac{\Delta^{2}}{E(x)\,E(\tilde x)}\Bigr)
           \im D_{0}^{R}\!\bigl(E(\tilde x)+E(x)\bigr)\,
           \tilde H_0\!\bigl(-E(x),E(\tilde x)\bigr)
       \Bigr\}\!,
\end{split}\eeq
with $E(x)=\sqrt{x^{2}+\Delta^{2}}$. We specialize now sharp photons, meaning $\gamma_{B}\!\to\!0$ allows us to express the imaginary parts as delta functions.

\begin{equation}
\im D_{0}^{R}(\omega)\Bigl|_{\gamma_B\to 0}
  \;=\;
  -\frac{1}{4}\Bigl[\delta\!\bigl(\omega-\nu_{0}\bigr)
                    -\delta\!\bigl(\omega+\nu_{0}\bigr)\Bigr].
\end{equation}
We now want to relate the condition for the non-zero value of the delta function in $E$ to the condition for $\tilde{x}$.
\beq\begin{split}
\im D_{0}^{R}\!\bigl(E(x)\!\pm\!E(\tilde x)\bigr)
  =-\frac{1}{4}\Bigl[
      \delta\!\bigl(E(x)\!\pm\!E(\tilde x)-\nu_{0}\bigr)
      -\delta\!\bigl(E(x)\!\pm\!E(\tilde x)+\nu_{0}\bigr)
     \Bigr].
\end{split}\eeq

Define $f_{s}(x,\tilde x)=E(x)+pE(\tilde x)+s\nu_{0}$,($s,p=\pm1$).
Zeros for the tilde variables obey $\tilde x^{2}+\Delta^{2}=\bigl(p\sqrt{x^{2}+\Delta^{2}}+s\,\nu_{0}\bigr)^{2}$ and
\(
\partial f_{s}/\partial \tilde x = p\tilde x/E(\tilde x).
\)
Hence
\[
\delta\!\bigl(E(x)+p E(\tilde x)-\nu_{0}\bigr)
  =\frac{E(\tilde x)}{\tilde x}\,\delta(\tilde x-\tilde x_{sp})\; ,
\]
Where we have defined
\beq\begin{split}
   \tilde x_{sp} = \sqrt{\bigl(\sqrt{x^{2}+\Delta^{2}}+ps\,\nu_{0}\bigr)^{2}-\Delta^{2}} 
\end{split}\eeq

%%%%%%%%%%%%%%%%%%%%%%%%%%%%%%%%%%%%%%%%%%%%%%%%%%%%%%%%%%%%%%%%%%%%%%%%
% Page 3 – algebra with photon kernels
%%%%%%%%%%%%%%%%%%%%%%%%%%%%%%%%%%%%%%%%%%%%%%%%%%%%%%%%%%%%%%%%%%%%%%%%
Let us now analyze what the delta functions mean. Physically when we have $\tilde E -E -\nu_0=0$ it can be interpreted as a cavity photon being absorbed such that if before the quasiparticle had energy $E$ it will then have energy $\tilde{E}+\nu_0$. Similarly for $\tilde E -E +\nu_0=0$ it will emit a photon and thus reduce its energy to $E-\nu_0$. Now since a superconductor or CDW state has a gap and the energies are only positive this can only occur if $E-\nu_0>0$.
Since the energies are positive we see that $\tilde E +E +\nu_0=0$ can never occur for positive cavity frequency. Moreover, the process with  $\tilde E +E -\nu_0=0$ can occur only for $\nu_0-E>0$. Going back to the integral in energy we can see this from the integral being expressed as $\int_0^{\nu_0}dE$ with vanishing width.  We may therefor focus only on the process with $\tilde E -E -\nu_0=0$ and $\tilde E -E +\nu_0=0$ and $\tilde E +E -\nu_0=0$. The distribution will then become:

\beq\begin{split}
\delta F\!\bigl(\sqrt{x^{2}+\Delta^{2}}\bigr)
  &= \frac{g_{0}^{2}\,D_{F}}{4\gamma}
     \int_{0}^{J} \!d\tilde x\;
     \Bigl[
        \Bigl(\!1+\eta\frac{\Delta^{2}}{E(x)\,E(\tilde x)}\Bigr)
       \frac{E(\tilde x)}{\tilde x} \tilde H_0\!\bigl(E(x),E(x)+\nu_{0}\bigr)\delta(\tilde x - \tilde x_+)
        \\[3pt]
     &\hspace{1cm}
        -\Bigl(\!1+\eta\frac{\Delta^{2}}{E(x)\,E(\tilde x)}\Bigr)
       \frac{E(\tilde x)}{\tilde x}  \tilde H_0\!\bigl(E(x),E( x)-\nu_{0}\bigr) \delta(\tilde x - \tilde x_-)\Theta(E(x)-\nu_0)
     \Bigr]\\[3pt]
     &\hspace{1cm}
        -\Bigl(\!1-\eta\frac{\Delta^{2}}{E(x)\,E(\tilde x)}\Bigr)
       \frac{E(\tilde x)}{\tilde x}  \tilde H_0\!\bigl(-E(x),\nu_{0}-E(x)\bigr) \delta(\tilde x - \tilde x_-)\Theta(\nu_0-E(x))
     \Bigr],
\end{split}\eeq
Simplifying we have
\beq\begin{split}
\delta F\!\bigl(\sqrt{x^{2}+\Delta^{2}}\bigr)
  &= \frac{g_{0}^{2}\,D_{F}}{4\gamma}
     \Bigl[
        \Bigl(\!E(x)+\nu_0+\eta\frac{\Delta^{2}}{E(x)}\Bigr)
       \frac{1}{\tilde x_+} \tilde H_0\!\bigl(E(x),E(x)+\nu_{0}\bigr)
        \\[3pt]
     &\hspace{1cm}
        -\Bigl(\!E(x)-\nu_0+\eta\frac{\Delta^{2}}{E(x)}\Bigr)
       \frac{1}{\tilde x_-}  \tilde H_0\!\bigl(E(x),E(x)-\nu_{0}\bigr) \Theta(E(x)-\nu_0)
     \Bigr]\\[3pt]
     &\hspace{1cm}
        -\Bigl(\! \nu_0-E(x)-\eta\frac{\Delta^{2}}{E(x)}\Bigr)
       \frac{1}{\tilde x_-}  \tilde H_0\!\bigl(-E(x),\nu_{0}-E(x)\bigr) \Theta(\nu_0-E(x))
     \Bigr]
\end{split}\eeq
We now write 
\beq\begin{split}
\int dE\,\nu(E)\,\frac{\delta F(E)}{E} &= \frac{g_{0}^{2}\,D_{F}^2}{4\gamma} \int_0^J dx \Bigl[
        \Bigl(\!1+\dfrac{\nu_0}{E(x)}+\eta\frac{\Delta^{2}}{E^2(x)}\Bigr)
       \frac{1}{\tilde x_+} \tilde H_0\!\bigl(E(x),E(x)+\nu_{0}\bigr)
        \\[3pt]
     &\hspace{1cm}
        -\Bigl(\!1-\dfrac{\nu_0}{E(x)}+\eta\frac{\Delta^{2}}{E^2(x)}\Bigr)
       \frac{1}{\tilde x_-}  \tilde H_0\!\bigl(E(x),E(x)-\nu_{0}\bigr) \Theta(E(x)-\nu_0)
     \Bigr]\\[3pt]
     &\hspace{1cm}
        -\Bigl(\! \dfrac{\nu_0}{E(x)}-1-\eta\frac{\Delta^{2}}{E^2(x)}\Bigr)
       \frac{1}{\tilde x_-}  \tilde H_0\!\bigl(-E(x),\nu_{0}-E(x)\bigr) \Theta(\nu_0-E(x))
     \Bigr],
\end{split}\eeq
Using now that
\(
\tilde H_0(\omega,\omega')=\bigl(F_{0}(\omega')-F_{0}(\omega)\bigr)
  \bigl(B_{\tcav}(\omega'-\omega)-B_{\tcry}(\omega'-\omega)\bigr)
\)
we can write 
\beq\begin{split}
    \tilde H_0\!\bigl(E(x),E(x)+\nu_{0}\bigr) = \bigl(F_{0}(E(x)+\nu_{0})-F_{0}(E(x))\bigr)
  \bigl(B_{\tcav}(\nu_0)-B_{\tcry}(\nu_0)\bigr) \\
  = \tilde B(\nu_0) (F_{0}(E(x)+\nu_{0})-F_{0}(E(x)))
\end{split}\eeq
We defined $\tilde B(\nu_0)=B_{\tcav}(\nu_0)-B_{\tcry}(\nu_0)$. Similarly we have:

\beq\begin{split}
    \tilde H_0\!\bigl(E(x),E(x)-\nu_{0}\bigr) = \bigl(F_{0}(E(x)-\nu_{0})-F_{0}(E(x))\bigr)
  \bigl(B_{\tcav}(-\nu_0)-B_{\tcry}(-\nu_0)\bigr) \\
  = \tilde B(\nu_0) (F_{0}(E(x))-F_{0}(E(x)-\nu_{0}))
\end{split}\eeq
and also 

\beq\begin{split}
    \tilde H_0\!\bigl(-E(x),\nu_{0}-E(x)\bigr) = \bigl(F_{0}(\nu_{0}-E(x))-F_{0}(-E(x))\bigr)
  \bigl(B_{\tcav}(\nu_0)-B_{\tcry}(\nu_0)\bigr) \\
  = \tilde B(\nu_0) (F_{0}(E(x))-F_{0}(E(x)-\nu_{0}))
\end{split}\eeq
%%%%%%%%%%%%%%%%%%%%%%%%%%%%%%%%%%%%%%%%%%%%%%%%%%%%%%%%%%%%%%%%%%%%%%%%
% Page 4 – high–$T$ approximation
%%%%%%%%%%%%%%%%%%%%%%%%%%%%%%%%%%%%%%%%%%%%%%%%%%%%%%%%%%%%%%%%%%%%%%%%
Approximate now a small gap $\Delta\ll\tcry$ so that 
$F_{0}(E/2T_{\rm cry})=\tanh\bigl(E/2T_{\rm cry}\bigr)
  \simeq E/2T_{\rm cry}$

\vspace{-0.5\baselineskip}
\begin{equation}
F_{0}\!\bigl(E(x)+\nu_{0}\bigr)-F_{0}\!\bigl(E(x)\bigr)
  \simeq \frac{\nu_{0}}{2T_{\rm cry}},\qquad
F_{0}(E)-F_{0}\!\bigl(E-\nu_{0}\bigr)\simeq \frac{\nu_{0}}{2T_{\rm cry}}\, .
\end{equation}
Therefore we have then

\beq\begin{split}
\int dE\,\nu(E)\,\frac{\delta F(E)}{E} &= \frac{g_{0}^{2}\,D_{F}^2 \nu_0 \tilde B(\nu_0)}{8 \gamma T_{cry}} \int_0^J dx \Bigl[
        \Bigl(\!1+\dfrac{\nu_0}{E(x)}+\eta\frac{\Delta^{2}}{E^2(x)}\Bigr)
       \frac{1}{\tilde x_+} 
        \\[3pt]
     &\hspace{1cm}
        -\Bigl(\!1-\dfrac{\nu_0}{E(x)}+\eta\frac{\Delta^{2}}{E^2(x)}\Bigr)
       \frac{1}{\tilde x_-}  \Theta(E(x)-\nu_0)
     \Bigr]\\[3pt]
     &\hspace{1cm}
        +\Bigl(\!1- \dfrac{\nu_0}{E(x)}+\eta\frac{\Delta^{2}}{E^2(x)}\Bigr)
       \frac{1}{\tilde x_-}   \Theta(\nu_0-E(x))
     \Bigr],
\end{split}\eeq
Let us now assume that $\Delta$ is small such that we can approximate
\beq\begin{split}
    \tilde x_+ = \sqrt{\bigl(\sqrt{x^{2}+\Delta^{2}}+\,\nu_{0}\bigr)^{2}-\Delta^{2}} =\sqrt{x^2+\Delta^2+\nu_0^2+2\sqrt{x^{2}+\Delta^{2}}\nu_0-\Delta^{2}} \approx  \sqrt{x^2+\nu_0^2+2\sqrt{x^{2}}\nu_0}= \abs{x+\nu_0}
\end{split}\eeq
similarly $\tilde x_- \approx \abs{x-\nu_0}$. The integrals we must calculate are then

\beq\begin{split}
\int dE\,\nu(E)\,\frac{\delta F(E)}{E} &\approx  \frac{g_{0}^{2}\,D_{F}^2 \nu_0 \tilde B(\nu_0)}{8 \gamma T_{cry}} \int_0^J dx \Bigl[
        \Bigl(\!1+\dfrac{\nu_0}{\sqrt{x^2+\Delta^2}}+\eta\frac{\Delta^{2}}{x^2+\Delta^2}\Bigr)
       \frac{1}{x+\nu_0} 
        \\[3pt]
     &\hspace{1cm}
        -\Bigl(\!1-\dfrac{\nu_0}{\sqrt{x^2+\Delta^2}}+\eta\frac{\Delta^{2}}{x^2+\Delta^2}\Bigr)
       \frac{1}{\abs{x-\nu_0}}  \Theta(E(x)-\nu_0)
     \Bigr]\\[3pt]
     &\hspace{1cm}
        +\Bigl(\!1-\dfrac{\nu_0}{\sqrt{x^2+\Delta^2}}+\eta\frac{\Delta^{2}}{x^2+\Delta^2}\Bigr)
       \frac{1}{\abs{x-\nu_0}}  \Theta(\nu_0-E(x))
     \Bigr],
\end{split}\eeq

The first integral of the form $\int dx~1/(x\pm\nu_0)$ will only give logarithmic corrections independent of $\Delta$ which we do not care about for the gap equation expansion. The second terms with  $\int dx \ \nu_0/(E(x))$ are second order in the small cavity frequency $\nu_0\ll J,T_{cry}$, since there is already a $\nu_0$ multiplying everything. We therefore need to only worry about the terms with the energy squared in the denominator. We want to find the lowest order in $\Delta$. We notice that the integrand of 
\beq\begin{split}
    \int_0^J \dd x \dfrac{\Delta^2}{x^2+\Delta^2}\dfrac{1}{\abs{x-\nu_0}} \Theta(E(x)-\nu_0)
\end{split}\eeq
has a divergence at $x=\nu_0$ which will give the main contribution for $\nu_0\gg \Delta$ and a coefficient in front of $\dfrac{\Delta^2}{\nu_0^2+\Delta^2}$ which has as a lowest order expansion in the gap only quadratic terms. There is no contribution for small $x$ since the Heaviside forbids so. Nevertheless, looking at the other integral we notice that 
\beq\begin{split}
    \int_0^J \dd x \dfrac{\Delta^2}{x^2+\Delta^2}\dfrac{1}{\abs{x-\nu_0}} \Theta(\nu_0-E(x))
\end{split}\eeq
has an integrand also with a divergent term but in this case because of the Heaviside function we are allowed to have $x$ small since $\nu_0-\Delta>0$, thus for small $x$ we may consider the contribution to the integral for small $x$ up to the smallest scale $\Delta$
\beq\begin{split}
    \int_0^\Delta \dd x \dfrac{\Delta^2}{x^2+\Delta^2}\dfrac{1}{\abs{x-\nu_0}} \approx \int_0^\Delta \dd x \dfrac{\Delta^2}{\Delta^2}\dfrac{1}{\abs{\nu_0}} = \dfrac{\Delta}{\nu_0}
\end{split}\eeq
A similar reasoning can be applied to the integral with $1/(x+\nu_0)$, the remaining part of the integral gives logarithmic contributions or quadratic in the gap. Thus keeping only the lowest order contributions that have a gap dependence, neglecting constants, we get:

\beq\begin{split}
\int dE\,\nu(E)\,\frac{\delta F(E)}{E} &\approx \eta \frac{g_{0}^{2}\,D_{F}^2 \tilde B(\nu_0)}{4 \gamma T_{cry}} \Delta + \mathcal{O}(\Delta ^{2})
\end{split}\eeq
We now fit this result together with the gap equation and the Ginzburg-Landau analysis, let us define

\[
I(\Delta,T_{\rm cry})
  \equiv
  \int_{\Delta}^{\sqrt{J^{2}+\Delta^{2}}}
       dE\,\Lambda(E)\,\frac{F_{0}(E)+\delta F(E)}{E}
  \;=\;\frac{1}{\kappa}.
\]

For $T_{\rm cry}=\tc$ (non-eq.\ critical temperature) one has
$I(\Delta\!=\!0,\tc)=1/\kappa$.
Expand about $(\Delta,T_{\rm cry})=(0,\tc)$:
\beq\begin{split}
I(\Delta,T_{\rm cry})
  &=I(0,\tc)
    +\left.\frac{\partial I}{\partial T_{\rm cry}}\right|_{\Delta=0}
      (T_{\rm cry}-\tc)
    +\left.\frac{\partial I}{\partial\Delta}\right|_{\Delta=0}
      \Delta
    +\tfrac12\left.\frac{\partial^{2} I}{\partial\Delta^{2}}\right|_{\Delta=0}
      \Delta^{2}
    +\cdots .
\end{split}\eeq

The gap equation becomes 
\[
\left.\frac{\partial I}{\partial T_{\rm cry}}\right|_{\substack{\Delta=0\\T_{\rm cry}=\tc}}
  (T_{\rm cry}-\tc)
  +\left.\frac{\partial I}{\partial\Delta}\right|_{\Delta=0}\Delta
  +\tfrac12\left.\frac{\partial^{2} I}{\partial\Delta^{2}}\right|_{\Delta=0}
     \Delta^{2}
  \;=\;0
\]
has no linear-in-$\Delta$ contribution from equilibrium terms (they start
at $\propto\Delta^{2}$). In fact, $\tfrac12\left.\frac{\partial^{2} I^{eq}}{\partial\Delta^{2}}\right|_{\Delta=0}= -\frac{7D_F\zeta(3)}{8\pi^2(\tc)^2}$ while $\left.\frac{\partial I}{\partial T_{cry}}\right|_{\Delta=0,\ T_{\rm cry}=\tc}= \frac{-D_F}{\tc}$ so that the gap equation becomes for $T_{\text{cry}}\approx \tc$ and $\Delta\approx 0$:
\begin{equation}
\Delta\left(\frac{\tc-T_{\rm cry}}{\tc}
  -\frac{7\zeta(3)}{8\pi^{2}}\,
   \frac{\Delta^{2}}{(\tc)^{2}}
  +\eta \frac{g_{0}^{2}\,D_{F}(B_{\tcav}(\nu_0)-B_{\tc}(\nu_0))}{4 \gamma \tc}  \Delta
  +\cdots
  \;\right)=\;0.
\end{equation}
The Landau functional can be obtained by integrating this and noting a minus sign is present 
\beq\begin{split}
    \mathcal F_{\mathrm{neq}}[\Delta]= \frac{T_{\rm cry}-\tc}{2\tc}\Delta^2+\eta \frac{g_{0}^{2}\,D_{F}(B_{\tc}(\nu_0)-B_{\tcav}(\nu_0))}{12 \gamma \tc}  \Delta^3+ \frac{7\zeta(3)}{8\pi^{2}}\,
   \frac{1}{4(\tc)^{2}}\Delta^4+\cdots
\end{split}\eeq
Noting that $D_F=2/\pi J$ we get 
\beq\begin{split}
    \mathcal F_{\mathrm{neq}}[\Delta]= \frac{T_{\rm cry}-\tc}{2\tc}\Delta^2+\eta \frac{g_{0}^{2}\,(N_{\tc}(\nu_0)-N_{\tcav}(\nu_0))}{3\pi \gamma \tc}  \Delta^3+ \frac{7\zeta(3)}{8\pi^{2}}\,
   \frac{1}{4(\tc)^{2}}\Delta^4+\cdots
\end{split}\eeq
The solution of the gap equation can be approximated as
\beq
\Delta[\tcry]\approx\frac{\pi\gamma J}{g_0^2}\frac{\eta}{N_{\tcav}(\nu_0)-N_{\tc}(\nu_0)}(\tcry-\tc).
\eeq
Note that it grows linearly with $|\tcry-\tc|$ and has opposite slopes for SC and CDW (determined by $\eta$). This is in contrast to equilibrium where the gap grows as $\sqrt{\tc-\tcry}$. 
\bibliography{cavitytrans.bib}